\documentclass[a4paper,11pt]{article}
\pdfoutput=1
\usepackage{jheppub}
\usepackage{graphicx,color}
\usepackage{amsmath}
\usepackage{array}
\newcolumntype{P}[1]{>{\centering\arraybackslash}p{#1}}
\allowdisplaybreaks
\RequirePackage{ifpdf}
\usepackage{amsmath}
\usepackage{mathtools}

\usepackage{autobreak}
\usepackage{jheppub}
\usepackage{pstricks}
\usepackage[final]{pdfpages}
\usepackage{ifpdf}
\usepackage{slashed}

\usepackage{hyperref}

\usepackage{color}
\usepackage{graphics}

\usepackage{etoolbox}
\usepackage{fixmath}

\usepackage{caption}
\usepackage{subcaption}
\usepackage{amsfonts}

\usepackage{multirow}
\usepackage{epstopdf}

\usepackage{relsize}
\usepackage{float}

\usepackage[normalem]{ulem}

\usepackage{tikz}
\usetikzlibrary{positioning,arrows}
\usetikzlibrary{decorations.pathmorphing}
\usetikzlibrary{decorations.markings}
\usetikzlibrary{shapes.geometric}
\usepackage{endnotes}
\tikzset{
    vector/.style={decorate, decoration={snake}, draw},
    provector/.style={decorate, decoration={snake,amplitude=2.5pt}, draw},
    antivector/.style={decorate, decoration={snake,amplitude=-2.5pt}, draw},
    fermion/.style={draw=black,
      postaction={decorate},decoration={markings,mark=at position .55
        with {\arrow[draw=black]{>}}}},
    fermionbar/.style={draw=black, postaction={decorate},
                       decoration={markings,mark=at position .55 with {\arrow[draw=black]{<}}}},
    fermionnoarrow/.style={draw=black},
    gluon/.style={decorate, draw=black,decoration={coil,amplitude=4pt, segment length=6pt}},
    scalar/.style={dashed,draw=black,
      postaction={decorate},decoration={markings,mark=at position .55
        with {\arrow[draw=black]{>}}}},
    scalarbar/.style={dashed,draw=black,
      postaction={decorate},decoration={markings,mark=at position .55
        with {\arrow[draw=black]{<}}}},
    scalarnoarrow/.style={dashed,draw=black},
    electron/.style={draw=black,
      postaction={decorate},decoration={markings,mark=at position .55
        with {\arrow[draw=black]{>}}}},
    bigvector/.style={decorate, decoration={snake,amplitude=4pt}, draw},
}
\title{Infrared structure of $SU(N)\times U(1)$ gauge theory to three loops}

\author{A.H. Ajjath, Pooja Mukherjee and V. Ravindran}
\emailAdd{
ajjathah@imsc.res.in,
poojamukherjee@imsc.res.in,
ravindra@imsc.res.in}

\affiliation{The Institute of Mathematical Sciences, HBNI, Taramani, Chennai 600113, India}

\preprint{IMSc/2019/12/15}

\abstract{
	\textcolor{black}{We study the infrared (IR) structure of $SU(N) \times U(1)$ (QCD $\times$ QED) gauge theory with $n_f$ quarks and $n_l$ leptons within the
framework of perturbation theory. In particular, we unravel the IR structure
of the form factors and inclusive real emission cross sections that contribute to inclusive
production of color neutral
states, such as a pair of leptons or single W/Z in Drell-Yan processes and a Higgs boson in bottom quark annihilation,
in Large Hadron Collider (LHC) in the threshold limit.
Explicit computation of the relevant form factors to third order and the use of Sudakov's $K+G$ equation
in $SU(N)\times U(1)$ gauge theory demonstrate the universality of the cusp anomalous dimensions ($A_I , I = q, b$).
The abelianization rules that relate $A_I$ of $SU(N)$ with those from $U(1)$ and $SU(N)\times U(1)$  can be used to
predict the soft distribution that results from the soft gluon emission subprocesses in the threshold limit.
Using the latter and the third order form factors, we can obtain the collinear anomalous dimensions ($B_I$ )
and the renormalisation constant $Z_b$ to third order in perturbation theory. The form factors,
the process independent soft distribution functions can be used to predict fixed and resummed inclusive
cross sections to third order in couplings and in leading logarithmic approximation respectively.
}}

\begin{document}
\allowdisplaybreaks[4]
\unitlength1cm
\maketitle
\flushbottom
\let\footnote=\endnote
\renewcommand*{\thefootnote}{\fnsymbol{footnote}}
 
\section{Introduction}
Precision studies at the Large Hadron Collider (LHC) is one of the thrust areas
in particle physics.  LHC has enormous potential \cite{Azzi:2019yne} to unravel the details of the standard
model (SM) and also to discover new physics.  This is possible due to large center of mass 
energy and high design luminosity options \cite{Schmidt:2016jra}.  The discovery of the Higgs boson \cite{Chatrchyan:2012xdj,Aad:2012tfa} 
at the LHC and the on-going studies to understand the nature of the Higgs bosons and its couplings
to the other SM particles provide opportunities to probe the SM in great detail \cite{deFlorian:2016spz}.   
In addition, both ATLAS and CMS collaborations have been engaged in plenty of high precision
measurements of variety of observables in the scattering reactions.  These include 
measurement of production cross sections of lepton pairs and vector bosons in Drell-Yan process,
pair of top quarks, Higgs bosons, jets etc.  These can be used to constrain 
parameters of beyond the SM (BSM) as well.
In general, theoretical predictions based on precise computations within the SM
scenarios can be compared against the data from various observables 
to look for tiny deviations that could hint for BSM.
For example, the mass of the W boson at the lowest order in electroweak theory 
can be predicted in terms of mass of the Z boson, fine structure and Fermi constants. 
Radiative corrections from the SM alter the predictions and are sensitive to parameters
of the SM as well as heavy states from BSM.  
While leading order process is often electroweak, radiative corrections from quantum chromodynamics (QCD)
dominate over those from the electroweak (EW) theory and they improve the predictions significantly. 
Efforts are also underway to include higher order corrections from EW theory.
In addition, the predictions are found to be sensitive to parton distributions of down type quarks which
leave large theoretical uncertainties. 
Hence, the on-going precise measurements of W boson mass
by ATLAS and CMS through Drell-Yan process are absolutely necessary to confirm the consistency of 
the SM and also to set constraints on the parameters of the BSMs.
Similarly, understanding the physics of top quarks provide opportunities to probe new physics. 
Observables related to top quark, being the heaviest particle in the SM, are expected
to be sensitive to new physics.  Hence, dedicated studies on the production cross sections
of top quarks have been topic of interest ever since it was discovered at the Tevtaron. 
For the top quarks, since the leading order process is through strong interaction, there is a large theoretical
uncertainty and hence higher order QCD corrections have been consistently 
included to stabilize the theoretical predictions.    

LHC being the hadron machine, QCD plays an important role
in all these studies.  Often, the leading order predictions suffer large theoretical uncertaintities   
and hence, we witnessed plethora of works that include higher order QCD corrections to
most of the observables that can be measured at the LHC.  Needless to say that
the state-of-the art computations not only provided most precise theoretical predictions for the
observables at the collider experiments but also
generated lot of interest among theorists to study the universal structure of the perturbative
series at high energies.  Computations of the multi loop and multi leg scattering 
amplitudes and cross sections in QCD provide laboratory to unravel the underlying infrared dynamics 
in terms of universal anomalous dimensions.  

The precision in the predictions from the dominant QCD corrections has reached the level that requires 
inclusion of corrections from the electroweak sector \textcolor{black}{as well}.   The fact that
the square of the strong coupling constant ($\alpha_s^2$) is comparable to fine structure constant $\alpha$ necessitates
the inclusion of effects from quantum electrodynamics (QED).   
In addition, electroweak logarithms in the Sudakov regions need to be included for a consistent prediction at the LHC.
In \cite{Dittmaier:2012kx,Dittmaier:2013hha, Frederix:2016ost}, predictions for the di-jet productions are improved by including electroweak
corrections.
One finds that the third order QCD effects
for the inclusive production rate for the Higgs bosons at the LHC are comparable to those from the 
electroweak sector.   Also, EW corrections play an important role in the W mass measurements 
through DY process.    While there are already several important works in this direction, 
there is a surge of efforts now towards estimating these corrections for the scattering processes 
at the LHC. 
Note that the next to leading order electro weak corrections to Drell-Yan process was computed in \cite{Dittmaier:2001ay,Baur:2004ig,CarloniCalame:2006zq,
Baur:2001ze,Baur:1997wa}.
Similarly, for the Higgs boson production, the dominant two loop effects from EW sector \cite{Aglietti:2006yd} plays important
role in the theoretical predictions.  

Unlike QCD, electroweak sector contains several heavy particles which can make the computations technically challenging.
The loop integrals as well as the phase space integrals involve massive particles 
making them hard to evaluate.  However, the subset of radiative corrections from QED resemble
those of QCD if lepton masses are set equal to zero, an approximation valid at high energies where the
quarks are treated light in most of the perturbative QCD computations.   

At hadron colliders, EW corrections affect the evolution of parton distribution functions as well as parton level cross sections. 
In \cite{deFlorian:2015ujt,deFlorian:2016gvk}, ${\cal O}(\alpha_s \alpha)$ as well as ${\cal O}(\alpha^2)$ corrections to the splitting function 
that govern the evolution of PDFs were obtained using the algorithm called abelianization which is incorporated in the determination of 
precise photon distributions in the proton within the LUXqed approach \cite{Ball:2014uwa,Harland-Lang:2014zoa,Dulat:2015mca}. 
For the Drell-Yan process, there have been continuous effort to obtain NNLO EW corrections as NLO EW corrections for both charged \cite{Dittmaier:2001ay,Baur:2004ig,CarloniCalame:2006zq} as well as neutral \cite{Baur:2001ze,Baur:1997wa} currents are known for a while. 
For example, works towards  NNLO EW corrections can be found in \cite{Actis:2006ra,Actis:2006rb,Actis:2006rc}.  
Mixed QCD and EW/QED effects are known in the pole approximation \cite{Dittmaier:2014koa,Dittmaier:2014qza}.
In \cite{Bonciani:2016wya,Bonciani:2016ypc}, master integrals for double real as well as 
two virtual corrections relevant for two loop QCD-EW were computed to
obtain predictions for W production at NNLO level in QCD-EW couplings.  
In \cite{deFlorian:2018wcj}, pure QED as well as QCD$\times$QED corrections at ${\cal O}(\alpha^2)$ and ${\cal O}(\alpha_s \alpha)$ respectively
for Drell-Yan process were obtained using abelianization to study the phenomenological importance of QED effects at LHC energies.  For earlier work on this can be found in \cite{Kilgore:2011pa}.   
NNLO QED as well as QCD$\times$QED corrections to Higgs production in bottom quark annihilation were obtained \cite{H:2019nsw} to estimate the
impact of QED radiative corrections.  More recently, NNLO EW-QCD effects to single vector boson production
were reported in \cite{Bonciani:2019nuy, Bonciani:2019jsw}.

\textcolor{black}{Thanks to a large number of perturbative results available in $SU(N)$ gauge theory, its
IR structure is well understood in terms of universal anomalous dimensions (see \cite{Catani:1998bh,Sterman:2002qn,
Becher:2009cu,Becher:2009qa,Gardi:2009zv, Almelid:2015jia,Almelid:2017qju}).
Recent results which point to the relevance of electroweak corrections} to the LHC observables provide the opportunity to understand the
underlying infrared structure of $U(1)$ gauge theory with massless fermions.  
The infrared structure of amplitudes with mixed gauge groups at two loops was reported in \cite{Kilgore:2013uta}. 
In \cite{deFlorian:2015ujt, deFlorian:2016gvk,deFlorian:2018wcj}, abelianizations provide a useful tool to 
obtain NLO QED effects for the splitting functions of parton distribution functions and NNLO QED corrections to inclusive cross section
for the Drell-Yan process. One finds that the abelianization can be used to relate ultraviolate and infrared anomalous dimensions of QCD with those
of QED.  In \cite{H:2019nsw}, explicit computation of form factors of vector and scalar operators in QED and QCD$\times$QED set up to
second order in perturbation theory demonstrates the
usefulness of abelianization.  In addition, the results on inclusive cross sections for DY process, such as di-lepton or $W/Z$ 
productions
and also for Higgs production in bottom quark annihilation support this procedure of abelianization up to NNLO level in QED as well as in QCD$\times$QED.  
Hence, it is tempting to apply the abelianization to obtain results beyond two loops and also beyond NNLO level for QED and QCD$\times$QED
for di-lepton or $W/Z$ production in light quark annihilation and for the Higgs boson production in bottom quark annihilation. 
In this paper, we perform this exercise at three loop level in QED and  QCD$\times$QED to find out the scope of abelianization.  In addition,
the explicit computations at three loop level provide valuable informations on the universality of cusp, collinear and soft anomalous dimensions
up to third order in couplings both in QED as well as QCD$\times$QED.  We use Sudakov's K plus G (K+G) equation to study the infrared structure of the 
three loop form factors and the validity of abelianization.  We derive the third order corrections in QED and QCD$\times$QED to 
soft distribution function resulting from those parton level subprocesses where at least one real soft gluon is present. 
Using these form factors and the soft distribution functions and exploiting 
the universal property of the inclusive Drell-Yan production, we obtain the infrared safe
parton level soft plus virtual contributions to third in QED and QCD$\times$QED.  We also derive 
the resummed threshold enhanced contribution the inclusive DY production up to next to next to next to leading logarithmic
N$^3$LL approximation. \textcolor{black}{In the following, instead of restricting to $SU(3) \times U(1)$, we study $SU(N) \times U(1)$ gauge theory where it is transperent to
understand abelianisation relations between $SU(N)$ and $U(1)$ gauge theories.  Setting $N=3$, one can easily obtain
the corresponding results in QCD $\times$ QED gauge theory.  In addition, the IR structure of the former goes through
for QCD $\times$ QED straightforwardly.  
Hence, in the rest of the paper we use QCD $\times$ QED
interchangebly with $SU(N)\times U(1)$ without loss of generality}. 
 
\section{Theoretical framework}
\textcolor{black}{We work with the gauge theory which is invariant under the gauge group $SU(N) \times U(1)$. The gauge group $SU(N)$ corresponds to QCD
which describes the strong interaction while $U(1)$ (QED) describes the electromagnetic
interaction. 
The Lagrangian of $SU(N) \times U(1)$ is given as,
\begin{equation}
\mathcal{L} = \bar{\psi^{i}}\bigg(i\gamma_{\mu}D^{\mu}_{ij} - m\delta_{ij}\bigg)\psi^{j} - \frac{1}{4}\mathcal{G}^{a}_{\mu\nu}\mathcal{G}^{a \mu\nu} - \frac{1}{4}\mathcal{F}_{\mu\nu}\mathcal{F}^{\mu\nu}
 - \frac{1}{2 \xi}\bigg(\partial^{\mu}G_{\mu}^a\bigg)^2 \,.
\end{equation}
Here $\xi$ is the gauge fixing parameter and $\psi^{k}$ denotes the fermionic field in the fundamental representation of the $SU(N)$ group with $k = 1,\cdots,N$. The covariant derivative $D_{ij}^{\mu} = \partial^{\mu}\delta_{ij} - i g_s \big(T^{c}\big)_{ij}G^{\mu}_c - i e A^{\mu}\delta_{ij}$. The gluonic and photonic field strength tensors are given respectively as,
\begin{align*}
\mathcal{G}^{a}_{\mu\nu} &= \partial_{\mu}G_{\nu}^{a} - \partial_{\nu}G_{\mu}^{a} + i g_s f^{abc}G_{\mu}^{b}G_{\nu}^{c} \,, \nonumber\\
\mathcal{F}_{\mu\nu} &= \partial_{\mu}A_{\nu} - \partial_{\nu}A_{\mu} \,,
\end{align*} 
where the gluon gauge fields $G_{\mu}^{a}$ with $a=1,\cdots,N^2-1$ and the photon gauge fields $A_{\mu}$  belong to the adjoint representation}. We use the standard perturbation theory to compute
various quantities in this theory in powers of coupling constants defined by $a_s=g_s^2/16 \pi^2$ and $a_e=e^2/16 \pi^2$ 
where $g_s$ and $e$ are strong and electromagnetic coupling constants respectively.
Since we are interested in quantities in the high energy limit, 
both the quarks and leptons are treated massless throughout. We use dimensional 
regularisation to perform higher order computations and $\overline {MS}$ 
to renormalise the fields and the couplings in this theory.  In dimensional regularisation the 
space time dimension is taken to be $d=4+\varepsilon$.  The field as well as  coupling 
constant renormalisation constants contain 
poles in $\varepsilon$ in the vicinity of  $d=4$ space time dimensions due to ultraviolet (UV) divergences.  
Higher order radiative corrections are often sensitive to soft divergences
due to massless gluons of $SU(N)$ and massless photons of $U(1)$ and also to collinear divergences due to the presence of
(almost) massless quarks and leptons.  These are called infrared (IR) divergences and they also show up as poles in 
$\varepsilon$ in dimensional regularisation.       

We begin with the renormalisation of the coupling constants when 
both the interactions are simultaneously present.       
Let us denote the renormalisation constant $Z_{a_c}, c=s, e$ for the QCD and QED coupling constants 
respectively.  Then the unrenormalised coupling constants $\hat a_c, c=s,e$ will be
related to the renormalised ones  through $Z_{a_c}$ as 
\begin{align} 
\frac{\hat{a}_c}{\big(\mu^2 \big)^{\frac{\varepsilon}{2}}} S_{\varepsilon} =
\frac{a_c(\mu_R^2)}{\big( \mu_{R}^2 \big)^{\frac{\varepsilon}{2}}} ~ Z_{a_c}\left(a_s(\mu_{R}^2),a_e(\mu_R^2),
\varepsilon\right)\,,
\end{align}
where $ a_c=\{a_s,a_e\}$.  
Here, $S_{\varepsilon} \equiv \exp[\left(\gamma_E-\ln4\pi\right)\frac{\varepsilon}{2}]$
is the phase-space factor in $d$-dimensions, $\gamma_{E} = 0.5772...$ is the 
Euler-Mascheroni constant and $\mu$ is an arbitrary
mass scale introduced to make $\hat a_s$ and $\hat a_e$ dimensionless in  $d$-dimensions and
$\mu_R$ is the renormalisation scale.   The fact that bare coupling constants $\hat a_c$ is 
independent of the renormalisation scale $\mu_R$ results in renormalisation group equations
for the couplings $a_c(\mu_R^2)$:
\textcolor{black}{\begin{eqnarray}\label{RGEa}
\mu_R^2 {d \over d\mu_R^2} \ln Z_{a_c} = \frac{\varepsilon}{2} + \beta_{a_c}(a_s(\mu_R^2),a_e(\mu_R^2))\,.
\end{eqnarray}}
In the perturbation theory with both the interactions active, the beta functions 
$\beta_{a_c}$ can be expanded in powers of $a_s$ as well as $a_e$: 
\begin{eqnarray}
\label{beta}
\beta_{a_s} = -\sum_{i,j=0}^\infty \beta_{ij} a_s^{i+2} a_e^{j}
\,,\quad \quad 
\beta_{a_e} = -\sum_{i,j=0}^\infty  \beta'_{ij} a_e^{j+2} a_s^{i}\,.
\end{eqnarray}
\textcolor{black}{The explicit calculation of $\beta$-function begins with renormalising the Lagrangian. This involves  renormalising fields, the Lagrangian. This involves  renormalising fields, couplings, masses and gauge-fixing parameter. Hence we redefine the fields as,
\begin{equation}
\psi = Z_2^{1/2}\psi_r,\quad G_{\mu}^{a} = Z_3^{1/2}G_{\mu r}^{a},\quad A_{\mu} = Z_{3\gamma}^{1/2}A_{\mu r} \,, 
\end{equation}
and the parameters as,
\begin{equation}
g_s=Z_g g_r,\quad e = Z_{e} e_{r},\quad m=Z_m m_{r}, \quad  \xi = Z_3\xi_r \,,
\end{equation}
where the constants $Z_2,Z_3,Z_{3\gamma}$ are called the fermion-field, gluon-field and photon-field renormalisation constants while the constants $Z_g$, $Z_e$ and $Z_m$ are called the coupling-constant and mass renormalisation constant respectively. All the renormalized fields, masses and parameters are designated with an subscript $r$. Inserting all of these into the Lagrangian and collecting all the terms involving $\delta Z$, where $\delta Z = Z - 1$ for any renormalisation constants, we get the counterterm Lagrangian. Now let us look at the counterterm for the four-gluon vertex to compute the
renormalisation constant $Z_{a_s}$. The leading order starts at $a_s^2$ and hence the QED contributions to $Z_{a_s}$ will always be proportional to $a_s^2$.
Similarly for the QED $\beta$-function, the counterterm for the fermion-fermion-photon vertex is,
\begin{equation}
\sim e_r \bar{\psi_r}\gamma_\mu A^{\mu}_r\psi_r(Z_2Z_{3\gamma}^{1/2}Z_e -1) \equiv e_r \bar{\psi_r}\gamma_\mu A^{\mu}_r\psi_r (Z_{1e} - 1),
\end{equation}
which implies 
\begin{equation}
Z_e = Z_{1e}/(Z_2 \sqrt{Z_{3\gamma}}).
\end{equation}
The Ward identity, which is derived using the conservation of the electromagnetic current, demands that $Z_{1e} = Z_2$ to all orders in perturbation theory.  This in turn suggests that $Z_{a_e}$ is fully determined by $Z_{3\gamma}$ which starts 
at order $a_e$ and hence the QCD corrections
to $Z_{a_e}$ will always be proportional to $a_e$. 
}
Substituting \textcolor{black}{Eq}.(\ref{beta}) in \textcolor{black}{Eq}.(\ref{RGEa}) and solving 
for the renormalisation constants $Z_{a_c}$ up to third order, we obtain
\begin{align}
 Z_{a_s} &= 1 + a_s \Big( \frac{2\beta_{00}}{\varepsilon} \Big)
 + a_s a_e \Big( \frac{\beta_{01}}{\varepsilon} \Big)
 + a_s a_e^2 \left( {\frac{2 \beta'_{00} \beta_{01}}{3 \varepsilon^2}}
                  + {\frac{2 \beta_{02}}{3 \varepsilon}} \right)
 + a_s^2
\left({4 \beta_{00}^{2} \over \varepsilon^2}+{\beta_{10}\over \varepsilon}\right)
\nonumber
\\&
 + a_s^2 a_e \left({4 \beta_{00} \beta_{01} \over \varepsilon^2 }
                  + {2 \beta_{11} \over {3 \varepsilon}} \right)
 + a_s^3 \left( {8 \beta_{00}^3 \over \varepsilon^3}
              + {14 \beta_{00} \beta_{10} \over {3 \varepsilon^2}}
              + {2 \beta_{20}\over {3 \varepsilon}} \right) + \cdot \cdot \cdot \,,
\nonumber\\
 Z_{a_e} &= 1 + a_e \Big( \frac{2\beta'_{00}}{\varepsilon} \Big)
 + a_e a_s \Big( \frac{\beta'_{10}}{\varepsilon} \Big)
 + a_e a_s^2 \left( {\frac{2 \beta_{00} \beta'_{10}}{3 \varepsilon^2}}
                  + {\frac{2 \beta'_{20}}{3 \varepsilon}} \right)
 + a_e^2
\left({4 \beta_{00}^{'2} \over \varepsilon^2}+{\beta'_{01}\over \varepsilon}\right)
\nonumber
\\&
 + a_e^2 a_s \left({4 \beta'_{00} \beta'_{10} \over \varepsilon^2 }
                  + {2 \beta'_{11} \over {3 \varepsilon}} \right)
 + a_e^3 \left( {8 \beta_{00}^{'3} \over \varepsilon^3}
              + {14 \beta'_{00} \beta'_{01} \over {3 \varepsilon^2}}
              + {2 \beta'_{02}\over {3 \varepsilon}} \right) + \cdot \cdot \cdot \,.
\end{align}
We have used the symbol $\cdot \cdot \cdot$ to denote the missing higher order terms of the order 
$a_s^i a_e^j, i+j> 3$ throughout. 
While, these constants are sufficient to obtain UV finite observables, the 
UV divergences resulting from composite operators in the theory beyond leading order require
additional overall renormalisation constants.  These constants are expanded in  power series
expansions of both $a_s$ as well as $a_e$.  Similarly, if the fields of QCD and QED couple
to external fields, then the corresponding couplings are renormalised by separate renormalisation constants.  One such example that we need to study in the present paper is Yukawa coupling
that describes the coupling of a Higgs boson with the bottom quarks in this theory.
If we denote the bare Yukawa coupling by $\hat \lambda_b$, then the corresponding 
renormalisation constant $Z^b_{\lambda}$ 
relates this to the renormalised one $\lambda_b$ by
\begin{align} \label{UV1}
\frac{\hat{\lambda}_b}{\big(\mu^2 \big)^{\frac{\varepsilon}{2}}} S_{\varepsilon} = 
\frac{\lambda_b(\mu_R^2)}{\big( \mu_{R}^2 \big)^{\frac{\varepsilon}{2}}} ~ Z^b_{\lambda} \left(a_s(\mu_{R}^2),a_e(\mu_R^2),
\varepsilon\right)\,, 
\end{align}
where $ a_c=\{a_s,a_e\}$.
The renormalisation constant $Z_{\lambda}^{b}(a_s, a_e) $ satisfies the renormalisation group equation:
\begin{eqnarray}
\mu_R^2 {d \over d\mu_R^2} \ln Z^{b}_\lambda = \frac{\varepsilon}{4} + \gamma_{b} (a_s(\mu_R^2),a_e(\mu_R^2))\,,
\end{eqnarray}
whose solution in terms of the anomalous dimensions $\gamma^{(i,j)}_b$ and $\beta_{ij},\beta'_{ij}$ 
up to three loops is found to be 
\begin{align} \label{Zlam}
Z^b_{\lambda} (& a_s, a_e,\varepsilon ) = 
1  + a_s \Big\{ \frac{1}{\varepsilon}\big(2 \gamma^{(1,0)}_b \big) \Big\} 
   + a_e \Big\{ \frac{1}{\varepsilon}\big(2 \gamma^{(0,1)}_b \big) \Big\}    
%
   + a_s^2 \Big\{ \frac{1}{\varepsilon^2} \Big( 2 \big(\gamma^{(1,0)}_b \big)^2 + 2 \beta_{00} \gamma^{(1,0)}_b \Big)
 \nonumber\\&
                + \frac{1}{\varepsilon} \gamma^{(2,0)}_b \Big\} 
  + a_e^2 \Big\{ \frac{1}{\varepsilon^2} \Big( 2 \big(\gamma^{(0,1)}_b \big)^2 + 2 \beta_{00}' \gamma^{(0,1)}_b \Big)
                + \frac{1}{\varepsilon} \gamma^{(0,2)}_b \Big\}
   + a_s  a_e \Big\{ \frac{1}{\varepsilon^2}\Big(4 \gamma^{(1,0)}_b \gamma^{(0,1)}_b \Big) 
   \nonumber\\&
   +\frac{1}{\varepsilon}\big(\gamma^{(1,1)}_b \big) \Big\}  
  + a_s^3 \Big\{ \frac{1}{\varepsilon^3} \Big( \frac{4}{3}\big(\gamma^{(1,0)}_b \big)^3
          + 4 \beta_{00} \big( \gamma^{(1,0)}_b \big)^2
          + \frac{8}{3} \beta_{00}^2 \big( \gamma^{(1,0)}_b \big) \Big)
\nonumber\\&
     +  \frac{1}{\varepsilon^2} \Big( 2 \gamma^{(1,0)}_b \gamma^{(2,0)}_b 
          + \frac{4}{3} \beta_{10} \gamma^{(1,0)}_b 
          + \frac{4}{3} \beta_{00} \gamma^{(2,0)}_b \Big)
     +  \frac{1}{\varepsilon} \Big( \frac{2}{3} \gamma^{(3,0)}_b \Big) \Big\}
  + a_e^3 \Big\{ \frac{1}{\varepsilon^3} \Big( \frac{4}{3}\big(\gamma^{(0,1)}_b \big)^3
  \nonumber\\&
          + 4 \beta_{00}' \big( \gamma^{(0,1)}_b \big)^2
          + \frac{8}{3} \beta_{00}^{'2} \big( \gamma^{(0,1)}_b \big) \Big)
     +  \frac{1}{\varepsilon^2} \Big( 2 \gamma^{(0,1)}_b \gamma^{(0,2)}_b 
          + \frac{4}{3} \beta_{01}' \gamma^{(0,1)}_b 
          + \frac{4}{3} \beta_{00}' \gamma^{(0,2)}_b \Big)
          \nonumber\\&
     +  \frac{1}{\varepsilon} \Big( \frac{2}{3} \gamma^{(0,3)}_b \Big) \Big\}
  + a_s a_e^2 \Big\{ \frac{1}{\varepsilon^3} \Big( 
            4 \gamma^{(1,0)}_b \big( \gamma^{(0,1)}_b \big)^2
          + 4 \beta_{00}' \gamma^{(1,0)}_b \gamma^{(0,1)}_b \Big)
\nonumber\\&
     +  \frac{1}{\varepsilon^2} \Big( 
            2 \gamma^{(0,1)}_b \gamma^{(1,1)}_b 
          + 2 \gamma^{(1,0)}_b \gamma^{(0,2)}_b 
          + \frac{4}{3} \beta_{10}' \gamma^{(0,1)}_b 
          + \frac{2}{3} \beta_{00}' \gamma^{(1,1)}_b \Big)
     +  \frac{1}{\varepsilon} \Big(
          \frac{2}{3} \gamma^{(1,2)}_b \Big) \Big\}
\nonumber\\&
  + a_s^2 a_e \Big\{ \frac{1}{\varepsilon^3} \Big( 
            4 \big(\gamma^{(1,0)}_b \big)^2 \gamma^{(0,1)}_b 
          + 4 \beta_{00} \gamma^{(1,0)}_b \gamma^{(0,1)}_b \Big)
     +  \frac{1}{\varepsilon^2} \Big( 
            2 \gamma^{(0,1)}_b \gamma^{(2,0)}_b 
          + 2 \gamma^{(1,0)}_b \gamma^{(1,1)}_b 
          \nonumber\\&
          + \frac{4}{3} \beta_{01} \gamma^{(1,0)}_b 
          + \frac{2}{3} \beta_{00} \gamma^{(1,1)}_b \Big)
     +  \frac{1}{\varepsilon} \Big(
          \frac{2}{3} \gamma^{(2,1)}_b \Big) \Big\}
+ \cdot \cdot \cdot \,.
\end{align}
Note that while the UV singularities factorize through $Z^b_{\lambda}$, singularities from QCD and QED
mix from two loops onwards.

Having expanded the renormalisation constants of $a_s,a_e$ and $\lambda_b$ in powers of
$a_s$ and $a_e$, our next task is to determine the constants $\beta_{ij},\beta'_{ij}$ 
that appear in $Z_{a_c},c=s,e$ as well as 
$\gamma_b^{(i,j)}$ up to three loops in QED and QCD$\times$QED.  The text book approach to this 
is to compute relevant loop corrections to the truncated \textcolor{black}{$n$-point} off-shell Green's functions 
of the fermions and gauge fields in the regularised theory.  Alternatively 
\cite{Ravindran:2005vv,Ravindran:2006cg}, 
one can determine them by studying the on-shell form factors of certain local/composite operators
in the theory.  For example, computing the form factors of vector as well as scalar operators
made up of fermionic fields up to three loops in QCD$\times$QED, and  exploiting 
their universal infrared structure using Sudakov's K+G equation, we demonstrate how we can obtain most of these 
constants to the desired accuracy.
In the process, we confirm some of the results for these constants which are already known
in the literature. \textcolor{black}{For QCD, $\beta_{i0}$ and  $\gamma^{(i,0)}_b$ are 
known to five loops \cite{Chetyrkin:2017bjc,Luthe:2017ttg,Herzog:2017ohr,Baikov:2016tgj}}.
In the following, we elaborate on how we determine them and also discuss the reliability of
abelianization at three loop level.
\section{Sudakov Formalism}
\subsection{Form factors}\label{sec:FF}
In quantum field theory, form factor (FF) of a composite operator is defined by its matrix element
between on-shell states.  Given a composite operator ${\cal O}(x)$, the form factor
in the Fourier space is found to be
\begin{eqnarray}
\label{FFdef}
F_{{\cal O}}(q^2) (2 \pi)^4 \delta^{(4)}(q+p_1-p_2) = \int d^4 y e^{i q\cdot y} \langle p_2 | {\cal O}(y) |p_1 \rangle
\,.
\end{eqnarray}
We restrict ourselves to two composite operators namely
\begin{eqnarray}
{\cal O}^\mu(y) = \overline \psi(y) \gamma^\mu \psi(y), \quad \quad {\cal O}(y) =\overline \psi(y) \psi(y)\,.
\end{eqnarray}
The corresponding form factors are denoted by $F_q$ and $F_b$ respectively.
$F_q$ contributes to di-lepton or W/Z production and $F_b$ \textcolor{black}{contributes} to production of Higgs bosons in bottom quark
annihilation.
In the Eq.(\ref{FFdef}), $\psi$ is the fermion field and the states $|p_i\rangle, i=1,2$ are on-shell fermionic states
with momenta $p_i$.  
We begin with the bare form factors $\hat{F}_I(Q^2,\mu^2,\varepsilon)$ where
$I=q,b$ and the invariant scale is defined by $Q^2=(p_1-p_2)^2$.  
They are calculable in perturbation theory in powers of $a_s$ and $a_e$ using dimensional
regularisation.
Both QCD as well as QED interactions are taken into account simultaneously. 
Beyond the leading order in perturbation theory, the FFs contain both UV, soft and collinear divergences.
UV divergences are removed by coupling constants as well as overall renormalisation constant 
namely $Z_{a_c},c=s,e$ and $Z^b_{\lambda}$ respectively.  
The soft divergences arise due to massless gauge bosons and the collinear ones are due to 
massless fermions.  Explicit computation of the form factors shows that the IR singularities, 
resulting from QCD and QED interactions factorize. 
In particular, they can be factored out using a universal IR counter term
denoted by $Z_{IR}(a_s,a_e,Q^2,\mu_R^2)$ and hence we can write  $\hat F_I$  
in dimensional regularisation as
\begin{eqnarray}
\hat F_I(Q^2,\mu^2,\varepsilon) = Z_{IR}(Q^2,\mu^2,\mu_R^2,\varepsilon) 
\hat F_I^{fin}(Q^2,\mu^2,\mu_R^2,\varepsilon)  \,,
\end{eqnarray}
where $Z_{IR}$ contains all the IR poles while $\hat F_I^{fin}$ is IR
finite.  In addition, both $\hat F_I$ and $\hat F_I^{fin}$  can be made UV finite after appropriate UV renormalisation.  
Differentiating both sides with respect to
$Q^2$, we obtain K+G equation for the form factors $\hat F_I$ as
\begin{align}\label{eq:KG}
Q^2\frac{d}{dQ^2}\ln \hat{F}_I(Q^2,\mu^2,\varepsilon)
= \frac{1}{2}\Big[&  K_I \Big(\{\hat a_c\}, \frac{\mu_R^2}{\mu^2}, \varepsilon \Big) 
+ G_I \Big(\{\hat a_c\}, \frac{Q^2}{\mu_R^2},\frac{\mu_R^2}{\mu^2},\varepsilon \Big) \Big]
\,,
\end{align}
where 
\begin{eqnarray}
K_I\left(\{\hat a_c\},{\mu_R^2 \over \mu^2},\varepsilon\right)
&=& 2 Q^2\frac{d}{dQ^2}\ln Z_{IR}(Q^2,\mu^2,\mu_R^2,\varepsilon)  \,,
\nonumber\\
G_I\left(\{\hat a_c\},{Q^2 \over \mu_R^2},{\mu^2 \over \mu_R^2},\varepsilon\right)
&=&2 Q^2\frac{d}{dQ^2}\ln \hat F^{fin}_I(Q^2,\mu^2,\mu_R^2,\varepsilon) \,.  
\end{eqnarray}
Following, \cite{Ravindran:2005vv,Ravindran:2006cg} we solve 
Sudakov K+G equation (\textcolor{black}{Eq}.(\ref{eq:KG})) order by order in perturbation theory. 

The radiative corrections resulting from QCD and QED interactions 
can not be factored out independently.  
In other words, if we factorize IR singularities from the
FFs, neither the IR singular function $Z_{IR}$ nor the finite FF, $\hat F_I^{fin}$,
can be written as a product of pure QCD and pure 
QED contributions.  
More specifically, there will be terms proportional to $a_s^i a_e^j$, where $i,j > 0$, 
which will not allow factorization of 
QCD and QED contributions for both $Z_{IR}$ and $\hat F_I^{fin}$.  
One finds that the constant $K_I$ will have IR poles in $\varepsilon$ from pure QED and pure QCD in every order in
perturbation theory and in addition, from 
\textcolor{black}{QCD$\times$QED} starting from ${\cal O} (a_s a_e)$.
The constants $G_I$s are IR finite and they get contributions from both QCD as well as QED and they
mix beyond leading order in perturbation theory.
Since, the IR singularities of FFs have dipole structure, $K_I$ will be independent of $Q^2$ while 
$G_I$s will be finite in $\varepsilon \rightarrow 0$ and also contain only logarithms in $Q^2$.  
The fact that $\hat{F}_I$ are renormalisation group (RG) invariant implies the sum $K_I+G_I$ is
also RG invariant.  This implies
\begin{eqnarray}
\label{eq:KSoln} 
\mu_R^2\frac{d}{d\mu_R^2}K_I \Big(
\{\hat a_c\},\frac{\mu_R^2}{\mu^2},\varepsilon \Big) = -
\mu_R^2\frac{d}{d\mu_R^2}G_I \Big(
\{\hat a_c\},\frac{Q^2}{\mu_R^2},\frac{\mu_R^2}{\mu^2},\varepsilon \Big) = -A_I(\{a_c(\mu_R^2)\} 
)\,,
\label{eq:GSoln} 
\end{eqnarray}   
where $A_I$ are the cusp anomalous dimensions.  Since the FFs are dependent on both the couplings
$a_s$ as well as $a_e$, the $A_I$ also depend on them.
The solutions to the above RG equations for $K_I$ can be obtained by expanding the 
cusp anomalous dimensions ($A_I$) in powers of renormalized coupling constants $a_s(\mu_R^2)$ and  
$a_e(\mu_R^2)$ as 
\begin{eqnarray}
A_I (\{a_c(\mu_R^2)\}) =  \sum_{i,j=0} a_s^i(\mu_R^2) a_e^j(\mu_R^2) A_I^{(i,j)} \,, \quad \quad A_I^{(0,0)}=0 \,,
\end{eqnarray}
and $K_I$ as
\begin{eqnarray}
\label{eq:Kexp}
K_I(\{\hat a_c\},\mu_R^2,\varepsilon)  =  \sum_{i,j=0}\hat{a}_s^i \hat{a}_e^j \Big( 
\frac{\mu_R^2}{\mu^2}\Big)^{(i+j) \frac{\varepsilon}{2}}S_{\varepsilon}^{(i+j)} K_I^{(i,j)}(\varepsilon) \,,
\quad \quad K_I^{(0,0)} =0 \,,
\end{eqnarray}
where $A_I^{(i,0)}$ and $A_I^{(0,i)}$ result from pure QCD and pure QED interactions respectively and $A_I^{(i,j)}, i,j>0$ from 
\textcolor{black}{QCD$\times$QED}.   
The constants $K_I^{(i,j)}$ in Eq.(\ref{eq:Kexp}) can be obtained 
using Eq.(\ref{eq:KSoln}) and RG equations for the couplings  $a_s$ and $a_e$, \textcolor{black}{in terms of the cusp anomalous dimensions}.  They are listed in the Appendix \ref{ap:KFF} .
Since $G_I$s depend on the finite part of the FFs, they do not contain any IR singularities but 
depend only on $Q^2$ and hence we expand them as  
\begin{align}
G_I \Big( \{\hat{a_c}\}, \frac{Q^2}{\mu_R^2},\frac{\mu_R^2}{\mu^2},\varepsilon \Big) 
&= G_I(\{a_c(Q^2)\},1,\varepsilon) 
+ \int_{\frac{Q^2}{\mu_R^2}}^{1} \frac{d\lambda^2}{\lambda^2} A_I(\{a_c(\lambda^2\mu_R^2)\} 
) 
\,,
\end{align}
where the first term results from the boundary condition on each $G_I$ at $\mu_R^2=Q^2$.
Again expanding $A_I$ in powers of $a_s$ and $a_e$ 
and using RG equations for QCD and QED couplings, we obtain 
\begin{align}
 \int_{\frac{Q^2}{\mu_R^2}}^{1} \frac{d\lambda^2}{\lambda^2} & A_I(\{a_c(\lambda^2\mu_R^2)\}) 
= \sum_{i,j} \hat{a}_s^i \hat{a}_e^j \Big( \frac{\mu_R^2}{\mu^2} \Big)^{(i+j) \frac{\varepsilon}{2}} 
S_{\varepsilon}^{(i+j)} \Big[ \Big( \frac{Q^2}{\mu_R^2} \Big)^{(i+j) \frac{\varepsilon}{2}} - 1 \Big] K_I^{(i,j)}(\varepsilon) \,.
\end{align}
The next step is to integrate Eq.(\ref{eq:KG}) over $Q^2$ to obtain
the solution to K+G equation for $\hat F_I$.  For this, we substitute the solutions of $K_I$ and $G_I$ in 
the right hand side of Eq.(\ref{eq:KG}) 
along with the expansion for $G_I(a_s(Q^2),a_e(Q^2),1,\varepsilon)$ given by 
\begin{eqnarray}\label{Gexp}
G_I( \{a_c(Q^2)\}, 1, \varepsilon ) = \sum_{i,j}a_s^i(Q^2)a_e^j(Q^2) G_I^{(i,j)}(\varepsilon) \,.
\end{eqnarray}
We thus obtain,
\begin{eqnarray}
\ln \hat{F}_I
= \sum_{i,j}\hat{a}_s^i \hat{a}_e^j\Big( \frac{Q^2}{\mu^2} \Big)^{(i+j)\frac{\varepsilon}{2}}S_{\varepsilon}^{(i+j)}\hat {\mathcal{L}}_{F_I}^{(i,j)}(\varepsilon) \,,
\end{eqnarray}
where,
\begin{align} \label{LFo}
\hat {\mathcal{L}}_{F_I}^{(1,0)} &= \frac{1}{\varepsilon^2}\Big( -2A_I^{(1,0)} \Big) + \frac{1}{\varepsilon}\Big( G_I^{(1,0)}(\varepsilon) \Big) \,, 
\nonumber \\
\hat {\mathcal{L}}_{F_I}^{(0,1)} &= \frac{1}{\varepsilon^2}\Big( -2A_I^{(0,1)} \Big) + \frac{1}{\varepsilon}\Big( G_I^{(0,1)}(\varepsilon) \Big) \,,
\nonumber\\
\hat {\mathcal{L}}_{F_I}^{(2,0)} &= \frac{1}{\varepsilon^3}\Big( \beta_{00} A_I^{(1,0)} \Big) 
           + \frac{1}{\varepsilon^2}\Big( -\frac{1}{2}A_I^{(2,0)} 
           - \beta_{00} G_I^{(1,0)}(\varepsilon) \Big)
           + \frac{1}{2\varepsilon}\Big( G_I^{(2,0)}(\varepsilon) \Big) \,,
\nonumber\\
\hat {\mathcal{L}}_{F_I}^{(0,2)} &= \frac{1}{\varepsilon^3}\Big( \beta_{00}' A_I^{(0,1)} \Big) 
           + \frac{1}{\varepsilon^2}\Big( -\frac{1}{2}A_I^{(0,2)} 
           - \beta_{00}' G_I^{(0,1)}(\varepsilon) \Big)
           + \frac{1}{2\varepsilon}\Big( G_I^{(0,2)}(\varepsilon) \Big) \,,
\nonumber \\
\hat {\mathcal{L}}_{F_I}^{(1,1)} &= \frac{1}{\varepsilon^2}\Big( -\frac{1}{2}A_I^{(1,1)} \Big) + \frac{1}{2\varepsilon}\Big( G_I^{(1,1)}(\varepsilon) \Big) \,,
\nonumber\\           
\hat {\mathcal{L}}_{F_I}^{(3,0)} &= 
 \frac{1}{\varepsilon^4} \Big(-\frac{8}{9} \beta_{00}^2 A_I^{(1,0)}  \Big) 
 +\frac{1}{\varepsilon^3}\Big(\frac{8}{9} \beta_{00} A_I^{(2,0)} 
 + \frac{2}{9} \beta_{10} A_I^{(1,0)}
 + \frac{4}{3} \beta_{00}^2 G_I^{(1,0)} \Big) 
\nonumber\\&           
 + \frac{1}{\varepsilon^2}\Big( -\frac{2}{9} A_I^{(3,0)} 
 - \frac{1}{3} \beta_{10} G_I^{(1,0)}
 -\frac{4}{3} \beta_{00} G_I^{(2,0)} \Big)           
 + \frac{1}{3 \varepsilon} \Big( G_I^{(3,0)} \Big)\,,
\nonumber\\
\hat {\mathcal{L}}_{F_I}^{(1,2)} &= 
 \frac{1}{\varepsilon^3}\Big(\frac{4}{9} \beta'_{00} A_I^{(1,1)} 
 + \frac{2}{9} \beta'_{10} A_I^{(0,1)} \Big) 
 + \frac{1}{\varepsilon^2}\Big( -\frac{2}{9} A_I^{(1,2)} 
 - \frac{1}{3} \beta'_{10} G_I^{(0,1)}
 -\frac{2}{3} \beta'_{00} G_I^{(1,1)} \Big) 
  \nonumber\\&
 + \frac{1}{3 \varepsilon} \Big(G_I^{(1,2)} \Big) \,,
\nonumber\\           
\hat {\mathcal{L}}_{F_I}^{(2,1)} &= 
 \frac{1}{\varepsilon^3}\Big(\frac{4}{9} \beta_{00} A_I^{(1,1)} 
 + \frac{2}{9} \beta_{01} A_I^{(1,0)} \Big) 
 + \frac{1}{\varepsilon^2}\Big( -\frac{2}{9} A_I^{(2,1)} 
 - \frac{1}{3} \beta_{01} G_I^{(1,0)}
 -\frac{2}{3} \beta_{00} G_I^{(1,1)} \Big)  
  \nonumber\\&
 + \frac{1}{3 \varepsilon} \Big(G_I^{(2,1)} \Big)     \,,      
\nonumber\\    
\hat {\mathcal{L}}_{F_I}^{(0,3)} &= 
 \frac{1}{\varepsilon^4} \Big(-\frac{8}{9} \beta_{00}^{'2} A_I^{(0,1)}  \Big) 
 +\frac{1}{\varepsilon^3}\Big(\frac{8}{9} \beta'_{00} A_I^{(0,2)} 
 + \frac{2}{9} \beta'_{01} A_I^{(0,1)}
 + \frac{4}{3} \beta_{00}^{'2} G_I^{(0,1)} \Big)
\nonumber\\&           
 + \frac{1}{\varepsilon^2}\Big( -\frac{2}{9} A_I^{(0,3)} 
 - \frac{1}{3} \beta'_{01} G_I^{(0,1)}
 -\frac{4}{3} \beta'_{00} G_I^{(0,2)} \Big)
 + \frac{1}{3 \varepsilon} \Big(G_I^{(0,3)} \Big) 
 \,.         
\end{align}
Our next task is to compute the FFs to third order in the couplings of QCD, QED and \textcolor{black}{QCD$\times$QED}.
The method of this computation is well documented in the literature \cite{Baikov:2009bg,Gehrmann:2010ue} and applied to 
several of the form factors \cite{Ahmed:2015qpa,Ahmed:2016vgl,Ahmed:2016qjf} in QCD.
Following \cite{Baikov:2009bg,Gehrmann:2010ue}, we computed $\hat F_I$ for $I=q,b$ up to third order in QCD,QED and \textcolor{black}{QCD$\times$QED}.
The form factors $\hat F_I, I=q,b$ thus obtained in the present paper   
up to three loop level are listed in the Appendix \ref{ap:g6}.  We use them to extract
the cusp anomalous dimensions ($A_I^{(i,j)}$) by comparing them against Eq.(\ref{LFo}).  
From the one loop result for $\hat {\mathcal{L}}_{F_I}^{(1,0)}$
we obtain $\{G_I^{(1,0)},A_I^{(1,0)}(\varepsilon)\}$ and 
from the result for $\hat {\mathcal{L}}_{F_I}^{(0,1)}$
we get $\{G_I^{(0,1)},A_I^{(0,1)}(\varepsilon)\}$.
Substituting these one loop results for $A_I$ and $G_I(\varepsilon)$ 
along the two loop results for 
$\hat {\mathcal{L}}_{F_I}^{(2,0)}$ and $\hat {\mathcal{L}}_{F_I}^{(0,2)}$ in Eq.(\ref{LFo}),
we obtain $\{\beta_{00},G_I^{(2,0)}(\varepsilon),A_I^{(2,0)}\}$ and $\{\beta_{00}',G_I^{(0,2)}
(\varepsilon),
A_I^{(0,2)}\}$ respectively.  We continue this procedure \textcolor{black}{with} three loop
results of \textcolor{black}{FF} to determine 
$\{\beta_{10},G_I^{(3,0)}(\varepsilon),A_I^{(3,0)}\}$ and  
$\{\beta'_{01},G_I^{(0,3)}(\varepsilon),A_I^{(0,3)}\}$.  From the \textcolor{black}{QCD$\times$QED} two and three
loops results for the FFs, we obtain 
$\{G_I^{(1,1)}(\varepsilon),A_I^{(1,1)}\}$ and
$\{\beta_{01},G_I^{(2,1)}(\varepsilon),A_I^{(2,1)}\}$,   
\textcolor{black}{$\{\beta'_{10},G_I^{(1,2)}(\varepsilon),A_I^{(1,2)}\}$} respectively.  This way 
we can obtain all the cusp anomalous dimensions, beta function coefficients 
and $G_I(\varepsilon)$s up to three loops.
We find $A_q^{(i,j)}=A_b^{(i,j)}$ up to three loops in QCD,QED and \textcolor{black}{QCD$\times$QED}
demonstrating the universal nature of these constants. However, we find that the constants
$G_I(\varepsilon)$s depend on the type of form factors.
\textcolor{black}{The constants $A_I^{(i,0)}$ are known till three loops in \cite{Moch:2005tm} 
and here in Appendix \ref{ap:g2} we enlist the new ones $A_I^{(1,2)},A_I^{(2,1)}$ along with the existing ones in \cite{H:2019nsw}}.
The $\beta$s (see \cite{Cieri:2018sfk} for the leading order ones) are given by 
\begin{align}
\beta_{00} &= \frac{11}{3} C_A - \frac{4}{3} n_f T_F\,, \quad
\beta'_{00} = -\frac{4}{3} \Big( N \sum_q e_q^2 + \sum_l e_l^2\Big)\,, \quad
\beta_{01} = -2 \left( \sum_q e_q^2 \right) \,, \quad
\nonumber\\ 
\beta'_{01} &= -4 \left( N \sum_q e_q^4 + \sum_l e_l^4\right)\,,\quad
\beta_{10} = \left( {34 \over 3} C_A^2  - {20 \over 3} C_A n_f T_F - 4 C_F n_f T_F \right)   \,,\nonumber\\
\beta'_{10} &= -4 C_F \left( N \sum_q e_q^2 \right) \,.
\end{align}
Here, $C_A=N$ is the adjoint Casimir of $SU(N)$ and the fundamental Casimir
is $C_F=(N^2-1)/2 N$, $T_F=1/2$ and
$n_f$($n_l$) is the number of active quark flavors (leptons).  The electric charge of quark $q$ is denoted by $e_q$ while  $e_l$ refers to the electric charge of the lepton $l$.

Our next task is to investigate the structure of the constants $G_I^{(i,j)}(\varepsilon)$ 
following the observation made in \cite{Ravindran:2004mb} for the $G_I^{(i,0)}, I=q,b,g$ in QCD, namely 
we expand $G_I^{(i,j)}(\varepsilon)$ around $\varepsilon=0$ in terms of
collinear ($B_I^{(i,j)}$),  soft ($f_I^{(i,j)}$) and UV ($\gamma_I^{(i,j)}$) anomalous dimensions as
\begin{align} \label{G}
G_I^{(i,j)}(\varepsilon) &= 2(B_I^{(i,j)} - \gamma_I^{(i,j)} ) + f_I^{(i,j)} +  \sum_{k=0}\varepsilon^k g_{I,ij}^{k} \,,
\end{align}
with
\begin{align}
 g_{I,10}^0 &= 0 \,, \quad  g_{I,01}^0 = 0 \,, \quad g_{I,11}^0 = 0 \,,
\quad
 g_{I,20}^0 = - 2 \beta_{00} g_{I,10}^1 \,, \quad  g_{I,02}^0 = - 2 \beta_{00}' g_{I,01}^1 \,,
\nonumber\\
 g_{I,30}^0 &= - 2 \left( \beta_{10} g_{I,10}^1 + \beta_{00} \left( 2 \beta_{00} g_{I,10}^2 
                       + g_{I,20}^1 \right) \right) \,,
\quad
 g_{I,21}^0 = - 2  \beta_{01} g_{I,10}^1 - \beta_{00}  g_{I,11}^1  \,,  
\nonumber\\
 g_{I,03}^0 &= - 2 \left( \beta_{01}' g_{I,01}^1 + \beta_{00}' \left( 2 \beta_{00}' g_{I,01}^2 
                       + g_{I,02}^1 \right) \right) \,,
\quad
 g_{I,12}^0 = - 2  \beta_{10}' g_{I,01}^1 - \beta_{00}'  g_{I,11}^1 \,.
\end{align}
It was found in the context of QCD up to three loops and \textcolor{black}{in} QED and \textcolor{black}{QCD$\times$QED} up to two loops that
the constants $B_I$, $f_I$ depend only on the type of external states, not on the operator.  In other words,
the constants $B_q$($f_q$) and $B_b$($f_b$) extracted from $F_q$ and $F_b$ respectively were found to be the same and similarly
$B_g$s ($f_g$s) extracted from $G_{\mu \nu}^a G^{\mu \nu a}$ and $G_{\mu \nu}^a \tilde G^{\mu \nu a}$ were also same
(see \cite{Gehrmann:2014vha,Ahmed:2015qpa}).
Hence, we expect that $B_q^{(i,j)}=B_b^{(i,j)}$ and $f_q^{(i,j)} = f_b^{(i,j)}$ in QED as well as in \textcolor{black}{QCD$\times$QED}.  
However, the anomalous dimensions
for $\gamma_b$ and $\gamma_q$ will be different because they {\color{black} originate} from the UV sector.  Since ${\cal O}_q$ is
a conserved operator, $\gamma_q$ is identically zero to all orders in both $a_s$ and $a_e$ and this is not
the case for $\gamma_b$ which gets contribution from $a_s$ as well as $a_e$ in the perturbative expansion.   
Using the fact that $B_I$ and $f_I$ are operator independent and that $\gamma_q=0$ to all orders, we can obtain
$\gamma_b$ up to three loops in QCD, QED and in \textcolor{black}{QCD$\times$QED} by computing $G_b(\varepsilon)-G_q(\varepsilon)$.    
Thus we obtain $\gamma_b^{(i,0)},\gamma_b^{(0,i)}$ for $i=1,2,3$ from pure QCD , QED and $\gamma_b^{(1,1)}, 
\gamma_b^{(2,1)}$ and $\gamma_b^{(1,2)}$ from \textcolor{black}{QCD$\times$QED} and they are listed in the Appendix \ref{ap:g3}.
Substituting these anomalous dimensions in Eq.(\ref{Zlam}), we obtain 
$Z^b_{\lambda}$ to third order in the couplings.
\subsection{Soft distribution function}
In the following, we show how we can determine \textcolor{black}{collinear} and soft anomalous dimensions from the
soft distribution function.
Unlike $A_I^{(i,j)}$, the 
other anomalous dimensions $B_I^{(i,j)}$, $f_I^{(i,j)}$ and \textcolor{black}{$\gamma_I^{(i,j)}$ 
($\gamma^{(i,j)}_q$ is zero)} can not be disentangled from $\hat F_q$ and $\hat F_b$ alone.  
In order to disentangle $B_I^{(i,j)}$ and $f_I^{(i,j)}$,  we study the partonic cross sections
resulting from soft gluon and soft photon emissions alone as they are sensitive to only $f_I^{(i,j)}$.
The process independent part of 
soft gluon/photon contributions in the real emission sub-processes can be obtained 
following the method described in \cite{Ravindran:2005vv,Ravindran:2006cg},  
where it was demonstrated up to three loops in QCD, the soft distribution function, denoted by $\Phi_I$, 
for the inclusive cross section for producing a colorless state
can be computed using the \textcolor{black}{FFs} and partonic sub-process cross sections involving real emissions of gluons.
For the QED and \textcolor{black}{QCD$\times$QED} we use the respective \textcolor{black}{FFs} and the inclusive cross sections involving photons
as well as gluons that contribute in the soft limit.
In the case of QCD, the soft distribution functions 
were found to be dependent on cusp ($A_I$) and soft ($f_I$) anomalous dimensions, where
$I=q,b,g$.  \textcolor{black}{Up to three loops in QCD, one finds $\Phi_b=\Phi_q=C_F/C_A \Phi_g$
\cite{Ravindran:2005vv,Ravindran:2006cg,Ahmed:2014cla},where $\Phi_g$ was found in \cite{Anastasiou:2014vaa,Li:2014afw}} .
\textcolor{black}{This relation is expected to hold beteween quark and gluon soft distribution functions
because they are defined by the expectation value of certain gauge invariant bi-local quark and gluon operators computed
between on-shell quark and gluons fields.  The Wilson lines made up of guage fields make these bi-local operators gauge invariant.
(see \cite{Sterman:1986aj,Bauer:2000ew, Bauer:2000yr, Bauer:2001yt, Beneke:2002ph, Becher:2014oda,Catani:1989ne}}.
Using the partonic sub-processes of either DY process 
($\hat \sigma_{q \overline q}$)
or the Higgs boson production in bottom quark annihilation  
($\hat \sigma_{b \overline b}$) normalized by the square of the 
bare form factor $\hat F_q$ or $\hat F_b$, we can obtain $\Phi_I$, $i=q,b$. 
$\Phi_I$s  are found to be functions of the scaling variable 
$z = q^2/s$ and $q^2$ is invariant mass square of the
lepton pair for the DY and $q^2=m_h^2$ for the Higgs production.   
Note that $Q^2$ introduced in the form factors is related to $q^2$ by
$Q^2=-q^2$.
We write,
\begin{equation}
\label{phidef}
\mathcal{C} \exp\Big( 2 \Phi_{I}(z)\Big)=
{\hat \sigma_{I\overline I}(z)
\over
Z_I^2 \big| \hat F_I \big|^2 }\,,\quad \quad\quad I = q,b
\end{equation}
where $Z_q=1$ for the ${\cal O}_q$ as $\gamma_q=0$ and $Z^b_{\lambda}$ can be obtained using
Eq.(\ref{Zlam}) in terms of $\gamma_b^{(i,j)}$ known up to three loops in QCD, QED and \textcolor{black}{QCD$\times$QED}. 
The symbol $\mathcal{C}$ refers to ``ordered exponential'' which has the following expansion:
\begin{eqnarray}
\mathcal{C}e^{f(z)} = \delta(1-z) + \frac{1}{1!}f(z) + \frac{1}{2!}(f\otimes f)(z) + \cdots
\quad \quad \quad .
\end{eqnarray}
Here $\otimes$ denotes the Mellin convolution and $f(z)$ is a distribution of the kind $\delta(1-z)$ and $\mathcal{D}_i$. The plus distribution $\mathcal{D}_i$ is defined as,
\begin{equation}
 \mathcal{D}_i = \Bigg( \frac{\ln^i (1-z)}{(1-z)} \Bigg)_+ \,.
\end{equation}
In \cite{H:2019nsw}, \textcolor{black}{we computed  UV finite $\hat \sigma_{I\overline I}$ up to NNLO in QCD, QED and QCD$\times$QED , the two loop corrected bare FFs and the overall renormalisation constant $Z^b_{\lambda}$ to obtain the soft distribution function 
 $\Phi_I$ up to second order in $a_s$ from QCD, in $a_e$ from QED and in $a_s a_e$ from \textcolor{black}{QCD$\times$QED}}.   
We found in \cite{H:2019nsw} that the soft distribution functions extracted from two different processes satisfy a remarkable
relation, namely $\Phi_q=\Phi_b$ up to second order in both the couplings 
$a_s$ and $a_e$ demonstrating the universality among QCD, QED and \textcolor{black}{QCD$\times$QED} results.
It was found in \cite{Ahmed:2014cla} that this relation is valid up to three loops in QCD.   
Hence, we propose that this relation continues to hold true up to three loops \textcolor{black}{even} in QED and in \textcolor{black}{QCD$\times$QED}.
Our next task is to predict $\Phi_q$ (equivalently $\Phi_b$) to third order in QED and \textcolor{black}{QCD$\times$QED}.

Following \cite{Ravindran:2005vv,Ravindran:2006cg, Ahmed:2014cla}
we express the soft distribution function $\Phi_I$ in terms of cusp ($A_I$) 
and soft ($f_I$) anomalous dimensions order by order in perturbation theory. \textcolor{black}{It was also shown in \cite{Ravindran:2005vv,Ravindran:2006cg}, 
that $\Phi_I$  satisfy Sudakov K+G equation analogous to  FF \textcolor{black}{owing to} universal IR structure of these quantities:}
\begin{align}\label{eq:KGphi}
q^2\frac{d}{dq^2}\Phi_I = \frac{1}{2} \Big[ \overline{K}_I \Big(& \{\hat{a}_c\}, \frac{\mu_R^2}{\mu^2},\varepsilon,z \Big) 
+ \overline{G}_I \Big( \{\hat{a}_c\},\frac{q^2}{\mu_R^2},\frac{\mu_R^2}{\mu^2},\varepsilon,z \Big) \Big] \,,
\end{align}
where, \textcolor{black}{$\overline{K}_I$} contains all the IR singularities and the IR finite part is denoted by \textcolor{black}{$\overline{G}_I$}. 
One finds that the \textcolor{black}{RG invariance of $\Phi_I$} leads to
\begin{align}
\mu_R^2\frac{d}{d\mu_R^2}\overline{K}_I = - \mu_R^2\frac{d}{d\mu_R^2}\overline{G}_I = 
A_I(\{a_c(\mu_R^2)\})\delta(1-z) \,.
\end{align}
From the explicit results \cite{Ahmed:2014cla} computed up to second order in QCD, QED and \textcolor{black}{QCD$\times$QED}, we had shown that 
the anomalous dimension $A_I$ are identical to the cusp anomalous dimension
that appears in the form factors $\hat F_I$ confirming the universality of IR structure of   
the underlying gauge theory(ies).  In other words,  
$A_I$ are universal and they govern the evolution of both $K_I,G_I$ and $\overline K_I,
\overline G_I$.

Following the method described in \cite{Ravindran:2005vv,Ravindran:2006cg}, we obtain
\begin{align}\label{Phi}
\Phi_I(\{\hat{a}_c\},
q^2,\mu^2, &\varepsilon,z) = \sum_{i,j}\hat{a}_s^i\hat{a}_e^j\Big(\frac{q^2(1-z)^2}{\mu^2}\Big)^{(i+j)\frac{\varepsilon}{2}}
S_{\varepsilon}^{(i+j)}
\Big(\frac{(i+j)\varepsilon}{1-z}\Big)\hat{\phi}_I^{(i,j)}(\varepsilon)  \,,
\end{align} 
where,
\begin{eqnarray}
\hat{\phi}_I^{(i,j)}(\varepsilon) = \frac{1}{(i+j)\varepsilon}\Big[ \overline{K}_I^{(i,j)}(\varepsilon) + \overline{G}_I^{(i,j)}(\varepsilon) \Big]\,.
\end{eqnarray} 
To obtain $\overline G_I^{(ij)}$, we first expand $\overline{G}_I(\{a_c(q^2)\},1,\varepsilon,z)$ 
in terms of $\overline{G}_I^{(i,j)}(\varepsilon)$ and relate the latter  to $\overline {\cal G}_I^{(i,j)}$ as
\begin{align}
\label{eq:oGI}
\overline{G}_I(\{a_c(q^2)\},1,\varepsilon,z) &=
\sum_{i,j}\hat{a}_s^i\hat{a}_e^j \Big(\frac{q_z^2}{\mu^2}\Big)^{(i+j) \frac{\varepsilon}{2}}  S_\varepsilon^{(i+j)} \overline{G}_I^{(i,j)}(\varepsilon) 
\nonumber\\
&= \sum_{i,j}a_s^i\big(q_z^2\big) a_e^j\big(q_z^2 \big) \overline{\mathcal{G}}_I^{(i,j)}(\varepsilon) \,,
\end{align}
where $q_z^2 = q^2(1-z)^2$ and 
the IR finite $\overline{\mathcal{G}}^{(i,j)}_I (\varepsilon)$ can be expanded 
(following \cite{Ravindran:2005vv,Ravindran:2006cg}) 
as
\begin{align}
\mathcal{\overline G}_I^{(i,j)}(\varepsilon) &= -f_I^{(i,j)} + \sum_{k=0}\varepsilon^k \mathcal{\overline G}_{I,ij}^{(k)}  \,.
\end{align}
Up to third order, one finds
\begin{align}
\mathcal{\overline G}_{I,10}^{(0)} &= 0\,, 
 \qquad
 \mathcal{\overline G}_{I,01}^{(0)} = 0\,,
 \qquad
 \mathcal{\overline G}_{I,11}^{(0)} = 0\,,
 \qquad \quad
 \mathcal{\overline G}_{I,20}^{(0)} = -2 \beta_{00} \mathcal{\overline G}_{I,10}^{(1)}\,,
 \quad 
 \mathcal{\overline G}_{I,02}^{(0)} = -2 \beta_{00}' \mathcal{\overline G}_{I,01}^{(1)} \,,
 \nonumber\\
 \mathcal{\overline G}_{I,30}^{(0)} &= - 2 \beta_{10} \mathcal{\overline G}_{I,10}^{(1)} 
                                      - 2 \beta_{00} \mathcal{\overline G}_{I,20}^{(1)} 
                                      - 4 \beta_{00}^2 \mathcal{\overline G}_{I,10}^{(2)} \,,
\quad
 \mathcal{\overline G}_{I,03}^{(0)} = - 2 \beta_{01}' \mathcal{\overline G}_{I,01}^{(1)} 
                                     - 2 \beta_{00}' \mathcal{\overline G}_{I,02}^{(1)} 
                                     - 4 \beta_{00}^{'2} \mathcal{\overline G}_{I,01}^{(2)} \,,
 \nonumber\\
 \mathcal{\overline G}_{I,21}^{(0)} &= - 2 \beta_{01} \mathcal{\overline G}_{I,10}^{(1)} 
                                      -   \beta_{00} \mathcal{\overline G}_{I,11}^{(1)} \,, 
\quad \qquad \qquad \qquad
 \mathcal{\overline G}_{I,12}^{(0)} = - 2 \beta_{10}' \mathcal{\overline G}_{I,01}^{(1)} 
                                      -   \beta_{00}' \mathcal{\overline G}_{I,11}^{(1)} \,. 
\end{align}
As already mentioned, $\Phi_I$ is known to third order in QCD and to second order 
in pure QED and in mixed QCD$\times$QED. To determine third order contribution to $\Phi_I$
in QED and in mixed QCD$\times$QED,
we require the constants $\{f^{(0,3)}_I,f^{(1,2)}_I,f^{(2,1)}_I\}$ and $\{\overline{\cal G}^{(0)}_{I,12},\overline{\cal G}^{(0)}_{I,21},\overline{\cal G}^{(0)}_{I,03}\}$. 
They can be obtained by computing
the N$^3$LO contributions to Drell-Yan production taking into account QED and QCD$\times$QED effects.  
While this is beyond the scope of our present work, we predict these constants from the corresponding
ones in QCD using certain relations that relate QCD cusp anomalous dimension with the corresponding ones in 
QED and QCD$\times$QED.  We find these relations owing to the fact that
the cusp anomalous dimensions in QCD, QED and QCD$\times$QED can be extracted unambiguously from  the form 
factors of vector and scalar operators order by order in perturbation theory.
Interestingly, in QCD up to three loops, the cusp $A_I^{(i,0)}$, the soft $f_I^{(i,0)}$ and the constants
$\overline{\cal G}_{I,ij}^{(k)}$ contain identical set of color factors, namely at one loop, we have $\{C_F\}$, at two loops
$\{C_F C_A,C_F T_f n_f\}$ and at three loops $\{C_F C_A^2, C_F C_A T_f n_f, C_F^2 T_f n_f,C_F T_f^2 n_f^2\}$.  In other
words, the soft distribution function $\Phi_I$ in QCD demonstrate uniform color factor structure 
at every order in perturbation theory.  This is
in contrast to the constants $A_I, B_I, \gamma_I$ and $G_I$ that contribute to the form factors, which contain different sets of
color factors.  Hence, a uniform and unambiguous relations between QCD, QED and mixed QCD$\times$QED do not exist 
for the latter ones.  Section \ref{sec:ab} is devoted to study of these transformation rules in detail.  
Assuming that uniform color and charge factor structure for $\Phi_I$ in pure QED and mixed QCD$\times$QED
holds true to third order, we apply the relations that relate their cusp anomalous dimensions, 
to predict $f_I^{(i,j)}$, $\overline {\cal G}_{I,ij}^{(k)}$
and hence the entire $\Phi_I$ to third order from the corresponding ones in QCD.  This can be validated only by
explicit computation which is reserved for future publication. 
Now that we have $f_q^{(i,j)}$ to third order, it is straightforward to obtain $B_q^{(i,j)}$ in Eq.(\ref{G}) to the same
accuracy in QCD,QED and \textcolor{black}{QCD$\times$QED}  from
the explicit results on $G^{(i,j)}_q$ as $\gamma^{(i,j)}_q=0$.
Similarly, substituting $f_b^{(i,j)}$ and $\gamma_b^{(i,j)}$ in $G^{(i,j)}_b$ (Eq.(\ref{G}))
we can determine $B_b^{(i,j)}$ up to third order in QCD, QED and \textcolor{black}{QCD$\times$QED}.
We find $B_q^{(i,j)}$ obtained from $G_q^{(i,j)}$ is identical to $B_b^{(i,j)}$ from $G_b^{(i,j)}$, namely $B_q^{(i,j)}=B_b^{(i,j)}$ up to
third order in QCD,QED and \textcolor{black}{QCD$\times$QED}.  \textcolor{black}{The constants $f_I$  and $B_I$  for the pure QCD case are known to three loops in \cite{Moch:2005tm,Moch:2004pa,Vogt:2004mw} and the new ones along with the pre-existing ones are enlisted in 
Appendix \ref{ap:g4} and \ref{ap:g5} respectively} .

\section{Soft-Virtual and Resummed cross sections}
\textcolor{black}{The results obtained so far have two important  phenomenological 
implications. Firstly third order threshold predictions for DY (di-lepton or W/Z production) as well as
for Higgs boson production in bottom quark annihilation in QED and \textcolor{black}{QCD$\times$QED}. Secondly,
the  threshold enhanced resummed predictions in the $N$ Mellin space .}

We begin with the threshold predictions for DY and Higgs boson productions.  Denoting the mass-factorised finite cross-section by
$\Delta^{SV}_I$ and following \cite{Ravindran:2005vv,Ravindran:2006cg}, we find
\textcolor{black}{\begin{eqnarray}
\label{eq:DelSV}
\Delta^{SV}_I(\tau,Q^2,\mu_R^2,\mu_F^2) = Z_I^2(\mu_R^2) |{\hat F}_I(Q^2)|^2 \delta(1-z) \otimes {\cal C} e^{2 \Phi_I(q^2)} \otimes \Gamma_{II}
(\mu_F^2)
\otimes \Gamma_{\overline I \overline I}(\mu_F^2) \,,
\end{eqnarray}}
where $\Gamma_{II}=\Gamma_{\overline I\overline I}$ are \textcolor{black}{Altarelli-Parisi (AP) kernels \cite{ Altarelli:1977zs} that are required to
remove the \textcolor{black}{initial state collinear} singularities}.  
The scale $\mu_F$ is called factorization scale at which
collinear singularities are removed from the partonic cross sections.  
\textcolor{black}{In the above equation, we drop all the regular terms after the convolutions
are performed to obtain only threshold contributions, often called soft plus virtual contributions (SV). } 
In above equation for $\Delta^{SV}_I$, 
the soft and collinear singularities arising from gluons/photons/fermions in the virtual sub-processes 
are \textcolor{black}{guaranteed} to cancel against those from the real sub-processes
when all the degenerate states are summed up, thanks to the KLN theorem \cite{Kinoshita:1962ur, Lee:1964is}.
The remaining initial state collinear singularities are removed by 
mass factorization 
kernels, namely the AP kernels which satisfy renormalisation group equations
\begin{eqnarray}
\mu_F^2 {d \over d \mu_F^2}\Gamma_{II}(z,\mu_F^2) = {1 \over 2} P_{II}(\mu_F^2) \
 \Gamma_{II}(\mu_F^2)\,,
\end{eqnarray}
where $P_{II}(z,\mu_F^2)$ are \textcolor{black}{AP splitting functions known upto three loop level in pure QCD \cite{Moch:2004pa,Vogt:2004mw}}.
In \cite{deFlorian:2015ujt,deFlorian:2016gvk}, these splitting functions up to NNLO level, both in QED and \textcolor{black}{QCD$\times$QED},
were obtained using the abelianization procedure.
The splitting functions that we have obtained \cite{Ahmed:2014cla} by demanding finite-ness of the
mass factorised cross section, agreed with those in \cite{deFlorian:2015ujt}. 
One finds AP splitting functions can be expressed in terms of distribution and regular functions as follows:
\begin{eqnarray}
P_{II} (z,\mu_F^2) = 2 \left({ A_I(\{a_c(\mu_F^2)\}) \over (1-z)_+} +  B_I(\{a_c(\mu_F^2))\}~ \delta(1-z)\right) + P_{reg,II}(z,\mu_F^2)\,.
\end{eqnarray}
Since we are interested only in threshold corrections to \textcolor{black}{finite mass-factorised partonic cross-section} $\Delta_I$, it is sufficient to drop $P_{II}^{reg}$
in $P_{II}$ when computing $\Delta_I^{SV}$ using Eq.(\ref{eq:DelSV}).
Hence, we need only $A_I$ and $B_I$ from $P_{II}$ to obtain $\Gamma_{II}$. Using $Z_I$, \textcolor{black}{${\hat F}_I$}, $\Phi_I$
and $\Gamma_{II}$ to third order in QCD,QED and \textcolor{black}{QCD$\times$QED}, we can readily obtain $\Delta^{SV}_I$ to third order.  
We expand $\Delta^{SV}_I$ as
\begin{eqnarray} \label{DeltaSV}
\Delta^{SV}_I(z,q^2,\mu_F^2) = \sum_{i,j=0} a_s^i(\mu_R^2) a_e^j(\mu_R^2) \Delta^{SV,(i,j)}_I(z,q^2,\mu_F^2,\mu_R^2)\,,
\end{eqnarray}
and present the results for $\Delta^{SV,(i,j)}_I$ in the Appendix \ref{ap:g8}
up to third order in QED and \textcolor{black}{QCD$\times$QED} for $I=q,b$.  Up to two loops, our results for SV agree with
those obtained earlier \cite{deFlorian:2018wcj,H:2019nsw} and results at third order are our predictions using 
the \textcolor{black}{factorisation} properties of the scattering cross section and the universal 
structure of the soft distribution function.  

In the following we exploit these properties to systematically resum certain class of logarithms to all orders
in perturbation theory.  
In QCD, it is well known that threshold 
logarithms of the kind ${\cal D}_i(z)$ spoil the reliability of the fixed order perturbation theory when $z=q^2/\hat s$ 
is closer to threshold namely $z \rightarrow 1$.  These logarithms originate from the soft distribution functions after
the cancellation of soft and collinear singularities against those from the \textcolor{black}{FF} and the AP (mass factorisation) kernels.     
These logarithms when convoluted with the appropriate parton distribution functions to compute the production cross sections 
at the hadronic level, can enhance the cross section provided the latter also dominates.  In other words, the interplay between
perturbative (threshold logarithms) and non-perturbation (parton distribution functions) terms enhance the cross section
at every order in perturbation theory questioning the truncation of the perturbative series.  The resolution to
this problem was provided in  \cite{Sterman:1986aj,Catani:1989ne} which proposed a systematic way of reorganising the
perturbative series through a procedure called threshold resummation.  Working in Mellin space parametrised by a complex
variable $N$, one can resum the order one terms of the form \textcolor{black}{$a_s \beta_{00} \log N $ or $a_e \beta_{00}' \log N$} to all orders
in perturbation theory.  Following \cite{Ravindran:2005vv,Ravindran:2006cg}, it is straightforward to obtain $z$ space
result that is required to obtain resummed result in the Mellin space.   In order to get the $z$ space result, we 
write the soft distribution function $\Phi_I$ as  
\begin{eqnarray}
\label{resumu}
\Phi_I &=& \left(\int_{\mu_F^2}^{q_z^2} {d \lambda^2 \over \lambda^2 (1-z)} A_I\left(a_s(\lambda^2),a_e(\lambda^2)\right)  
+ {1 \over (1-z) } \frac{D_I}{2} \left (a_s(q_z^2),a_e(q_z^2)\right) \right)_+
\nonumber\\
&&+\sum_{i,j=1}^\infty \hat a_s^i \hat a_e^j S_\varepsilon^{i+j} \left({\mu_F^2 \over \mu^2} \right)^{(i+j){\varepsilon \over 2}}
{1 \over (1-z)_+} \overline K_I^{(i,j)} (\varepsilon)
\nonumber\\
&& +\sum_{i,j=1}^\infty a_s^i a_e^j S_\varepsilon^{i+j}  \left({q^2 \over \mu^2} \right)^{(i+j){\varepsilon \over 2}} \delta(1-z)
\hat \phi_I^{(i,j)} (\varepsilon)\,,
\end{eqnarray}
where $D_I$ is related to $\overline G_I$ given in Eq.(\ref{eq:oGI}) through
\begin{eqnarray}
D_I \left (a_s(q_z^2),a_e(q_z^2)\right)= 2 
\overline{G}_I\left(\{a_c(q^2)\},1,\varepsilon,z\right) |_{\varepsilon =0}  \,.
\end{eqnarray}
Up to third order, they are listed in the Appendix \ref{ap:g7}.
The first term in the above expression is finite as $\varepsilon \rightarrow 0$ while the second and third terms contain 
singularities that cancel against those from the AP kernels and the \textcolor{black}{FF}.  
Substituting \textcolor{black}{Eq}.(\ref{resumu}) in \textcolor{black}{Eq}.(\ref{eq:DelSV}), and taking \textcolor{black}{a Mellin transform}, we obtain
\begin{eqnarray}\label{eq:res}
\Delta^{res}_{I,N}(q^2) &=& \int_0^1 d\tau \tau^{N-1} \Delta^{SV}_I (\tau,q^2)\nonumber\\
&=& C_{I,0}(q^2,\mu_R^2,\mu_F^2) \exp \left(G_{I,N}(q^2,\mu_R^2,\mu_F^2)\right)\,,
\end{eqnarray}
where
\begin{equation}
\!\!G_{I,N}\!\!=\!\!\int_0^1 dz z^{N-1}
\left(\int_{\mu_F^2}^{q_z^2} {d \lambda^2 \over \lambda^2 (1-z)} 2A_I\left(a_s(\lambda^2),a_e(\lambda^2)\right)  
+ {1 \over 1-z } D_I \left (a_s(q_z^2),a_e(q_z^2)\right) \right)_+ \!\!\,
\end{equation}  
where $C_{I,0}$ gets contributions \textcolor{black}{from} terms proportional to $\delta(1-z)$ \textcolor{black}{in FF} and the soft distribution function.
All the $\mu_R$ and $\mu_F$ dependent logarithms in $C_0$ come from renormalisation constant and the AP kernels after
the poles in $\varepsilon$ cancel against the form factors and soft distribution functions.   
We have presented $C_{I,0}$ in the \textcolor{black}{Appendix \ref{ap:g9} }for $I=q,b$.

\section{Abelianisation procedure}\label{sec:ab}
In a series of works \cite{deFlorian:2015ujt,deFlorian:2016gvk,deFlorian:2018wcj}, 
the second order contributions to Altarelli-Parisi splitting functions and inclusive cross section for 
Drell-Yan production in QED and in mixed QCD$\times$QED were obtained from the existing QCD results 
using certain transformation rules  
that relate color and charge factors for the relevant Feynman diagrams.
These transformation rules (also called as the Abelianisation rules) were found to hold true \cite{H:2019nsw} for the FFs of vector and scalar operators and the 
inclusive cross section for the production of a Higgs boson in bottom quark annihilation.  
In addition, in \cite{H:2019nsw},  the infrared structure of QED and mixed QCD$\times$QED at NNLO level were studied thereby obtaining the same set of transformation rules
for the anamolous dimensions 
in QED  as well as in QCD$\times$QED from the QCD ones. 
In the following, we discuss, in detail, the existence of such transformation rules at the three loop level.

The explicit computation of FFs to third order shows that for the color factors $C_F^3$, $C_F^2C_A$, $C_F C_A^2$ and 
$C_F n_f^2T_F^2$ in pure QCD, there exists  definite transformation rules that relate FFs, $\Delta_I^{SV}$ and anomalous dimensions of pure QED and mixed QCD$\times$QED to pure QCD at the third order. However for the color factor $C_F^2n_f T_F$ which arises from topologies with single fermion loop, there is no one-to-one mapping from pure QCD to pure QED at $a_e^3$ and to mixed QCD$\times$QED.
We present a general set of such rules obtained from the explicit calculation of FFs in Table [\ref{tab:tablea}].
\begin{table}[h!]
  \begin{center}
  \begin{small}
    \begin{tabular}{|P{2.0cm}||P{3.5cm}|P{3.7cm}||P{3.7cm}|} 
      \hline
      QCD ($a_s^3$)& QCD$\times$QED ($a_s^2 a_e$) & QCD$\times$QED ($a_s a_e^2$)  & QED ($a_e^3$)\\
     \hline \hline 
        
        $C_F^3$  & $3C_F^2e_I^2$&$3 C_F e_I^4$ & $e_I^6$ \\ 
        \cline{1-4}
       $C_A C_F^2$&$C_A C_F e_I^2$& 0  & 0  \\
       \cline{1-4}
	    $C_F^2n_fT_F$& $ a\hspace{2mm} C_FT_F\big(\sum\limits_{q}e_q^2 \big)+ b \hspace{2mm} C_F n_fT_Fe_I^2$ & $ a \hspace{1mm} C_Fe_I^2\big(N\sum\limits_{q}e_q^2 \big) + b \hspace{1mm} C_Fe_I^2\big(N\sum\limits_{q}e_q^2 + \sum\limits_{l}e_l^2\big)$ &$a\hspace{2mm} e_I^2\big(N\sum\limits_{q}e_q^4+ \sum\limits_{l}e_l^4\big)+b\hspace{2mm} e_I^4\big(N\sum\limits_{q}e_q^2 + \sum\limits_{l}e_l^2\big)$ \\
       \cline{1-4}
       $C_FC_A^2$&0 &0 & 0\\
       \cline{1-4}
      $C_FC_A n_fT_F$&0 & 0 &0\\
       \cline{1-4}
       $C_Fn_f^2T_F^2$&0 &0 & $e_I^2\big(N\sum\limits_{q}e_q^2+ \sum\limits_{l}e_l^2\big)^2$\\
      \hline 
    \end{tabular}
    \end{small}
  \end{center}
  \caption{The index $q$ is summed over all the quark charges and index $l$ is summed over all the lepton charges. Moreover index $I=q,b$ corresponding to Drell-Yan pair production and Higgs production in bottom quark annihilation.}
\label{tab:tablea}
\end{table} 
The  coefficients $\{a,b\}$  against the corresponding  color factors depend on the contribution from relevant topologies and 
are dependent on the FFs. Similar set of transformation rules were obtained for $\Delta_I^{SV}$ and anomalous dimensions 
with different coefficients $\{a',b'\}$. However, strikingly for the cusp anomalous dimension, we find that there exists 
a one-to-one mapping from pure QCD to those of mixed QCD$\times$QED and to pure QED for all the color factors. This happens because the contributions from $C_F n_f T_F e_I^2$, {\color{black}$C_F e_I^2  \big(N \sum_{q}e_q^2 + \sum_l e_l^2\big)$}  and $e_I^4(N \sum_{q}e_q^2 + \sum_l e_l^2)$  are absent in $A_I^{(2,1)}$, {\color{black}$A_I^{(1,2)} $} and $A_I^{(3,0)}$ respectively. Thus for the cusp anomalous dimension, the color factor $C_F^2n_f T_F$ maps to 
{\color{black} $C_FT_F\big(\sum\limits_{q}e_q^2\big)$} in $a_s^2a_e$,
{\color{black} $C_F e_I^2\big(N\sum\limits_{q}e_q^2\big)$ in $a_sa_e^2$ }
and to $e_I^2\big(N\sum\limits_{q}e_q^4+ \sum\limits_{l}e_l^4\big)$ in $a_e^3$ as can be seen in Appendix \ref{ap:g2}.

With these observations at hand it is possible to understand the reason behind the transformation rules. We begin with the transformation rule for $C_F^3$. 
This color factor arises from those diagrams where no fermion or gluon loops are present. In the below we show some Feynman diagrams which leads to $C_F^3$ color factor.  
 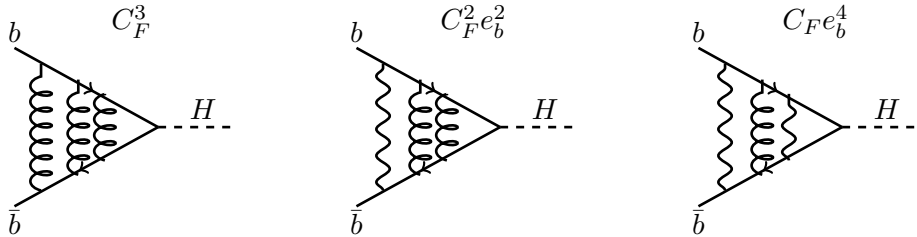
\begin{figure}[h!]
  \begin{tikzpicture}[line width=1 pt, scale=0.7]
  \hspace{1.5cm}
\draw[fermion] (-2.7,1.5) -- (0,0);
\draw[fermion] (0,0) -- (-2.7,-1.5);
\draw[gluon] (-1.5,-0.9) -- (-1.5,0.9);
\draw[gluon] (-2.2,-1.2) -- (-2.2,1.2);
\draw[gluon] (-1.0,-0.63) -- (-1.0,0.63);
\draw[scalarnoarrow] (0,0) -- (1.5,0);
\node at (-2.7,1.8) {$b$};
\node at (-2.7,-1.8) {$\bar{b}$};
\node at (0.85,0.35) {$H$};
\node at (-0.5,2) {$C_F^3$};
\hspace{4.5cm}
\draw[fermion] (-2.7,1.5) -- (0,0);
\draw[fermion] (0,0) -- (-2.7,-1.5);
\draw[gluon] (-1.5,-0.9) -- (-1.5,0.9);
\draw[vector] (-2.2,-1.2) -- (-2.2,1.2);
\draw[gluon] (-1.0,-0.63) -- (-1.0,0.63);
\draw[scalarnoarrow] (0,0) -- (1.5,0);
\node at (-2.7,1.8) {$b$};
\node at (-2.7,-1.8) {$\bar{b}$};
\node at (0.85,0.35) {$H$};
\node at (-0.5,2) {$C_F^2e_b^2$};
\hspace{4.5cm}
\draw[fermion] (-2.7,1.5) -- (0,0);
\draw[fermion] (0,0) -- (-2.7,-1.5);
\draw[gluon] (-1.5,-0.9) -- (-1.5,0.9);
\draw[vector] (-2.2,-1.2) -- (-2.2,1.2);
\draw[vector] (-1.0,-0.63) -- (-1.0,0.63);
\draw[scalarnoarrow] (0,0) -- (1.5,0);
\node at (-2.7,1.8) {$b$};
\node at (-2.7,-1.8) {$\bar{b}$};
\node at (0.85,0.35) {$H$};
\node at (-0.5,2) {$C_Fe_b^4$};
\end{tikzpicture}
\caption{An example of Feynman diagram which contributes to the color factors $C_F^3, C_F^2e_b^2,  C_Fe_b^4$.} 
\end{figure}
 \vspace{-\lineskip}
The numerical factor of three at $a_s^2a_e$ order accounts  for the number of ways a gluon field can be replaced by a photon field in a pure QCD Feynman diagram. 
For instance at the $a_s^2a_e$ order, the factor of three against $C_F^2e_I^2$ corresponds to the number of ways a gluon can be replaced by a photon for a particular pure QCD diagram. 
Having found the reason we anticipate that the  transformation rule for $C_F^3$ can be generalised for higher orders. For this, let $N_1$ denote the number of gluon fields in any pure QCD diagram, then the number of ways $N_2$ photon fields  and $N_1-N_2$ gluon fields can be arranged in any QCD$\times$QED diagram is 
\begin{equation}
C_f^{N_1} \rightarrow \frac{N_1!}{(N_1-N_2)!N_2!}C_f^{N_1-N_2}(e_q^2)^{N_2}
\end{equation}
Next we discuss the color factor $C_A^2C_F$, which arises from diagrams involving three gluon vertices.
Such diagrams are  absent in the mixed case as well as in pure QED case due to the absence of self-interaction vertices. Similarly the color factor $C_F n_f^2 T_F^2$ corresponds to diagrams shown in Table [\ref{tab:tableb}] and due to trace-less property of Gell-Mann matrices they will be absent in mixed QCD$\times$QED. Hence, for higher loop configurations, 
we can anticipate that diagrams where the internal lines are connected only by gluon loops or only by fermion loops will always be absent in the mixed case. 
But for the color factor $C_F^2n_fT_F$, the rules are not so definite and cannot be extended unambiguously to higher orders. 
This is due to the fact that at the third order some of the topologies, which mainly come from the single fermion loop configurations lack the aforementioned one-to-one mapping from QCD to QED and to QCD$\times$QED. In Table [\ref{tab:tableb}] we show some of those configurations which  demonstrate the ambiguous transformations. 
\begin{small}
\begin{table}[H]
  \begin{center}
    \begin{tabular}{|P{3.0cm}||P{3.0cm}|P{3.0cm}||P{3.0cm}|} 
      \hline
      $a_s^3$&  $a_s^2 a_e$ & $a_s a_e^2$  & $a_e^3$\\
     \hline 
     \hline 
     \begin{minipage}{2\textwidth} 
        \begin{figure}[H]
         \begin{tikzpicture}[line width=1 pt, scale=0.65]
\draw[fermion] (-2.5,1.5) -- (0,0);
\draw[fermion] (0,0) -- (-2.5,-1.5);
\draw[gluon] (-1.9,-1.17) -- (-1.9,-0.19);
\draw[fermion] (-1.9,0.00) circle (.2);
\draw[gluon] (-1.9,0.19) -- (-1.9,1.17);
\draw[gluon] (-1.0,-0.63) -- (-1.0,0.63);
\draw[scalarnoarrow] (0,0) -- (1.5,0);
\node at (-2.7,1.6) {$b$};
\node at (-2.7,-1.6) {$\bar{b}$};
\node at (0.85,0.35) {$H$};
\node at (-0.5,2.5) {\scriptsize{$C_F^2n_fT_F$}};
\end{tikzpicture}
\end{figure}
\end{minipage}
 &  
 \begin{minipage}{2\textwidth} 
        \begin{figure}[H]
         \begin{tikzpicture}[line width=1 pt, scale=0.65]
\draw[fermion] (-2.5,1.5) -- (0,0);
\draw[fermion] (0,0) -- (-2.5,-1.5);
\draw[gluon] (-1.9,-1.17) -- (-1.9,-0.19);
\draw[fermion] (-1.9,0.00) circle (.2);
\draw[gluon] (-1.9,0.19) -- (-1.9,1.17);
\draw[vector] (-1.0,-0.63) -- (-1.0,0.63);
\draw[scalarnoarrow] (0,0) -- (1.5,0);
\node at (-2.7,1.6) {$b$};
\node at (-2.7,-1.6) {$\bar{b}$};
\node at (0.85,0.35) {$H$};
\node at (-0.5,2.5) {\scriptsize{$C_Fn_fT_Fe_b^2$}};
\end{tikzpicture}
\end{figure}
\end{minipage}
&
\begin{minipage}{2\textwidth} 
        \begin{figure}[H]
         \begin{tikzpicture}[line width=1 pt, scale=0.65]
\draw[fermion] (-2.5,1.5) -- (0,0);
\draw[fermion] (0,0) -- (-2.5,-1.5);
\draw[vector] (-1.9,-1.17) -- (-1.9,-0.19);
\draw[fermion] (-1.9,0.00) circle (.2);
\draw[vector] (-1.9,0.19) -- (-1.9,1.17);
\draw[gluon] (-1.0,-0.63) -- (-1.0,0.63);
\draw[scalarnoarrow] (0,0) -- (1.5,0);
\node at (-2.7,1.6) {$b$};
\node at (-2.7,-1.6) {$\bar{b}$};
\node at (0.85,0.35) {$H$};
\node at (-0.5,2.5) {\scriptsize{$C_Fe_b^2\big(N\sum\limits_{q}e_q^2 + \sum\limits_{l}e_l^2 \big)$ \qquad}};
\end{tikzpicture}
\end{figure}
\end{minipage} &
\begin{minipage}{2\textwidth} 
        \begin{figure}[H]
         \begin{tikzpicture}[line width=1 pt, scale=0.65]
\draw[fermion] (-2.5,1.5) -- (0,0);
\draw[fermion] (0,0) -- (-2.5,-1.5);
\draw[vector] (-1.9,-1.17) -- (-1.9,-0.19);
\draw[fermion] (-1.9,0.00) circle (.2);
\draw[vector] (-1.9,0.19) -- (-1.9,1.17);
\draw[vector] (-1.0,-0.63) -- (-1.0,0.63);
\draw[scalarnoarrow] (0,0) -- (1.5,0);
\node at (-2.7,1.6) {$b$};
\node at (-2.7,-1.6) {$\bar{b}$};
\node at (0.85,0.35) {$H$};
\node at (-0.5,2.5) {\scriptsize{$e_b^4\big(N\sum\limits_{q}e_q^2+ \sum\limits_{l}e_l^2 \big)$ \qquad}};
\end{tikzpicture}
\end{figure}
\end{minipage} \\ 
      \cline{1-4}
\begin{minipage}{2\textwidth} 
        \begin{figure}[H]
         \begin{tikzpicture}[line width=1 pt, scale=0.6]
\draw[fermion] (-2.7,1.7) -- (0,0);
\draw[fermion] (0,0) -- (-2.7,-1.7);
\draw[gluon] (-2.10,-1.35) -- (-2.10,-0.55);
\draw[fermion] (-2.3,0.00) circle (.6);
\draw[gluon] (-2.10,0.55) -- (-2.10,1.35);
\draw[gluon] (-2.55,0.53) -- (-2.55,-0.5);
\draw[scalarnoarrow] (0,0) -- (1.5,0);
\node at (-2.9,1.6) {$b$};
\node at (-2.9,-1.6) {$\bar{b}$};
\node at (0.85,0.35) {$H$};
\node at (-0.5,2.4) {\scriptsize{$C_F^2 n_f T_F$}};
\end{tikzpicture}
\end{figure}
\end{minipage}        
&
\begin{minipage}{2\textwidth} 
        \begin{figure}[H]
         \begin{tikzpicture}[line width=1 pt, scale=0.6]
\draw[fermion] (-2.7,1.7) -- (0,0);
\draw[fermion] (0,0) -- (-2.7,-1.7);
\draw[gluon] (-2.10,-1.35) -- (-2.10,-0.55);
\draw[fermion] (-2.3,0.00) circle (.6);
\draw[gluon] (-2.10,0.55) -- (-2.10,1.35);
\draw[vector] (-2.55,0.53) -- (-2.55,-0.5);
\draw[scalarnoarrow] (0,0) -- (1.5,0);
\node at (-2.9,1.6) {$b$};
\node at (-2.9,-1.6) {$\bar{b}$};
\node at (0.85,0.35) {$H$};
\node at (-0.5,2.5) {\scriptsize{$C_F T_F\big(\sum\limits_{q}e_q^2\big)$ \qquad}};
\end{tikzpicture}
\end{figure}
\end{minipage}  
&
\begin{minipage}{2\textwidth} 
        \begin{figure}[H]
         \begin{tikzpicture}[line width=1 pt, scale=0.6]
\draw[fermion] (-2.7,1.7) -- (0,0);
\draw[fermion] (0,0) -- (-2.7,-1.7);
\draw[vector] (-2.10,-1.35) -- (-2.10,-0.55);
\draw[fermion]  (-2.3,0.00) circle (.6);
\draw[vector] (-2.10,0.55) -- (-2.10,1.35);
\draw[gluon] (-2.55,0.53) -- (-2.55,-0.5);
\draw[scalarnoarrow] (0,0) -- (1.5,0);
\node at (-2.9,1.6) {$b$};
\node at (-2.9,-1.6) {$\bar{b}$};
\node at (0.85,0.35) {$H$};
\node at (-0.5,2.5) {\scriptsize{$ C_Fe_b^2\big(N\sum\limits_{q}e_q^2 \big)$ \qquad}};
\end{tikzpicture}
\end{figure}
\end{minipage}   
&
\begin{minipage}{2\textwidth} 
        \begin{figure}[H]
         \begin{tikzpicture}[line width=1 pt, scale=0.6]
\draw[fermion] (-2.7,1.7) -- (0,0);
\draw[fermion] (0,0) -- (-2.7,-1.7);
\draw[vector] (-2.10,-1.35) -- (-2.10,-0.55);
\draw[fermion]  (-2.3,0.00) circle (.6);
\draw[vector] (-2.10,0.55) -- (-2.10,1.35);
\draw[vector] (-2.55,0.53) -- (-2.55,-0.5);
\draw[scalarnoarrow] (0,0) -- (1.5,0);
\node at (-2.9,1.6) {$b$};
\node at (-2.9,-1.6) {$\bar{b}$};
\node at (0.85,0.35) {$H$};
\node at (-0.5,2.5) {\scriptsize{$e_b^2\big(N\sum\limits_{q}e_q^4 + \sum\limits_{l}e_l^4 \big)$\qquad}};
\end{tikzpicture}
\end{figure}
\end{minipage}   \\
       \cline{1-4}
\begin{minipage}{2\textwidth} 
        \begin{figure}[H]
         \begin{tikzpicture}[line width=1 pt, scale=0.6]
\draw[fermion] (-2.7,1.7) -- (0,0);
\draw[fermion] (0,0) -- (-2.7,-1.7);
\draw[gluon] (-2.05,-1.3) -- (-2.05,-0.49);
\draw[fermion] (-2.1,0.00) circle (.5);
\draw[gluon] (-2.05,0.49) -- (-2.05,1.3);
\draw[gluon] (-2.6,0.0) -- (-1.60,0.0);
\draw[scalarnoarrow] (0,0) -- (1.5,0);
\node at (-2.9,1.6) {$b$};
\node at (-2.9,-1.6) {$\bar{b}$};
\node at (0.85,0.35) {$H$};
\node at (-0.5,3) {\scriptsize{$\big(C_F^2 - \frac{C_AC_F}{2}\big)n_fT_F$\qquad}};
\end{tikzpicture}
\end{figure}
\end{minipage}        
&
\begin{minipage}{2\textwidth} 
        \begin{figure}[H]
         \begin{tikzpicture}[line width=1 pt, scale=0.6]
\draw[fermion] (-2.7,1.7) -- (0,0);
\draw[fermion] (0,0) -- (-2.7,-1.7);
\draw[gluon] (-2.05,-1.3) -- (-2.05,-0.49);
\draw[fermion] (-2.1,0.00) circle (.5);
\draw[gluon] (-2.05,0.49) -- (-2.05,1.3);
\draw[vector] (-2.6,0.0) -- (-1.60,0.0);
\draw[scalarnoarrow] (0,0) -- (1.5,0);
\node at (-2.9,1.6) {$b$};
\node at (-2.9,-1.6) {$\bar{b}$};
\node at (0.85,0.35) {$H$};
\node at (-0.5,3) {\scriptsize{$C_FT_F\big(\sum\limits_{q}e_q^2 \big)$}};
\end{tikzpicture}
\end{figure}
\end{minipage}  
&
\begin{minipage}{2\textwidth} 
        \begin{figure}[H]
         \begin{tikzpicture}[line width=1 pt, scale=0.6]
\draw[fermion] (-2.7,1.7) -- (0,0);
\draw[fermion] (0,0) -- (-2.7,-1.7);
\draw[vector] (-2.05,-1.3) -- (-2.05,-0.49);
\draw[fermion] (-2.1,0.00) circle (.5);
\draw[vector] (-2.05,0.49) -- (-2.05,1.3);
\draw[gluon] (-2.6,0.0) -- (-1.60,0.0);
\draw[scalarnoarrow] (0,0) -- (1.5,0);
\node at (-2.9,1.6) {$b$};
\node at (-2.9,-1.6) {$\bar{b}$};
\node at (0.85,0.35) {$H$};
\node at (-0.6,3) {\scriptsize{$ C_Fe_b^2\big(N\sum\limits_{q}e_q^2  \big)$ \qquad}};
\end{tikzpicture}
\end{figure}
\end{minipage}   
&
\begin{minipage}{2\textwidth} 
        \begin{figure}[H]
         \begin{tikzpicture}[line width=1 pt, scale=0.6]
\draw[fermion] (-2.7,1.7) -- (0,0);
\draw[fermion] (0,0) -- (-2.7,-1.7);
\draw[vector] (-2.05,-1.3) -- (-2.05,-0.49);
\draw[fermion] (-2.1,0.00) circle (.5);
\draw[vector] (-2.05,0.49) -- (-2.05,1.3);
\draw[vector] (-2.6,0.0) -- (-1.60,0.0);
\draw[scalarnoarrow] (0,0) -- (1.5,0);
\node at (-2.9,1.6) {$b$};
\node at (-2.9,-1.6) {$\bar{b}$};
\node at (0.85,0.35) {$H$};
\node at (-0.5,3) {\scriptsize{$e_b^2\big(N\sum\limits_{q}e_q^4 + \sum\limits_{l}e_l^4\big)$}};
\end{tikzpicture}
\end{figure}
\end{minipage}   \\
\cline{1-4}
       \begin{minipage}{2\textwidth} 
        \begin{figure}[H]
        \hspace{-.3cm}
         \begin{tikzpicture}[line width=1 pt, scale=0.6]
        \draw[fermion] (-3.5,2.5) -- (0,0);
\draw[fermion] (0,0) -- (-1.5,-1.2);
\draw[gluon] (-2.5,0.73) -- (-2.5,1.8);
\draw[fermion] (-2.5,0.50) circle (.2);
\draw[gluon]  (-2.5,0.29) -- (-2.5,-1.05);
\draw[fermion] (-2.5,-1.05) -- (-2.9,-2.2);
\draw[fermion] (-1.5,-1.2) -- (-2.5,-1.05);
\draw[gluon] (-1.5,-1.2) -- (-2.9,-2.2);
\draw[fermion] (-2.9,-2.2) -- (-3.5,-2.5);
\draw[scalarnoarrow] (0,0) -- (1.5,0);
\node at (-3.7,1.6) {$b$};
\node at (-3.7,-1.6) {$\bar{b}$};
\node at (0.85,0.35) {$H$};
\node at (-1.5,3.6) {\scriptsize{$\big(C_F^2 - \frac{C_AC_F}{2}\big)n_fT_F$}};
\end{tikzpicture}
\end{figure}
\end{minipage}
&
\begin{minipage}{2\textwidth} 
        \begin{figure}[H]
        \hspace{-.3cm}
         \begin{tikzpicture}[line width=1 pt, scale=0.6]
        \draw[fermion] (-3.5,2.5) -- (0,0);
\draw[fermion] (0,0) -- (-1.5,-1.2);
\draw[gluon] (-2.5,0.73) -- (-2.5,1.8);
\draw[fermion] (-2.5,0.50) circle (.2);
\draw[gluon]  (-2.5,0.29) -- (-2.5,-1.05);
\draw[fermion] (-2.5,-1.05) -- (-2.9,-2.2);
\draw[fermion] (-1.5,-1.2) -- (-2.5,-1.05);
\draw[vector] (-1.5,-1.2) -- (-2.9,-2.2);
\draw[fermion] (-2.9,-2.2) -- (-3.5,-2.5);
\draw[scalarnoarrow] (0,0) -- (1.5,0);
\node at (-3.7,1.6) {$b$};
\node at (-3.7,-1.6) {$\bar{b}$};
\node at (0.85,0.35) {$H$};
\node at (-1.5,3.6) {\scriptsize{$C_F n_f T_F e_b^2$}};
\end{tikzpicture}
\end{figure}
\end{minipage}
& 
\begin{minipage}{2\textwidth} 
        \begin{figure}[H]
        \hspace{-.3cm}
         \begin{tikzpicture}[line width=1 pt, scale=0.6]
        \draw[fermion] (-3.5,2.5) -- (0,0);
\draw[fermion] (0,0) -- (-1.5,-1.2);
\draw[vector] (-2.5,0.73) -- (-2.5,1.8);
\draw[fermion] (-2.5,0.50) circle (.2);
\draw[vector]  (-2.5,0.29) -- (-2.5,-1.05);
\draw[fermion] (-2.5,-1.05) -- (-2.9,-2.2);
\draw[fermion] (-1.5,-1.2) -- (-2.5,-1.05);
\draw[gluon] (-1.5,-1.2) -- (-2.9,-2.2);
\draw[fermion] (-2.9,-2.2) -- (-3.5,-2.5);
\draw[scalarnoarrow] (0,0) -- (1.5,0);
\node at (-3.7,1.6) {$b$};
\node at (-3.7,-1.6) {$\bar{b}$};
\node at (0.85,0.35) {$H$};
\node at (-1.5,3.6) {\scriptsize{$C_Fe_b^2\big(N\sum\limits_{q}e_q^2 + \sum\limits_{l}e_l^2\big)$}};
\end{tikzpicture}
\end{figure}
\end{minipage}
& 
\begin{minipage}{2\textwidth} 
        \begin{figure}[H]
        \hspace{-.3cm}
         \begin{tikzpicture}[line width=1 pt, scale=0.6]
        \draw[fermion] (-3.5,2.5) -- (0,0);
\draw[fermion] (0,0) -- (-1.5,-1.2);
\draw[vector] (-2.5,0.73) -- (-2.5,1.8);
\draw[fermion] (-2.5,0.50) circle (.2);
\draw[vector]  (-2.5,0.29) -- (-2.5,-1.05);
\draw[fermion] (-2.5,-1.05) -- (-2.9,-2.2);
\draw[fermion] (-1.5,-1.2) -- (-2.5,-1.05);
\draw[vector] (-1.5,-1.2) -- (-2.9,-2.2);
\draw[fermion] (-2.9,-2.2) -- (-3.5,-2.5);
\draw[scalarnoarrow] (0,0) -- (1.5,0);
\node at (-3.7,1.6) {$b$};
\node at (-3.7,-1.6) {$\bar{b}$};
\node at (0.85,0.35) {$H$};
\node at (-1.5,3.6) {\scriptsize{$e_b^4\big(N\sum\limits_{q}e_q^2 + \sum\limits_{l}e_l^2\big)$}};
\end{tikzpicture}
\end{figure}
\end{minipage}\\     
       \cline{1-4}
       \begin{minipage}{2\textwidth} 
        \begin{figure}[H]
        \hspace{-.3cm}
         \begin{tikzpicture}[line width=1 pt, scale=0.6]
        \draw[fermion] (-3.5,2.5) -- (0,0);
\draw[fermion] (0,0) -- (-3.5,-2.5);
\draw[gluon] (-2.5,-1.8) -- (-2.5,-0.73);
\draw[fermion] (-2.5,-0.50) circle (.2);
\draw[gluon]  (-2.5,0.29) -- (-2.5,-0.29);
\draw[fermion] (-2.5,0.50) circle (.2);
\draw[gluon] (-2.5,0.73) -- (-2.5,1.8);
\draw[scalarnoarrow] (0,0) -- (1.5,0);
\node at (-3.7,1.6) {$b$};
\node at (-3.7,-1.6) {$\bar{b}$};
\node at (0.85,0.35) {$H$};
\node at (-0.5,3) {\scriptsize{$C_Fn_f^2T_F^2$}};
\end{tikzpicture}
\end{figure}
\end{minipage}
&
\begin{minipage}{2\textwidth} 
        \begin{figure}[H]
        \hspace{-.3cm}
         \begin{tikzpicture}[line width=1 pt, scale=0.6]
        \draw[fermion] (-3.5,2.5) -- (0,0);
\draw[fermion] (0,0) -- (-3.5,-2.5);
\draw[gluon] (-2.5,-1.8) -- (-2.5,-0.73);
\draw[fermion] (-2.5,-0.50) circle (.2);
\draw[vector]  (-2.5,0.29) -- (-2.5,-0.29);
\draw[fermion] (-2.5,0.50) circle (.2);
\draw[gluon] (-2.5,0.73) -- (-2.5,1.8);
\draw[scalarnoarrow] (0,0) -- (1.5,0);
\node at (-3.7,1.6) {$b$};
\node at (-3.7,-1.6) {$\bar{b}$};
\node at (0.85,0.35) {$H$};
\node at (-0.5,3) {0};
\end{tikzpicture}
\end{figure}
\end{minipage} 
& 
\begin{minipage}{2\textwidth} 
        \begin{figure}[H]
        \hspace{-.3cm}
         \begin{tikzpicture}[line width=1 pt, scale=0.6]
        \draw[fermion] (-3.5,2.5) -- (0,0);
\draw[fermion] (0,0) -- (-3.5,-2.5);
\draw[vector] (-2.5,-1.8) -- (-2.5,-0.73);
\draw[fermion] (-2.5,-0.50) circle (.2);
\draw[gluon]  (-2.5,0.29) -- (-2.5,-0.29);
\draw[fermion] (-2.5,0.50) circle (.2);
\draw[vector] (-2.5,0.73) -- (-2.5,1.8);
\draw[scalarnoarrow] (0,0) -- (1.5,0);
\node at (-3.7,1.6) {$b$};
\node at (-3.7,-1.6) {$\bar{b}$};
\node at (0.85,0.35) {$H$};
\node at (-0.5,3) {0};
\end{tikzpicture}
\end{figure}
\end{minipage}
& 
\begin{minipage}{2\textwidth} 
        \begin{figure}[H]
     \hspace{-.3cm}
         \begin{tikzpicture}[line width=1 pt, scale=0.6]
        \draw[fermion] (-3.5,2.5) -- (0,0);
\draw[fermion] (0,0) -- (-3.5,-2.5);
\draw[vector] (-2.5,-1.8) -- (-2.5,-0.73);
\draw[fermion] (-2.5,-0.50) circle (0.2);
\draw[vector]  (-2.5,0.29) -- (-2.5,-0.29);
\draw[fermion] (-2.5,0.50) circle (.2);
\draw[vector] (-2.5,0.73) -- (-2.5,1.8);
\draw[scalarnoarrow] (0,0) -- (1.5,0);
\node at (-3.7,1.6) {$b$};
\node at (-3.7,-1.6) {$\bar{b}$};
\node at (0.85,0.35) {$H$};
\node at (-0.5,3) {\scriptsize{$e_b^2\big(N\sum\limits_{q}e_q^2 + \sum\limits_{l}e_l^2 \big)^2$ \qquad \qquad}};
\end{tikzpicture}
\end{figure}
\end{minipage}\\
       \hline 
    \end{tabular}
  \end{center}
  \caption{Flavour and charge distributions for some pure QCD, pure QED and QCD$\times$QED loop configurations. Color factors are obtained for the $b\bar{b}$ channel after conjugating with born amplitude and taking the color average up to an overall $\frac{1}{N}$ factor.}
  \label{tab:tableb}
\end{table}
\end{small}
From the table we can see that the coefficient corresponding to color factor $C_F^2 n_f T_F$ in
QCD splits and gives rise to two different color-charge (charge) factor contributions in
QCD$\times$QED (QED). This accounts for the fact that QCD is flavor blind whereas QED is not and hence the above transformations.
\\
 Thus at higher orders, more of such loop configurations will open up which will lead to ambiguous mapping from QCD to QED as well as to QCD$\times$QED. Such one-to-many mappings are observed even in case of UV (Appendix \ref{ap:g3})  and collinear anomalous dimensions (Appendix \ref{ap:g5}) as well as for $\Delta_{I}^{SV}$ (Appendix \ref{ap:g8}). But we find that the cusp anomalous dimension is free from such ambiguity and it could be due to the fact that it contains the soft gluon or photon contribution
which is universal. 

In summary, we infer that the abelianisation procedure \cite{H:2019nsw,deFlorian:2018wcj} which succeeded in giving definite color transformation rules at the two loop level without explicit calculation fails at the three loop level. At two loop level 
in QCD$\times$QED, the single fermion loop diagrams do not contribute. Hence, taking the abelian limit of the pure QCD result is straightforward. On the other hand, in the case of two loop QED, although the single fermion loop diagrams contribute, 
still switching off 
diagrams involving three gluon vertices was sufficient to reproduce pure QED results from pure QCD ones. But at three loops, closed fermion loop configurations map to different charge-color factors in QED and QCD$\times$QED and hence taking abelian limit of the pure QCD FF results does not produce pure QED as well as QCD$\times$QED results. So, the fact that the coefficients $\{a,b\}$ for  QCD$\times$QED and pure QED color factors can only be fixed by explicit calculation, limits the use of abelianisation procedure beyond 
NNLO.
\section{Summary and Conclusions}
In this paper, we have studied the infrared structure of a theory which is invariant 
under $SU(N) \times U(1)$ (QCD $\times$ QED), containing $n_f$ number
of quarks and $n_l$ number of leptons with their respective anti-particles. We have treated all the quarks and 
leptons massless throughout.  We considered two inclusive reactions at hadron colliders, namely 
production of a pair of leptons through quark anti-quark annihilation 
and production of a Higgs boson in bottom quark annihilation 
as theoretical laboratories.  We used the parton model throughout.  
In the parton model, one factorises the the
hadron cross section into IR safe partonic cross sections and 
parton distribution functions.  The former ones being computable order by order in perturbation theory,
are expanded in double series expansion of the gauge couplings $a_s$ and $a_e$ of QCD and QED respectively 
to include 
radiative corrections.  The computation of these corrections beyond leading order
in perturbation theory provides ample opportunity to understand both UV and IR structures of the underlying gauge theory.
In our case, the computation of IR safe partonic cross sections can help to understand the contributions
from pure virtual and real Feynman diagrams and virtual-real diagrams both from QCD and QED.  While
UV divergences go away after including appropriate renormalisation constants, we are left with soft and collinear
divergences in virtual and real subprocesses.  In order to shed light on the IR structure, we have restricted
ourselves to the computation where only soft and virtual contributions to the partonic subprocess are included
in a IR safe way.  In particular, we have computed those contributions that result from threshold region alone.
This is achieved by appropriately combining entire pure virtual contributions with the soft part of the
certain partonic subprocesses where at least one real radiation is present and with the AP kernels
computed  in the threshold limit.  Each part of the computation involves careful study of its IR structure.
The form factors are shown to satisfy K+G equations up to third order in perturbation theory. The renormalisation
group invariance of the form factors can be used to obtain the universal cusp anomalous dimensions $A_I$
of the underlying theory in powers
of both $a_s$ and $a_e$ to third order.  We find that the abelianization relations hold among 
the coefficients $A_I^{(i,j)}$ at various orders in couplings
irrespective of $I$.  Assuming the universal structure of the single poles of the form factors, we determined for the
first time the renormalisation constant for the Yukawa coupling at third order both in pure QED and \textcolor{black}{QCD$\times$QED}.   
Using the abelianization rules 
obtained from the results of $A_I^{(i,j)}$, we have determined the constants 
$f_I^{(0,i)}, i=1,2,3$ and $f_I^{(1,1)},f_I^{(2,1)},f_I^{(1,2)}$ of QED and \textcolor{black}{QCD$\times$QED}. 
These are our predictions for $f_I$ in  QED and \textcolor{black}{QCD$\times$QED} up to third order irrespective of $I$.  
From the knowledge of $f_I^{(i,j)}$ up to third order, we can
determine the corresponding $B_I^{(i,j)}$ for QED and \textcolor{black}{QCD$\times$QED}.  Interestingly, we find that 
abelianization rules that we obtained at third order for $A_I^{(i,j)}$ do not work for $B_I$.  
Since we have explicitly computed the form factors up to third order,  
it is easy to find that certain color factors of the form factors at third order in QCD can come 
from different kinds of topologies while these topologies in pure QED and \textcolor{black}{QCD$\times$QED} cases can give 
different charge and color-charge factors.  In other words, there is 
no one to one mapping between color factors of form factors in QCD and charge or charge-color factors
in QED or \textcolor{black}{QCD$\times$QED}.  \textcolor{black}{However, these topologies do not contribute to $A_I$ allowing us to find
consistent abelianization rules for them, but the UV renormalisation constants and collinear anomalous dimensions $B_I$
in QCD get contributions from them}.  Hence, we fail to find consistent abelianization rules for
them.  In summary, we have determined $\beta$ function coefficients, cusp and soft anomalous dimensions 
from the \textcolor{black}{FFs} computed to third order in QCD,QED and \textcolor{black}{QCD$\times$QED}. We have predicted
the soft distribution function $\Phi_I$ 
by applying the abelianization rules on the corresponding ones in QCD  
and extracted collinear anomalous dimension $B_I$ from the explicit results of the form factors. 
Using these ingredients,  we have obtained the soft plus virtual cross section to
third order and resummed cross section to N$^3$LL accuracy in QED and \textcolor{black}{QCD$\times$QED}.
   
\section{Acknowledgements}
We thank T. Ahmed, P. Banerjee,  A. Chakraborty, P.K. Dhani, G. Ferrera, N. Rana and A. Vicini, for useful
discussions.  We sincerely thank T. Gehrmann for providing us three loop master integrals required for the present 
work. We are also grateful to Andrea Autieri for notifying some corrections in the paper.
\section{Appendix}
\appendix
\section{\texorpdfstring{$K_{I}^{(i,j)}$}{Kij}s in the form factor} \label{ap:KFF}
The constants $K_I^{(i,j)}$ in the form factor are given by,
\begin{small}
\begin{flalign}\label{eq:KFF}
K_I^{(1, 0)} &= \frac{1}{\varepsilon}\Big(-2A_I^{(1, 0)}\Big) 
\,, \quad \qquad
K_I^{(0, 1)} = \frac{1}{\varepsilon}\Big(-2A_I^{(0, 1)}\Big)
\,, \quad \qquad
K_I^{(1, 1)} = \frac{1}{\varepsilon}\Big(-A_I^{(1, 1)}\Big) 
\,, \nonumber\\ 
K_I^{(2, 0)} &= \frac{1}{\varepsilon^2}\Big(2\beta_{00} A_I^{(1, 0)}\Big) + \frac{1}{\varepsilon}\Big( -A_I^{(2,0)}\Big) 
\,, \quad
K_I^{(0, 2)} = \frac{1}{\varepsilon^2}\Big(2\beta_{00}' A_I^{(0, 1)} \Big) + \frac{1}{\varepsilon}\Big( -A_I^{(0, 2)}\Big) \,, \qquad
\nonumber\\
K_I^{(3,0)} & = \frac{1}{\varepsilon^3}\Big(-\frac{8}{3} \beta_{00}^2 A_I^{(1, 0)} \Big) 
              + \frac{1}{\varepsilon^2}
      \Big( \frac{8}{3} \beta_{00} A_I^{(2,0)} + \frac{2}{3} \beta_{10} A_I^{(1,0)} \Big)
      + \frac{1}{\varepsilon} \Big( -\frac{2}{3} A_I^{(3,0)} \Big)  \,,
\nonumber\\
K_I^{(1, 2)} &= \frac{1}{\varepsilon^2}\Big(  \frac{4}{3} \beta'_{00} A_I^{(1,1)} 
                                      + \frac{2}{3} \beta'_{10} A_I^{(0,1)} \Big)
               + \frac{1}{\varepsilon}\Big( - \frac{2}{3} A_I^{(1,2)} \Big)
\,, \nonumber\\
K_I^{(2, 1)} &= \frac{1}{\varepsilon^2}\Big(  \frac{4}{3} \beta_{00} A_I^{(1,1)} 
                                      + \frac{2}{3} \beta_{01} A_I^{(1,0)} \Big)
               + \frac{1}{\varepsilon}\Big( - \frac{2}{3} A_I^{(2,1)} \Big)
\,, \nonumber\\
K_I^{(0,3)} & = \frac{1}{\varepsilon^3}\Big(-\frac{8}{3} \beta_{00}^{'2} A_I^{(0, 1)} \Big) 
              + \frac{1}{\varepsilon^2}
      \Big( \frac{8}{3} \beta'_{00} A_I^{(0,2)} + \frac{2}{3} \beta'_{01} A_I^{(0,1)} \Big)
      + \frac{1}{\varepsilon} \Big( -\frac{2}{3} A_I^{(0,3)} \Big)  \,.
\end{flalign}
\end{small}

\section{Form Factor} \label{ap:g}

\newcommand{\f}{{\cal F}}\label{ap:g6}
\noindent
The unrenormalized form factor ($\hat{F}_I$) can be written as follows in the perturbative expansion of unrenormalized strong coupling constant ($\hat{a}_s$)
and unrenormalized fine structure constant ($\hat{a}_e$)
\begin{small}
\begin{flalign}
\label{FFexp}
 \hat{F}_I &= 1 + \hat{a}_s \Big(\frac{Q^2}{\mu^2}\Big)^{\frac{\varepsilon}{2}} {\cal S}_{\varepsilon} \Big[ C_F \f_1^I \Big]
              + \hat{a}_e \Big(\frac{Q^2}{\mu^2}\Big)^{\frac{\varepsilon}{2}} {\cal S}_{\varepsilon} \Big[ e_I^2 \f_1^I \Big]
              + \hat{a}_s^2 \Big(\frac{Q^2}{\mu^2}\Big)^{\varepsilon} {\cal S}_{\varepsilon}^2 \Big[ C_F^2 \f_{2,0}^I + C_A C_F \f_{2,1}^I
\nonumber\\& 
              + C_F n_f T_F \f_{2,2}^I \Big]
              + \hat{a}_s \hat{a}_e  \Big(\frac{Q^2}{\mu^2}\Big)^{\varepsilon} {\cal S}_{\varepsilon}^2 \Big[ 2 C_F e_I^2 \f_{2,0}^I \Big]
              + \hat{a}_e^2 \Big(\frac{Q^2}{\mu^2}\Big)^{\varepsilon} {\cal S}_{\varepsilon}^2 \Big[ e_I^4 \f_{2,0}^I + e_I^2 \Big( N \sum_q e_q^2 \Big) \f_{2,2}^I \Big] 
\nonumber\\&             
              + \hat{a}_s^3 \Big(\frac{Q^2}{\mu^2}\Big)^{\frac{3\varepsilon}{2}} {\cal S}_{\varepsilon}^3 
\Big[ C_F^3 \f_{3,0}^I + C_F^2 C_A \f_{3,1}^I + C_F C_A^2 \f_{3,2}^I + C_F^2 n_f T_F \f_{3,3}^I + C_F C_A n_f T_F \f_{3,4}^I
\nonumber\\&
+ C_F n_f^2 T_F^2 \f_{3,5}^I  + C_F N_{F,V} \bigg(\frac{N^2-4}{N}\bigg) \f_{3,6}^I\Big]
              + \hat{a}_s^2 \hat{a}_e \Big(\frac{Q^2}{\mu^2}\Big)^{\frac{3\varepsilon}{2}} {\cal S}_{\varepsilon}^3 
\Big[ 3 C_F^2 e_I^2 \f_{3,0}^I + C_FC_A e_I^2 \f_{3,1}^I \nonumber\\&+ C_F T_F\Big(\sum_q e_q^2 \Big)\f_{3,3a}^I
+ C_F n_f T_F e_I^2 \f_{3,3b}^I + 12 C_F e_I \bigg(\sum_q e_q \bigg) \f_{3,6}^I\Big]
\nonumber\\&
              + \hat{a}_s \hat{a}_e^2 \Big(\frac{Q^2}{\mu^2}\Big)^{\frac{3\varepsilon}{2}} {\cal S}_{\varepsilon}^3 
\Big[ 3 C_F e_I^4 \f_{3,0}^I + C_F e_I^2 \Big( N \sum_q e_q^2 \Big) \f_{3,3a}^I
	+ C_F e_I^2 \Big( N \sum_q e_q^2+ \sum_{l} e_l^2 \Big) \f_{3,3b}^I \Big]
           \nonumber\\&
            + \hat{a}_e^3 \Big(\frac{Q^2}{\mu^2}\Big)^{\frac{3\varepsilon}{2}} {\cal S}_{\varepsilon}^3 
\Big[ e_I^6 \f_{3,0}^I 

+ e_I^2 \Big( N \sum_q e_q^4 + \sum_l e_l^4\Big) \f_{3,3a}^I  + e_I^4 \Big( N \sum_q e_q^2 + \sum_l e_l^2 \Big) \f_{3,3b}^I  
\nonumber\\&
+ e_I^2\Big( N \sum_q e_q^2 + \sum_l e_l^2\Big)^2 \f_{3,5}^I 
+ 8 e_I^3 \bigg(N\sum_q e_q^3 + \sum_l e_l^3\bigg) \f_{3,6}^I\Big]
\end{flalign}
\end{small}
$I=q,b$ denotes the Drell-Yan pair production and the Higgs boson production in bottom quark annihilation, respectively. Here, $N_{F,V}$ corresponds to the charge weighted sum of the quark flavors \cite{Gehrmann:2010ue}. The Coefficients $\f_{i,j}^I$ for $i,j <3 $ are given in paper \cite{H:2019nsw}. The  $\f_{i,j}^q$ for DY at third order are given by,
\\
\begin{footnotesize}
\begin{align}
\mathcal{F}_{3,0}^q &= \frac{1}{\varepsilon^6}\bigg( \frac{-256}{3}\bigg)
+ \frac{1}{\varepsilon^5}\bigg( 192\bigg)
+ \frac{1}{\varepsilon^4}\bigg( -400 + 32 \zeta_2 \bigg)
+ \frac{1}{\varepsilon^3}\bigg( 664 + 24 \zeta_2 - \frac{800}{3}\zeta_3\bigg)
+ \frac{1}{\varepsilon^2}\bigg(-1030  
\nonumber\\ & \quad
- 154 \zeta_2 
+ \frac{426}{5} \zeta_2^2 + 552 \zeta_3\bigg)
+ \frac{1}{\varepsilon}\bigg( \frac{9073}{6} + 467\zeta_2 -\frac{1461}{10}\zeta_2^2
- \frac{4238}{3}\zeta_3 
+ \frac{428}{3} \zeta_2\zeta_3 - \frac{1288}{5} \zeta_5 \bigg) 
  \nonumber\\& \quad
  -\frac{53675}{24}
   - \frac{9095}{252}\zeta_2^3
   - \frac{13001}{12} \zeta_2 
 + \frac{12743}{40} \zeta_2^2  + 2669\zeta_3
+ 61\zeta_2\zeta_3 
- \frac{1826}{3} \zeta_3^2 + \frac{4238}{5} \zeta_5 \,, \nonumber\\
\vspace{8cm}
\mathcal{F}_{3,1}^q &=\frac{1}{\varepsilon^5}\bigg( -\frac{352}{3} \bigg)
+\frac{1}{\varepsilon^4}\bigg( \frac{3448}{9} - 32\zeta_2 \bigg)
+\frac{1}{\varepsilon^3}\bigg( 208\zeta_3 -\frac{25660}{27} + \frac{28}{3}\zeta_2 \bigg)
+\frac{1}{\varepsilon^2}\bigg( \frac{158554}{81} 
+ \frac{1487}{9}\zeta_2 
\nonumber\\& \quad
- \frac{332}{5}\zeta_2^2
- 840\zeta_3\bigg)
+\frac{1}{\varepsilon}\bigg( -\frac{1773839}{486} - \frac{38623}{54}\zeta_2 + \frac{9839}{36}\zeta_2^2 + \frac{6703}{3}\zeta_3
- \frac{430}{3}\zeta_2\zeta_3 + 284 \zeta_5 \bigg) 
\nonumber\\& \quad 
+  \frac{37684115}{5832}
 +  \frac{664325}{324}\zeta_2 - \frac{1265467}{2160}\zeta_2^2 - \frac{18619}{1260}\zeta_2^3 - \frac{96715}{18}\zeta_3 
+ \frac{46}{9}\zeta_2\zeta_3 + \frac{1616}{3}\zeta_3^2 - \frac{46594}{45}\zeta_5 \,,
\nonumber\\
\mathcal{F}_{3,2}^q &=\frac{1}{\varepsilon^4}\bigg( -\frac{3872}{81}\bigg)
+\frac{1}{\varepsilon^3}\bigg( \frac{52168}{243} - \frac{704}{27}\zeta_2 \bigg) 
-\frac{1}{\varepsilon^2}\bigg(\frac{161156}{243} + \frac{2212}{81}\zeta_2 +\frac{352}{45}\zeta_2^2 - \frac{6688}{27}\zeta_3 \bigg)
\nonumber\\&\quad
+ \frac{1}{\varepsilon}\bigg( \frac{3741128}{2187} 
+ \frac{68497}{243}\zeta_2 -\frac{1604}{15}\zeta_2^2 - \frac{24212}{27}\zeta_3 + \frac{176}{9}\zeta_2\zeta_3 + \frac{272}{3}\zeta_5\bigg)
-\frac{52268375}{13122} 
-\frac{6152}{189}\zeta_2^3
\nonumber\\&\quad
+ \frac{152059}{540}\zeta_2^2
 -\frac{1136}{9}\zeta_3^2
- \frac{767320}{729}\zeta_2
   + \frac{1341553}{486}\zeta_3 -\frac{710}{9}\zeta_2\zeta_3 + \frac{2932}{9}\zeta_5     \,,   \nonumber\\
\mathcal{F}_{3,3}^q &=\frac{1}{\varepsilon^5}\bigg(\frac{128}{3}\bigg)
-\frac{1}{\varepsilon^4} \bigg(\frac{1184}{9}\bigg)
+\frac{1}{\varepsilon^3} \bigg( \frac{8720}{27} + \frac{16}{3}\zeta_2 \bigg)
+\frac{1}{\varepsilon^2} \bigg( -\frac{51992}{81} - \frac{532}{9}\zeta_2 + \frac{1168}{9}\zeta_3\bigg) 
\nonumber\\&\quad
+\frac{1}{\varepsilon}\bigg( \frac{277730}{243}
+ \frac{5698}{27}\zeta_2 - \frac{337}{9}\zeta_2^2 - \frac{10228}{27}\zeta_3 \bigg)
 -\frac{2732173}{1458} - \frac{45235}{81}\zeta_2 + \frac{8149}{108}\zeta_2^2 
+ \frac{102010}{81}\zeta_3 
\nonumber\\&\quad
 - \frac{686}{9}\zeta_2\zeta_3  + \frac{556}{45}\zeta_5   \,,     
\nonumber \\ 
\mathcal{F}_{3,4}^q &=\frac{1}{\varepsilon^4}\bigg( \frac{2816}{81} \bigg)
+\frac{1}{\varepsilon^3} \bigg( \frac{256}{27}\zeta_2 -\frac{36064}{243} \bigg) 
+\frac{1}{\varepsilon^2}\bigg( \frac{109432}{243} + \frac{2528}{81} \zeta_2 - \frac{2048}{27}\zeta_3\bigg)
+\frac{1}{\varepsilon}\bigg( \frac{176}{5}\zeta_2^2 
\nonumber\\&\quad
- \frac{44108}{243}\zeta_2
-\frac{2495948}{2187} 
+ \frac{25744}{81}\zeta_3\bigg)
+ \frac{17120104}{6561} + \frac{442961}{729}\zeta_2
- \frac{2186}{27}\zeta_2^2 
- \frac{90148}{81}\zeta_3
\nonumber\\&\quad
+ \frac{736}{9}\zeta_2\zeta_3 
- \frac{416}{3}\zeta_5  \,,  \nonumber\\
\mathcal{F}_{3,5}^q &= \frac{1}{\varepsilon^4} \bigg( -\frac{512}{81} \bigg)
+\frac{1}{\varepsilon^3} \bigg( \frac{6016}{243}\bigg)
+\frac{1}{\varepsilon^2} \bigg( -\frac{1984}{27} - \frac{64}{9}\zeta_2\bigg)
+\frac{1}{\varepsilon}\bigg( \frac{399200}{2187} + \frac{752}{27}\zeta_2 - \frac{1088}{81}\zeta_3\bigg)
\nonumber\\&\quad
 -\frac{2710864}{6561} 
 - \frac{248}{3}\zeta_2 - \frac{332}{135}\zeta_2^2 + \frac{12784}{243}\zeta_3 \,, 
\nonumber\\
\mathcal{F}_{3,6}^q & = 4 - \frac{2}{5}\zeta_2^2 + 10\zeta_2 + \frac{14}{3}\zeta_3 - \frac{80}{3}\zeta_5 \,,
 \nonumber\\
 \mathcal{F}_{3,3a}^q &= \frac{1}{\varepsilon^3}\bigg( -\frac{32}{9}\bigg) + \frac{1}{\varepsilon^2}\bigg( \frac{656}{27} - \frac{128}{9}\zeta_3\bigg)
+\frac{1}{\varepsilon}\bigg( \frac{1472}{27}\zeta_3 -\frac{8012}{81} - 4\zeta_2 + \frac{64}{15}\zeta_2^2   \bigg)
+\bigg(\frac{76781}{243} 
+ \frac{82}{3}\zeta_2 
\nonumber\\&\quad
- \frac{736}{45}\zeta_2^2 
- \frac{14180}{81}\zeta_3 -16\zeta_2\zeta_3 - \frac{224}{9}\zeta_5\bigg)    \,,   \nonumber\\
\mathcal{F}_{3,3b}^q &= \frac{1}{\varepsilon^5}\bigg( \frac{128}{3} \bigg)
+ \frac{1}{\varepsilon^4}\bigg(-\frac{1184}{9}\bigg)
+ \frac{1}{\varepsilon^3} \bigg(\frac{8816}{27}
+ \frac{16}{3}\zeta_2\bigg) 
+\frac{1}{\varepsilon^2}\bigg( 144\zeta_3 -\frac{53960}{81} -\frac{532}{9}\zeta_2  \bigg) 
\nonumber\\&\quad
+\frac{1}{\varepsilon}\bigg(
\frac{301766}{243}
+ \frac{5806}{27}\zeta_2 - \frac{1877}{45}\zeta_2^2 - \frac{1300}{3}\zeta_3\bigg)
-\frac{3192859}{1458} - \frac{47449}{81}\zeta_2 
+ \frac{49577}{540}\zeta_2^2 + \frac{12910}{9}\zeta_3 
\nonumber\\&\quad
- \frac{542}{9}\zeta_2\zeta_3 
+ \frac{1676}{45}\zeta_5 \,.   
\end{align}
\end{footnotesize}
\\
The  FF $\f_{i,j}^b$ for bottom quark annihilation at third order are given by,
\begin{footnotesize}
\begin{align}
\f_{3,0}^b &= \frac{1}{\epsilon^6} \left(-\frac{256}{3}\right) + \frac{1}{\epsilon^4} \left(-64 + 32 \zeta_2\right) + \frac{1}{\epsilon^3} \left( 64 + 96 \zeta_2 -\frac{800}{3} \zeta_3 \right) 
+\frac{1}{\epsilon^2} \Big( -112  -104 \zeta_2 
+ \frac{426}{5} \zeta_2^2
\nonumber \\ & \quad
+ 240 \zeta_3 \Big)  
+ \frac{1}{\epsilon} \Big(\frac{476}{3}  + 188 \zeta_2 -\frac{252}{5} \zeta_2^2 -\frac{1568}{3} \zeta_3 + \frac{428}{3} \zeta_2 \zeta_3 - \frac{1288}{5} \zeta_5  \Big)
+ \Big(- \frac{385}{3} - \frac{1085}{3} \zeta_2 
\nonumber \\ & \quad
+ \frac{887}{10} \zeta_2^2 - \frac{9095}{252} \zeta_2^3 + 538 \zeta_3 + 202 \zeta_2 \zeta_3 - \frac{1826}{3} \zeta_3^2 + 676 \zeta_5 \Big) \,,
\nonumber\\
\f_{3,1}^b &=  \frac{1}{\epsilon^5} \left( -\frac{352}{3}\right) + \frac{1}{\epsilon^4} \left( \frac{1072}{9} - 32 \zeta_2 \right) + \frac{1}{\epsilon^3} \left(- \frac{4312}{27} - \frac{44}{3} \zeta_2 + 208 \zeta_3 \right)  
+ \frac{1}{\epsilon^2} \Big( \frac{18028}{81} 
+ \frac{1262}{9} \zeta_2
\nonumber \\ & \quad
- \frac{332}{5} \zeta_2^2 -540 \zeta_3 \Big) 
 + \frac{1}{\epsilon} \Big( -\frac{76024}{243} - \frac{10199}{27} \zeta_2 + \frac{31591}{180} \zeta_2^2 + \frac{3442}{3} \zeta_3 
 - \frac{430}{3} \zeta_2 \zeta_3 + 284 \zeta_5 \Big)
 \nonumber \\ & \quad
 + 
\frac{332065}{729} 
+ \frac{131161}{162} \zeta_2  - \frac{305831}{1080} \zeta_2^2 - \frac{18619}{1260} \zeta_2^3 - \frac{17273}{9} \zeta_3 
- \frac{1663}{18} \zeta_2 \zeta_3 + \frac{1616}{3} \zeta_3^2 - \frac{27829}{45} \zeta_5 \,,
 \nonumber\\
\f_{3,2}^b &=  \frac{1}{\epsilon^4} \left(- \frac{3872}{81} \right) +  \frac{1}{\epsilon^3} \left( \frac{26032}{243} - \frac{704}{27} \zeta_2 \right) 
+  \frac{1}{\epsilon^2} \left( -\frac{38828}{243} - \frac{2212}{81} \zeta_2 - \frac{352}{45} \zeta_2^2
+ \frac{6688}{27} \zeta_3 \right) 
 \nonumber \\ & \quad
+  \frac{1}{\epsilon} \Big( \frac{385325}{2187} + \frac{31966}{243} \zeta_2 - \frac{1604}{15} \zeta_2^2 
- \frac{17084}{27} \zeta_3 + \frac{176}{9} \zeta_2 \zeta_3 + \frac{272}{3} \zeta_5 \Big)  -\frac{1870897}{26244} 
- \frac{478157}{1458} \zeta_2
\nonumber \\ & \quad
+ \frac{100597}{540} \zeta_2^2 - \frac{6152}{189} \zeta_2^3 + \frac{306992}{243} \zeta_3- \frac{980}{9} \zeta_2 \zeta_3  - \frac{1136}{9} \zeta_3^2 + \frac{3472}{9} \zeta_5 \,,
\nonumber\\
\f_{3,3}^b &=  \frac{1}{\epsilon^5} \left( \frac{128}{3} \right) - \frac{1}{\epsilon^4} \left( \frac{320}{9} \right) + \frac{1}{\epsilon^3} \left( \frac{1664}{27} + \frac{16}{3} \zeta_2 \right) 
 - \frac{1}{\epsilon^2} \left( \frac{6920}{81} + \frac{424}{9} \zeta_2 - \frac{1168}{9} 
\zeta_3 \right) 
+ \frac{1}{\epsilon} \Big( \frac{31022}{243} 
\nonumber \\ & \quad
+ \frac{2944}{27} \zeta_2 - \frac{337}{9} \zeta_2^2 - \frac{7528}{27} \zeta_3 \Big) 
- \frac{307879}{1458} - \frac{16885}{81} \zeta_2 + \frac{15769}{270} \zeta_2^2  
+ \frac{55624}{81} \zeta_3 
- \frac{686}{9} \zeta_2 \zeta_3 + \frac{556}{45} \zeta_5 \,,
\nonumber\\
\f_{3,3a}^b &=   \frac{1}{\epsilon^3} \left(- \frac{32}{9}  \right) 
+ \frac{1}{\epsilon^2} \left( \frac{440}{27} - \frac{128}{9} \zeta_3 \right) 
+ \frac{1}{\epsilon} \left(- \frac{3638}{81} -4 \zeta_2 + \frac{64}{15} \zeta_2^2 + \frac{608}{27} \zeta_3 \right) 
 -\frac{51259}{486} 
+ \frac{55}{3} \zeta_2
\nonumber \\ & \quad
- \frac{304}{45} \zeta_2^2 - \frac{4460}{81} \zeta_3 - 16 \zeta_2 \zeta_3 - \frac{224}{9} \zeta_5    \,,
 \nonumber\\
\f_{3,3b}^b &=  \frac{1}{\epsilon^5} \left( \frac{128}{3} \right) - \frac{1}{\epsilon^4} \left( \frac{320}{9} \right) + \frac{1}{\epsilon^3} \left( \frac{1760}{27} + \frac{16}{3} \zeta_2 \right) 
 + \frac{1}{\epsilon^2} \left( -\frac{8240}{81} - \frac{424}{9} \zeta_2 + 144 \zeta_3 \right) 
+ \frac{1}{\epsilon} \Big( \frac{41936}{243} 
\nonumber \\ & \quad
+ \frac{3052}{27} \zeta_2 - \frac{1877}{45} \zeta_2^2 - \frac{904}{3} \zeta_3 \Big) 
  - \frac{230828}{729} - \frac{18370}{81} \zeta_2 + \frac{17593}{270} \zeta_2^2  
 + \frac{6676}{9} \zeta_3 
- \frac{542}{9} \zeta_2 \zeta_3 + \frac{1676}{45} \zeta_5    \,,                    
\nonumber \\ 
\f_{3,4}^b &=  \frac{1}{\epsilon^4} \left( \frac{2816}{81} \right) - \frac{1}{\epsilon^3} \Big( \frac{17056}{243} 
-\frac{256}{27} \zeta_2 \Big)  
+ \frac{1}{\epsilon^2} \left( \frac{27280}{243} + \frac{2528}{81} \zeta_2  - \frac{2048}{27} \zeta_3 \right) 
 + \frac{1}{\epsilon} \Big(
 -\frac{20132}{243} \zeta_2
  \nonumber \\ & \quad
  + \frac{176}{5} \zeta_2^2
   +  \frac{20560}{81} \zeta_3
 -\frac{361220}{2187}
  \Big) +
 \frac{1451329}{6561} + \frac{127142}{729} \zeta_2 
- \frac{7582}{135} \zeta_2^2 
- \frac{47524}{81} \zeta_3 
+ \frac{736}{9} \zeta_2 \zeta_3  
  \nonumber \\ & \quad
- \frac{416}{3} \zeta_5    \,,  
   \nonumber \\
\f_{3,5}^b &=  \frac{1}{\epsilon^4} \left( -\frac{512}{81} \right)  + \frac{1}{\epsilon^3} \left( \frac{2560}{243} \right) + \frac{1}{\epsilon^2} \left( -\frac{512}{27}  - \frac{64}{9} \zeta_2  \right) 
 + \frac{1}{\epsilon} \Big( \frac{76928}{2187} + \frac{320}{27} \zeta_2 
 - \frac{1088}{81} \zeta_3\Big) 
\nonumber \\& \quad
- \frac{438112}{6561} 
- \frac{64}{3}\zeta_2 - \frac{332}{135} \zeta_2^2 + \frac{5440}{243} \zeta_3 \,,
  \nonumber \\
\f_{3,6}^b &=0 \,.
\end{align}
\end{footnotesize}

\section{\texorpdfstring{Cusp anomalous dimension $A_{I}^{(i,j)}$}{Aij}s } \label{ap:g2}
The cusp anomalous dimensions $A_I^{(i,j)}$, $I = q,b$ up to three loop order are found to be,
\begin{small}
\begin{align}\label{eq:AFF}
A_I^{(1,0)} &= 4 C_F   
\,, \qquad
A_I^{(0,1)} = 4 e_I^2  
\,, \qquad
A_I^{(1,1)} = 0 \,,
\nonumber \\
A_I^{(2,0)} &=  \Big(8 C_A C_F \Big(\frac{67}{18} - \zeta_2 \Big) + 8C_Fn_fT_F\Big(-\frac{10}{9}\Big)\Big)\,, 
 \nonumber \\
A_I^{(0,2)} &= 8e_I^2 \Big( N\sum_{q} e_{q}^2 + \sum_l e_l^2\Big) \Big(- \frac{10}{9} \Big)  \,,
 \nonumber \\
A_I^{(3,0)} &=  C_A^2 C_F\Big(\frac{490}{3} - \frac{1072}{9} \zeta_2 
                       + \frac{176}{5} \zeta_2^2  + \frac{88}{3} \zeta_3\Big) 
  + C_A C_F n_f T_F \Big(-\frac{1672}{27} + \frac{320}{9} \zeta_2 
  \nonumber \\& \quad
  - \frac{224}{3} \zeta_3\Big) 
  + C_F^2 n_f T_F \Big( -\frac{220}{3} + 64 \zeta_3 \Big)
  + C_F n_f^2 T_F^2 \Big(-\frac{64}{27} \Big)
\,, \nonumber \\
A_I^{(1,2)} &=  C_F  e_I^2 \Big( N\sum_{q} e_{q}^2 \Big)  \Big( -\frac{220}{3} + 64 \zeta_3 \Big)\,,
\nonumber\\
A_I^{(2,1)} &=  C_F T_F  \Big( \sum_{q} e_{q}^2 \Big) \Big( -\frac{220}{3} + 64 \zeta_3 \Big)
\,, \nonumber \\
A_I^{(0,3)} &=  e_I^2  \Big( N\sum_{q} e_{q}^4 + \sum_l e_l^4\Big)  \Big( -\frac{220}{3} + 64 \zeta_3 \Big)
  + e_I^2  \Big( N\sum_{q} e_{q}^2 + \sum_l e_l^2\Big)^2  \Big(-\frac{64}{27} \Big) \,.
\end{align} 
\end{small}

\section{\texorpdfstring{UV anomalous dimensions $\gamma_{b}^{(i,j)}$}{gamij}s } \label{ap:g3}
The UV anomalous dimensions $\gamma_b^{(i,j)}$ up to three loop order are found to be,
\begin{small}
\begin{align}
\label{eq:gam}
 \gamma^{(1,0)}_b &= 3 C_F 
\,, \qquad
 \gamma^{(0,1)}_b = 3 e_b^2 
\,, \qquad
 \gamma^{(1,1)}_b  = 3 C_F e_b^2 \,,    
   \nonumber \\ 
 \gamma^{(2,0)}_b &=    \frac{3}{2} C_F^2 + \frac{97}{6} C_AC_F -\frac{10}{3} C_F n_f T_F 
\,, \quad
 \gamma^{(0,2)}_b =    \frac{3}{2} e_b^4 - \frac{10}{3} e_b^2 \Big( N\sum_{q}e_{q}^2+ \sum_l e_l^2\Big) \,,
\nonumber \\
 \gamma^{(3,0)}_b &=   C_A^2 C_F \Big( \frac{11413}{108} \Big)
          + C_A C_F^2 \Big(-\frac{129}{4} \Big)
          + C_A C_F n_f  T_F \Big( -\frac{556}{27} - 48 \zeta_3 \Big)
          + C_F^3 \Big( \frac{129}{2} \Big) \qquad
\nonumber \\& \quad
          + C_F n_f^2 T_F^2 \Big( -\frac{140}{27} \Big)
          + C_F^2 n_f T_F \Big( -46 + 48 \zeta_3 \Big) \,,
\nonumber \\ 
 \gamma^{(1,2)}_b &=   
           3 C_F e_b^4 \Big( \frac{129}{2} \Big)
            + C_F e_b^2 \Big( N\sum_{q}e_{q}^2+ \sum_l e_l^2\Big) \big( -1 \big) 
          + C_F e_b^2 \Big( N\sum_{q}e_{q}^2\Big) \Big( -45 + 48 \zeta_3 \Big) 
\,, \nonumber \\
 \gamma^{(2,1)}_b &=  
           C_A C_F e_b^2\Big(-\frac{129}{4} \Big)
          + 3 C_F^2 e_b^2 \Big( \frac{129}{2} \Big)
          + C_F n_f T_F e_b^2 \Big( -1 \Big) 
\nonumber \\& \quad
          + C_F T_F \Big(\sum_{q}e_{q}^2\Big) \Big( - 45 + 48\zeta_3 \Big) \,,
 \nonumber \\
 \gamma^{(0,3)}_b &=   
           e_b^6 \Big( \frac{129}{2} \Big)
          +  e_b^2 \Big( N\sum_{q}e_{q}^2 + \sum_l e_l^2\Big)^2 \Big( -\frac{140}{27} \Big)
          + e_b^4 \Big( N\sum_{q}e_{q}^2+ \sum_l e_l^2\Big) \Big( -1 \Big) 
\nonumber \\& \quad
          + e_b^2 \Big( N\sum_{q}e_{q}^4+ \sum_l e_l^4\Big) \Big( -45 + 48 \zeta_3 \Big) \,.
\end{align}
\end{small}

\section{\texorpdfstring{Soft anomalous dimensions $f_{I}^{(i,j)}$}{gamij}s } \label{ap:g4}
The soft anomalous dimensions $f_I^{(i,j)}$ up to three loop order are found to be,
\begin{small}
\begin{align}
\label{f}
f_I^{(1,0)} & = 0
 \,, \qquad
f_I^{(0,1)}  = 0
 \,, \qquad
f_I^{(1,1)}  = 0
\,, \nonumber \\
f_I^{(2,0)} &= C_A C_F\Big(-\frac{22}{3}\zeta_2  -28 \zeta_3 + \frac{808}{27} \Big) 
+ C_Fn_fT_F\Big(\frac{8}{3}\zeta_2 - \frac{224}{27}\Big) \,, 
\nonumber \\
f_I^{(0,2)} & =  e_I^2 \Big( N\sum_{q} e_{q}^2 + \sum_l e_l^2 \Big) 
        \Big(\frac{8}{3}\zeta_2 
       - \frac{224}{27}\Big) \,,
\nonumber \\
f_I^{(3,0)} &=  
  C_A^2 C_F \Big(\frac{136781}{729} 
     - \frac{12650}{81} \zeta_2 
     + \frac{352}{5} \zeta_2^2 
     - \frac{1316}{3} \zeta_3 
     + \frac{176}{3} \zeta_2 \zeta_3 + 192 \zeta_ 5 \Big) 
 \nonumber \\& \quad
 +  C_A C_F n_f T_F \Big(
      - \frac{23684}{729} 
      + \frac{5656}{81} \zeta_2 
      - \frac{192}{5} \zeta_2^2 
      + \frac{1456}{27} \zeta_3 \Big) 
 + C_F n_f^2 T_F^2 \Big(
  \nonumber \\& \quad
 -\frac{8320}{729}
        - \frac{160}{27}\zeta_2 
        + \frac{448}{27}\zeta_3 \Big)
 + C_F^2 n_f T_F \Big(
     - \frac{3422}{27} 
     + 8 \zeta_2 
     + \frac{64}{5} \zeta_2^2 
     + \frac{608}{9} \zeta_3  \Big)\,, \qquad
 \nonumber \\ 
f_I^{(1,2)} &= C_F e_I^2 \Big( N\sum_{q} e_{q}^2 \Big) 
           \Big(-\frac{3422}{27} + 8 \zeta_2 
          + \frac{64}{5} \zeta_2^2 
          + \frac{608}{9} \zeta_3 \Big) \,,
\nonumber \\
	f_I^{(2,1)} &=  C_F T_F \Big( \sum_{q} e_{q}^2 \Big)
     \Big( -\frac{3422}{27} 
       + 8 \zeta_2 
       + \frac{64}{5} \zeta_2^2 
       + \frac{608}{9} \zeta_3 \Big) \,,
\nonumber \\
f_I^{(0,3)} &=  e_I^2 \Big( N\sum_{q} e_{q}^2 + \sum_l e_l^2 \Big)^2
       \Big(-\frac{8320}{729}
        - \frac{160}{27}\zeta_2 
        + \frac{448}{27}\zeta_3 \Big)
 + e_I^2 \Big( N\sum_{q} e_{q}^4 + \sum_l e_l^4\Big) 
         \nonumber \\& \quad \quad
\Big(
     - \frac{3422}{27} 
     + 8 \zeta_2 
     + \frac{64}{5} \zeta_2^2 
     + \frac{608}{9} \zeta_3  \Big) \,.
\end{align}
\end{small}
\section{\texorpdfstring{Collinear anomalous dimensions $B_{I}^{(i,j)}$}{gamij}s } \label{ap:g5}
The collinear anomalous dimensions $B_I^{(i,j)}$ up to three loop order are found to be,
\begin{small}
\begin{align} \label{B}
B^{(1,0)}_I &= 3C_F 
\,, \qquad
B^{(0,1)}_I = 3 e_I^2 
\,, \qquad
B_I^{(1,1)}  = C_F e_I^2 \Big (3-24\zeta_2+48\zeta_3 \Big) \,,      
\nonumber \\
B_I^{(2,0)} &= \frac{1}{2} \Big \{ C_F^2 \big (3-24\zeta_2+48\zeta_3 \big) 
        + C_AC_F\Big(\frac{17}{3} + \frac{88}{3}\zeta_2 -24 \zeta_3  \Big) 
         \nonumber \\& \quad
+ C_Fn_fT_F\Big(-\frac{4}{3} -\frac{32}{3}\zeta_2 \Big) \Big\} \,, 
\nonumber\\
B_I^{(0,2)} &= \frac{1}{2} \Big \{ e_I^4 \big (3-24\zeta_2+48\zeta_3 \big) 
+ e_I^2 \Big( N \sum_{q} e_{q}^2 + \sum_l e_l^2 \Big) \Big(-\frac{4}{3} -\frac{32}{3}\zeta_2 \Big) \Big\} \,,
\nonumber\\
B_I^{(3,0)} & = 
  C_A^2 C_F \Big( -\frac{1657}{36} 
         + \frac{4496}{27} \zeta_2 
         - 2 \zeta_2^2 
         - \frac{1552}{9} \zeta_3 
         + 40 \zeta_5 \Big)
          \nonumber\\& \quad
 +  C_A C_F n_f T_F \Big(
 40- \frac{2672}{27} \zeta_2 
         + \frac{8}{5} \zeta_2^2 + \frac{400}{9} \zeta_3 \Big)
 +  C_F n_f^2 T_F^2 \Big( -\frac{68}{9}
        + \frac{320}{27} \zeta_2 
        - \frac{64}{9} \zeta_3 \Big)
         \nonumber\\& \quad
 + C_A C_F^2 \Big( \frac{151}{4}
         - \frac{410}{3} \zeta_2 
         - \frac{988}{15} \zeta_2^2
         + \frac{844}{3} \zeta_3 
         + 16 \zeta_2 \zeta_3 + 120 \zeta_5 \Big)
          \nonumber\\& \quad
 + C_F^3 \Big(\frac{29}{2} 
         + 18 \zeta_2 
         + \frac{288}{5} \zeta_2^2 + 68 \zeta_3 
         - 32 \zeta_2 \zeta_3
         - 240 \zeta_5 \Big)
         \nonumber\\& \quad
 + C_F^2 n_f T_F \Big(-46 + \frac{40}{3} \zeta_2 
        + \frac{464}{15} \zeta_2^2 
        - \frac{272}{3}\zeta_3 \Big) \,,
\nonumber\\
B_I^{(1,2)} & = 
  3 C_F e_I^4 \Big(\frac{29}{2} 
         + 18 \zeta_2 
         + \frac{288}{5} \zeta_2^2 + 68 \zeta_3 
         - 32 \zeta_2 \zeta_3 - 240 \zeta_5 \Big)
           \nonumber\\& \quad
 + C_F e_I^2 \Big(N \sum_{q} e_{q}^2 \Big) 
\quad \Big(-37
        -16 \zeta_2
       +48\zeta_3 \Big)
         \nonumber\\& \quad
 + C_F e_I^2 \Big(N \sum_{q} e_{q}^2 + \sum_l e_l^2 \Big) 
\quad \Big(-9
 + \frac{88}{3} \zeta_2 
        + \frac{464}{15} \zeta_2^2 
        - \frac{416}{3}\zeta_3 \Big) \,,
\nonumber\\
B_I^{(2,1)} & = 
  C_F C_A e_I^2 \Big( \frac{151}{4}
         - \frac{410}{3} \zeta_2 
         - \frac{988}{15} \zeta_2^2
         + \frac{844}{3} \zeta_3 
         + 16 \zeta_2 \zeta_3 + 120 \zeta_5 \Big)
          \nonumber\\& \quad
 + 3 C_F^2 e_I^2 \Big( \frac{29}{2} 
         + 18 \zeta_2 
         + \frac{288}{5} \zeta_2^2 + 68 \zeta_3 
         - 32 \zeta_2 \zeta_3 - 240 \zeta_5 \Big)
 + C_F T_F\Big( \sum_{q} e_{q}^2  \Big) 
           \nonumber\\&  \quad \quad
         \Big( -37 
           - 16 \zeta_2 + 48 \zeta_3 \Big)
 + C_F n_f T_F e_I^2 \Big(-9
         + \frac{88}{3}\zeta_2 
         + \frac{464}{15} \zeta_2^2 
         - \frac{416}{3} \zeta_3 \Big) \,,
\nonumber\\
B_I^{(0,3)} & = 
 e_I^2 \Big( N\sum_{q} e_{q}^2 + \sum_l e_l^2 \Big)^2 \Big( -\frac{68}{9}
        + \frac{320}{27} \zeta_2 
        - \frac{64}{9} \zeta_3 \Big)
 + e_I^6 \Big(\frac{29}{2} 
         + 18 \zeta_2 
         + \frac{288}{5} \zeta_2^2 
    \nonumber\\& \quad
+ 68 \zeta_3 
         - 32 \zeta_2 \zeta_3
         - 240 \zeta_5 \Big)
 + e_I^2 \Big(N \sum_{q} e_{q}^4 + \sum_l e_l^4 \Big) 
           \Big(-37 - 16 \zeta_2 + 48 \zeta_3 \Big)
            \nonumber\\& \quad
 + e_I^4\Big(N \sum_{q} e_{q}^2 + \sum_l e_l^2 \Big) \Big(-9
         + \frac{88}{3}\zeta_2 
         + \frac{464}{15} \zeta_2^2 
         - \frac{416}{3} \zeta_3 \Big) \,.
\end{align}
\end{small}

\section{\texorpdfstring{ $D_{I}^{(i,j)}$}{Dij}s in Soft distribution function } \label{ap:g7}
The constants $D_I^{(i,j)}$ in soft distributon functions up to three loop order are found to be,
\begin{small}
\begin{align}
\label{DI}
D^{(1,0)}_I &= 0 \,,  
\qquad
D^{(0,1)}_I = 0 \,,
\qquad
D_I^{(1,1)}  = 0 \,.     
\nonumber \\
D_I^{(2,0)} &=  C_F n_f T_F \Big(\frac{448}{27}
         - \frac{64}{3} \zeta_2 \Big)
     + C_F C_A \Big( -\frac{1616}{27} + 56\zeta_3 + \frac{176}{3}\zeta_2 \Big) 
\nonumber \\
D_I^{(0,2)} &=  e_I^2 \Big( N \sum_q e_q^2 + \sum_l e_l^2 \Big) \Big(\frac{448}{27}
         - \frac{64}{3} \zeta_2 \Big)
\,, \nonumber \\
D_I^{(3,0)} &=  
      C_F n_f^2 T_F^2 \Big(-\frac{14848}{729} + \frac{1280}{27}\zeta_3 
       + \frac{2560}{27}\zeta_2 \Big)
    + C_F C_A n_f T_F \Big(\frac{250504}{729} - \frac{4960}{9} \zeta_3  
\nonumber \\& \quad
- \frac{58784}{81} \zeta_2 
+ \frac{1472}{15} \zeta_2^2 \Big)
    + C_F C_A^2 \Big(-\frac{594058}{729} - 384 \zeta_5 + \frac{40144}{27}\zeta_3
       + \frac{98224}{81}\zeta_2 - \frac{352}{3}\zeta_2\zeta_3 
       \nonumber \\& \quad
       - \frac{2992}{15}\zeta_2^2 \Big)
    + C_F^2 n_f T_F \Big(\frac{6844}{27} - \frac{1216}{9}\zeta_3
          - 64\zeta_2 - \frac{128}{5}\zeta_2^2 \Big)
\,, \nonumber \\
D_I^{(1,2)} &=  
     C_F e_I^2 \Big( N \sum_q e_q^2  \Big)  \Big(\frac{6844}{27} - \frac{1216}{9}\zeta_3
          - 64\zeta_2 - \frac{128}{5}\zeta_2^2 \Big)
\,, \nonumber \\
D_I^{(2,1)} &=  
     C_F T_F \Big(\sum_q e_q^2  \Big) \Big(\frac{6844}{27} - \frac{1216}{9}\zeta_3
          - 64\zeta_2 - \frac{128}{5}\zeta_2^2 \Big)
\,, \nonumber \\
D_I^{(0,3)} &=  
      e_I^2 \Big( N \sum_q e_q^2 + \sum_l e_l^2 \Big)^2 \Big(-\frac{14848}{729} + \frac{1280}{27}\zeta_3 
       + \frac{2560}{27}\zeta_2 \Big)
       \,, \nonumber \\& \quad
    + e_I^2 \Big( N \sum_q e_q^4 + \sum_l e_l^4 \Big) \Big(\frac{6844}{27} - \frac{1216}{9}\zeta_3
          - 64\zeta_2 - \frac{128}{5}\zeta_2^2 \Big)
\,.
\end{align}
\end{small}

\section{ $\Delta_{I}^{(i,j),SV}$ for QCD, QED and QCD-QED up to N$^3$LO } \label{ap:g8}
Here we present the soft virtual cross-section $\Delta_{I}^{(i,j),SV}$ defined in Eq.(\ref{DeltaSV}) at the third order in the strong and electromagnetic coupling constants. The finite cross-section upto two loop is already available in Appendix C of \cite{H:2019nsw}. At the third order, $\Delta_{I}^{(i,j),SV}$ takes the following form:
\begin{small}
{\begin{align}
\label{eq:Del}
	\Delta_{I\bar{I}}^{(3,0),SV} =& \Big[ C_F^3 \Delta_{(3,1)}^{I\bar{I}} + C_F^2 C_A \Delta_{(3,2)}^{I\bar{I}} + C_F C_A^2 \Delta_{(3,3)}^{I\bar{I}} + C_F^2 n_f T_F \Delta_{(3,4)}^{I\bar{I}} + C_F C_A n_f T_F \Delta_{(3,5)}^{I\bar{I}} \nonumber\\& + C_F n_f^2 T_F^2 \Delta_{(3,6)}^{I\bar{I}} + C_FN_{F,V}\Big(\frac{N^2-4}{N}\Big)\Delta_{(3,7)}^{I\bar{I}} \Big] \,, \nonumber\\
 \Delta_{I\bar{I}}^{(0,3),SV} =& \Big[ e_I^6 \Delta_{(3,1)}^{I\bar{I}} + e_I^2 \Big( N \sum_q e_q^4 + \sum_l e_l^4\Big) \Delta_{(3,4a)}^{I\bar{I}}  + e_I^4 \Big( N \sum_q e_q^2 + \sum_l e_l^2\Big) \Delta_{(3,4b)}^{I\bar{I}}  + \nonumber\\&
 e_I^2\Big( N \sum_q e_q^2 + \sum_l e_l^2\Big)^2 \Delta_{(3,6)}^{I\bar{I}} + 8 e_I^3 \bigg(N\sum_q e_q^3 + \sum_l e_l^3\bigg)\mathrm{\Delta}_{(3,7)}^{I\bar{I}} \Big] \,,\nonumber\\
\Delta_{I\bar{I}}^{(2,1),SV} =& \Big[ 3 C_F^2 e_I^2 \Delta_{(3,1)}^{I\bar{I}} + C_F C_A e_I^2 \Delta_{(3,2)}^{I\bar{I}} + C_F T_F\Big(\sum_q e_q^2 \Big)\Delta_{(3,4a)}^{I\bar{I}}
+ C_F n_f T_F e_I^2 \Delta_{(3,4b)}^{I\bar{I}} \Big]\nonumber \,, \\
\Delta_{I\bar{I}}^{(1,2),SV} =& \Big[ 3 C_F e_I^4 \Delta_{(3,1)}^{I\bar{I}} +C_F e_I^2 \Big( N \sum_q e_q^2 \Big) \mathrm{\Delta}_{(3,4a)}^{I\bar{I}} + C_F e_I^2 \Big( N \sum_q e_q^2 + \sum_l e_l^2 \Big) \mathrm{\Delta}_{(3,4b)}^{I\bar{I}} \Big] \,.
\end{align}}
\end{small}
For ${I\bar{I}}$=$q\bar{q},b\bar{b}$ denotes the Drell-Yan pair production and the Higgs boson production in bottom quark annihilation, respectively. The coefficients for the above color factors are given as:

\begin{small}
\begin{align}
\mathrm{\Delta}_{(3,1)}^{q\bar{q}} &= \delta(1-z)\bigg( -\frac{184736}{315} \;{\zeta_2}^3
+ \frac{412}{5} \;{\zeta_2}^2
+ 80 \;{\zeta_2} {\zeta_3}
-\frac{130}{3} \;{\zeta_2} +\frac{10336}{3} \;{\zeta_3}^2 - 460 \;{\zeta_3}
+ 1328 \;{\zeta_5}
\nonumber\\&\quad
- \frac{5599}{6} \bigg)
+ \mathcal{D}_{0}\big( 12288\zeta_5
- 4096\zeta_3 - 6144 \zeta_2 \zeta_3 \big) + \mathcal{D}_{1}\bigg( 2044 -960\zeta_3 +2976\zeta_2
\nonumber\\&\quad
-\frac{14208}{5}\zeta_2^2 \bigg)
+\mathcal{D}_{2} \big( 10240\zeta_3\big)
+\mathcal{D}_{3} \big( -2048 -3072\zeta_2\bigg) + \mathcal{D}_{5} \big( 512\big)    \nonumber
\,, \nonumber \\
 \mathrm{\Delta}_{(3,2)}^{q\bar{q}} &=\delta(1-z) \bigg( -\frac{20816}{315} \;{\zeta_2}^3 -\frac{1664}{135} \;{\zeta_2}^2
+\frac{28736}{9} \;{\zeta_2} {\zeta_3} - \frac{13186}{27} \;{\zeta_2}
+\frac{3280}{3} \;{\zeta_3}^2 - \frac{20156}{9} \;{\zeta_3}
\nonumber\\&\quad
- \frac{39304}{9} \;{\zeta_5}
+ \frac{74321}{36}\bigg)
+\mathcal{D}_{0} \bigg( \frac{25856}{27} + \frac{26240}{9}\zeta_3 - \frac{12416}{27}\zeta_2 - 1472\zeta_2\zeta_3 + \frac{1408}{3}\zeta_2^2 \bigg)
\nonumber\\&\quad
+\mathcal{D}_{1}\bigg( -\frac{35572}{9}
 -5184\zeta_3
- \frac{11648}{9}\zeta_2 + \frac{3648}{5}\zeta_2^2\bigg)
+\mathcal{D}_{2}\bigg( -\frac{4480}{9} + 1344\zeta_3 + \frac{11264}{3}\zeta_2\bigg)
\nonumber\\&\quad
+ \mathcal{D}_{3} \bigg( \frac{17152}{9}
- 512\zeta_2 \bigg)
+\mathcal{D}_{4} \bigg( -\frac{7040}{9}\bigg)
\,, \nonumber \\
 \mathrm{\Delta}_{(3,3)}^{q\bar{q}} &=\delta(1-z) \bigg( \frac{13264}{315} \;{\zeta_2}^3 + \frac{14611 \
}{135} \;{\zeta_2}^2 - \frac{884}{3} \;{\zeta_2} {\zeta_3} + 843 \
{\zeta_2} - \frac{400}{3} \;{\zeta_3}^2 + \frac{82385}{81} \;{\zeta_3} - 204 \;{\zeta_5}
\nonumber\\& \quad
- \frac{1505881}{972} \bigg)
+\mathcal{D}_{0}\bigg( -\frac{594058}{729} - 384\zeta_5
+ \frac{40144}{27}\zeta_3 + \frac{98224}{81}\zeta_2 - \frac{352}{3}\zeta_2\zeta_3 - \frac{2992}{15}\zeta_2^2\bigg)
\nonumber\\& \quad
+\mathcal{D}_{1} \bigg( \frac{124024}{81} - 704\zeta_3 - \frac{12032}{9}\zeta_2
+\frac{704}{5}\zeta_2^2 \bigg)
+\mathcal{D}_{2} \bigg( -\frac{28480}{27} + \frac{704}{3}\zeta_2 \bigg)
+\mathcal{D}_{3}\bigg( \frac{7744}{27} \bigg)
\,, \nonumber \\
\mathrm{\Delta}_{(3,4)}^{q\bar{q}} &=\delta(1-z) \bigg( \frac{544}{135} \;{\zeta_2}^2
- \frac{11008}{9} \;{\zeta_2} {\zeta_3} +\frac{5264}{27} \;{\zeta_2}
+ \frac{7024}{9} \;{\zeta_3}
+ \frac{11072}{9} \;{\zeta_5}
-\frac{842}{3}\bigg)
\nonumber\\
&\quad
+\mathcal{D}_{0}\bigg(
-\frac{11456}{9}\zeta_3
-12 + \frac{3904}{27}\zeta_2 - \frac{2944}{15}\zeta_2^2 \bigg) + \mathcal{D}_{1} \bigg( \frac{8576}{9} + 2560\zeta_3 + \frac{4096}{9}\zeta_2\bigg)
\nonumber\\& \quad
+\mathcal{D}_{2}\bigg( \frac{1088}{9}
- \frac{4096}{3}\zeta_2 \bigg)
+ \mathcal{D}_{3}\bigg( -\frac{5120}{9} \bigg) + \mathcal{D}_{4} \bigg( \frac{2560}{9} \bigg)   \nonumber
\,, \nonumber \\
\mathrm{\Delta}_{(3,4a)}^{q\bar{q}} &=\delta(1-z) \bigg(\frac{3529}{9} -\frac{872}{9}\zeta_2 - \frac{96}{5}\zeta_2^2 - \frac{656}{3}\zeta_3 + \frac{128}{3}\zeta_2\zeta_3\bigg)
+\mathcal{D}_{0}\bigg( \frac{6844}{27} - \frac{1216}{9}\zeta_3 -64\zeta_2
\nonumber\\&\quad
- \frac{128}{5}\zeta_2^2\bigg)
+\mathcal{D}_{1} \bigg( -\frac{880}{3} + 256\zeta_3\bigg)
+\mathcal{D}_{2}\big( 64\big)  \nonumber
\,, \nonumber \\
\mathrm{\Delta}_{(3,4b)}^{q\bar{q}} &= \delta(1-z) \bigg( -\frac{6055}{9} +\frac{7880}{27}\zeta_2 + \frac{3136}{135}\zeta_2^2 + \frac{8992}{9}\zeta_3 - \frac{11392}{9}\zeta_2\zeta_3 +\frac{11072}{9}\zeta_5 \bigg)
\nonumber\\&\quad
+\mathcal{D}_0\bigg( -\frac{7168}{27}
- \frac{10240}{9}\zeta_3 + \frac{5632}{27}\zeta_2 - \frac{512}{3}\zeta_2^2 \bigg)
+\mathcal{D}_1\bigg( \frac{11216}{9} + 2304\zeta_3 + \frac{4096}{9}\zeta_2\bigg)
\nonumber\\&\quad
+\mathcal{D}_2 \bigg( \frac{512}{9} - \frac{4096}{3}\zeta_2\bigg)
+\mathcal{D}_3 \bigg( -\frac{5120}{9} \bigg)
+\mathcal{D}_4 \bigg( \frac{2560}{9} \bigg)  \nonumber
\,, \nonumber \\
\mathrm{\Delta}_{(3,5)}^{q\bar{q}} &=\delta(1-z)\bigg( 
\frac{221302}{243} - \frac{56264}{81}\zeta_2- \frac{12032}{81}\zeta_3 + \frac{416}{3}\zeta_2\zeta_3 - \frac{11512}{135}\zeta_2^2 -16 \zeta_5
\bigg) 
 \nonumber\\&\quad
+\mathcal{D}_{0}\bigg( \frac{250504}{729} - \frac{4960}{9}\zeta_3 - \frac{58784}{81}\zeta_2 
+ \frac{1472}{15}\zeta_2^2\bigg)
+\mathcal{D}_{1}\bigg( -\frac{65632}{81} + 768\zeta_2\bigg)
\nonumber\\&\quad
+\mathcal{D}_{2}\bigg( \frac{18496}{27} - \frac{256}{3}\zeta_2 \bigg)
+\mathcal{D}_{3}\bigg( -\frac{5632}{27}\bigg) \nonumber
\,, \nonumber \\
\mathrm{\Delta}_{(3,6)}^{q\bar{q}} &=\delta(1-z) \bigg( -\frac{28324}{243} + \frac{9664}{81}\zeta_2 + \frac{512}{27}\zeta_2^2 - \frac{5056}{81}\zeta_3\bigg)
+\mathcal{D}_{0} \bigg( -\frac{14848}{729} + \frac{1280}{27}\zeta_3 + \frac{2560}{27}\zeta_2\bigg)
\nonumber\\&\quad
+ \mathcal{D}_{1} \bigg( \frac{6400}{81}
- \frac{1024}{9}\zeta_2 \bigg)
+\mathcal{D}_{2}\bigg( -\frac{2560}{27}\bigg) + \mathcal{D}_{3} \bigg( \frac{1024}{27}\bigg)
\,, \nonumber\\
\mathrm{\Delta}_{(3,7)}^{q\bar{q}} &=
\Big(-\frac{4}{5} \;{\zeta_2}^2 + 20 \;{\zeta_2} + \frac{28}{3} \;{\zeta_3}
-\frac{160}{3} \;{\zeta_5} + 8 \Big) \,.
\end{align}
\end{small}
Similarly $\Delta^{b\bar{b}}_{(i,j)}$ for bottom quark annihilation at third order are given as,
\begin{small}
\begin{align}
\Delta^{b\bar{b}}_{(3,1)} &=
    \delta (1-z) \Big(  - \frac{184736}{315} {\zeta_2}^3 + \frac{152}{5} {\zeta_2}^2 - 64 {\zeta_2} {\zeta_3}
    - \frac{550}{3} {\zeta_2}
    + \frac{10336}{3} {\zeta_3}^2 - 1188 {\zeta_3} + 848 {\zeta_5}
\nonumber \\&
    + \frac{1078}{3} \Big)
    + \mathcal{D}_0  \Big( 12288\zeta_5
          - 1024\zeta_3
          - 6144\zeta_2\zeta_3 \Big)
    + \mathcal{D}_1 \Big( 256
          - 960\zeta_3
          + 1024\zeta_2
          - \frac{14208}{5}\zeta_2^2 \Big) \qquad \qquad
\nonumber \\&
    + \mathcal{D}_2 \Big( 10240 \zeta_3 \Big)
    + \mathcal{D}_3 \Big( - 512
          - 3072\zeta_2 \Big)
    + \mathcal{D}_5 \Big( 512 \Big) \nonumber
\,, \nonumber \\
\Delta^{b\bar{b}}_{(3,2)} &=
      \delta (1-z) \Big( - \frac{20816}{315} {\zeta_2}^3 - \frac{62468}{135} {\zeta_2}^2 + \frac{27872}{9} {\zeta_2} {\zeta_3}
    + \frac{22106}{27} {\zeta_2}
\nonumber \\
&
    + \frac{3280}{3} {\zeta_3}^2 - \frac{10940}{9} {\zeta_3} - \frac{37144}{9} {\zeta_5} - \frac{982}{3} \Big)
     + \mathcal{D}_0 \Big( \frac{6464}{27}
          + \frac{32288}{9}\zeta_3
          + \frac{6592}{27}\zeta_2
          - 1472\zeta_2\zeta_3
\nonumber \\&
          + \frac{1408}{3}\zeta_2^2 \Big)
     + \mathcal{D}_1 \Big(- \frac{544}{3} \quad
          - 5760\zeta_3
          - \frac{20864}{9}\zeta_2
          + \frac{3648}{5}\zeta_2^2 \Big)
     + \mathcal{D}_2 \Big( - \frac{10816}{9}
          + 1344\zeta_3
\nonumber \\&
          + \frac{11264}{3}\zeta_2 \Big)
     + \mathcal{D}_3 \Big( \frac{17152}{9}
          - 512\zeta_2 \Big)
     + \mathcal{D}_4 \Big(-\frac{ 7040}{9} \Big) \nonumber
\,, \nonumber \\
\Delta^{b\bar{b}}_{(3,3)} &=
      \delta (1-z) \Big( \frac{13264}{315} {\zeta_2}^3 + \frac{2528}{27} {\zeta_2}^2 - \frac{1064}{3} {\zeta_2} {\zeta_3}
    - 272 {\zeta_2} - \frac{400}{3} {\zeta_3}^2 - \frac{14212}{81} {\zeta_3}
\nonumber \\
&
    - 84 {\zeta_5} + \frac{68990}{81} \Big)
      + \mathcal{D}_0 \Big(- \frac{594058}{729}
          - 384\zeta_5
          + \frac{40144}{27}\zeta_3
          + \frac{98224}{81}\zeta_2
          - \frac{352}{3}\zeta_2\zeta_3
          - \frac{2992}{15}\zeta_2^2 \Big)
\nonumber \\&
      + \mathcal{D}_1 \Big( \frac{124024}{81}
          - 704\zeta_3
          - \frac{12032}{9}\zeta_2
          + \frac{704}{5}\zeta_2^2  \Big)
      + \mathcal{D}_2 \Big(  - \frac{28480}{27}
          + \frac{704}{3}\zeta_2 \Big)
      + \mathcal{D}_3 \Big( \frac{ 7744}{27} \Big) \nonumber
\,, \nonumber \\
\Delta^{b\bar{b}}_{(3,4)} &=
        \delta (1-z) \Big(\frac{24304}{135} {\zeta_2}^2 - \frac{11008}{9} {\zeta_2} {\zeta_3}
    - \frac{5200}{27} {\zeta_2} + \frac{8176}{9} {\zeta_3}
    + \frac{11072}{9} {\zeta_5} - \frac{140}{9}  \Big)
\nonumber \\&
       + \mathcal{D}_0 \Big(\frac{1684}{9}
          - \frac{11456}{9}\zeta_3
          - \frac{3008}{27}\zeta_2
          - \frac{2944}{15}\zeta_2^2  \Big)
       + \mathcal{D}_1 \Big(- \frac{368}{3}
          + 2560\zeta_3
          + \frac{6400}{9}\zeta_2 \Big) \qquad
\nonumber \\&
       + \mathcal{D}_2 \Big(\frac{3392}{9}
          - \frac{4096}{3}\zeta_2 \Big)
       + \mathcal{D}_3 \Big(- \frac{5120}{9} \Big)
       + \mathcal{D}_4 \Big(\frac{2560}{9} \Big) \nonumber
\,, \nonumber \\
\Delta^{b\bar{b}}_{(3,4a)} &=
        \delta (1-z) \Big( \frac{172}{9}
          - \frac{440}{9} \zeta_2 + \frac{64}{3}\zeta_3
          + \frac{128}{3} \zeta_2\zeta_3 \Big)
       + \mathcal{D}_0 \Big(\frac{6844}{27}
          - \frac{1216}{9}\zeta_3
          - 64\zeta_2
          - \frac{128}{5}\zeta_2^2  \Big) \qquad \quad
\nonumber \\&
       + \mathcal{D}_1 \Big(- \frac{880}{3}
          + 256\zeta_3 \Big)
       + \mathcal{D}_2 \Big( 64 \Big)\nonumber
\,, \nonumber \\
\Delta^{b\bar{b}}_{(3,4b)} &=
        \delta (1-z) \Big(- \frac{104}{3}
          + \frac{11072}{9}\zeta_5
          + \frac{7984}{9}\zeta_3
          - \frac{3880}{27}\zeta_2
          - \frac{11392}{9}\zeta_2\zeta_3
          + \frac{24304}{135}\zeta_2^2  \Big)
\nonumber \\&
       + \mathcal{D}_0 \Big(-\frac{1792}{27}
          - \frac{10240}{9}\zeta_3
          - \frac{1280}{27}\zeta_2
          - \frac{512}{3}\zeta_2^2  \Big)
       + \mathcal{D}_1 \Big(\frac{512}{3}
          + 2304\zeta_3
          + \frac{6400}{9}\zeta_2 \Big) \qquad \qquad
\nonumber \\&
       + \mathcal{D}_2 \Big(\frac{2816}{9}
          - \frac{4096}{3}\zeta_2 \Big)
       + \mathcal{D}_3 \Big(- \frac{5120}{9} \Big)
       + \mathcal{D}_4 \Big( \frac{2560}{9} \Big)\nonumber
\,, \nonumber \\
\Delta^{b\bar{b}}_{(3,5)} &=
       \delta (1-z) \Big( - \frac{13456}{135} {\zeta_2}^2 + \frac{416}{3} {\zeta_2} {\zeta_3}
+ \frac{6736}{81} {\zeta_2}
    + \frac{5104}{81} {\zeta_3} - 16 {\zeta_5} - \frac{23080}{81} \Big)
      + \mathcal{D}_0 \Big( \frac{250504}{729}
\nonumber\\&
          - \frac{4960}{9}\zeta_3
          - \frac{58784}{81}\zeta_2
          + \frac{1472}{15}\zeta_2^2 \Big)
      + \mathcal{D}_1 \Big( - \frac{65632}{81}
          + 768\zeta_2 \Big)
      +  \mathcal{D}_2 \Big( \frac{ 18496}{27}
          - \frac{256}{3}\zeta_2\Big)
\nonumber\\&
       +  \mathcal{D}_3 \Big(-\frac{5632}{27} \Big) \nonumber
\,, \nonumber \\
\Delta^{b\bar{b}}_{(3,6)} &=
       \delta (1-z) \Big( \frac{512}{27} {\zeta_2}^2 - \frac{128}{81} {\zeta_2} - \frac{4480}{81} {\zeta_3} + \frac{64}{27} \Big)
       + \mathcal{D}_0 \Big(- \frac{14848}{729}
          + \frac{1280}{27}\zeta_3
          + \frac{2560}{27}\zeta_2 \Big)
\qquad
      \nonumber \\&
       + \mathcal{D}_1 \Big(\frac{6400}{81}
          - \frac{1024}{9}\zeta_2 \Big)
       + \mathcal{D}_2 \Big( - \frac{2560}{27} \Big)
       + \mathcal{D}_3 \Big( \frac{1024}{27} \Big)\,,
\nonumber \\
\Delta_{(3,7)}^{b\bar{b}} &=0\,.
\end{align}
\end{small}

\section{ $C_{I,0}$ for QCD, QED and QCD$\times$QED up to N$^3$LO } \label{ap:g9}
Here we present $C_{I,0}$ in Eq.(\ref{eq:res}) with the following expansion in $a_s$ and $a_e$,
\begin{small}
\begin{align}
\label{eq:g0}
 C_{I,0} &= 1 + a_s  \Big[ C_F c_{1,1}^I \Big]
              + a_e  \Big[ e_I^2 c_{1,1}^I \Big]
              + a_s^2 \Big[ C_F^2 c_{2,1}^I + C_A C_F c_{2,2}^I + C_F n_f T_F c_{2,3}^I \Big]
\nonumber\\&   \quad          
              + a_s a_e  \Big[ 2 C_F e_I^2 c_{2,1}^I \Big]
              + a_e^2 \Big[ e_I^4 c_{2,1}^I + e_I^2 \Big( N \sum_q e_q^2 \Big) c_{2,3}^I \Big]
+a_s^3 \Big[ C_F^3 c_{3,1}^I + C_F^2 C_A c_{3,2}^I 
\nonumber\\&\quad
+ C_F C_A^2 c_{3,3}^I + C_F^2 n_f T_F c_{3,4}^I + C_F C_A n_f T_F c_{3,5}^I 
+ C_F n_f^2 T_F^2 c_{3,6}^I  + C_F N_{F,V} \Bigg( \frac{N^2 -4}{N} \Bigg) c_{3,7}^{I} \Big]
\nonumber\\& \quad
+ a_e^3 \Big[ e_I^6 c_{3,1}^I + e_I^2 \Big( N \sum_q e_q^4 + \sum_l e_l^4\Big) c_{3,4a}^I  + e_I^4 \Big( N \sum_q e_q^2 + \sum_l e_l^2\Big) c_{3,4b}^I
 + e_I^2\Big( N \sum_q e_q^2 
\nonumber\\& \quad
+ \sum_l e_l^2\Big)^2 c_{3,6}^I + 8 e_c^3 \bigg(N\sum_q e_q^3 + \sum_l e_l^3\bigg)c_{3,7}^{I} \Big]
+ a_s^2 a_e \Big[ 3 C_F^2 e_I^2 c_{3,1}^I + C_F C_A e_I^2 c_{3,2}^I
\nonumber\\& \quad
+ C_F T_F\Big(\sum_q e_q^2 \Big)c_{3,4a}^I
+ C_F n_f T_F e_I^2 c_{3,4b}^I \Big]
+ a_s a_e^2 \Big[ 3 C_F e_I^4 c_{3,1}^I + C_F e_I^2 \Big( N \sum_q e_q^2  \Big) c_{3,4a}^I
\nonumber\\& \quad
        + C_F e_I^2 \Big( N \sum_q e_q^2 + \sum_l e_l^2 \Big) c_{3,4b}^I \Big] \,.
\end{align}
\end{small}
As before $I$=$q,b$ for the Drell-Yan pair production and the Higgs boson production in bottom quark annihilation, respectively. For brevity we denote $\log \big(\frac{\mu_F^2}{\mu_R^2}\big) = L_{fr}$ and $\log \big(\frac{q^2}{\mu_R^2}\big) = L_{qr}$. The coefficients for the above color factors are,
\begin{small}
\begin{align}
c_{1,1}^q &= \bigg\{-16 + 8 \zeta_2 - \bigg( 6 \bigg) L_{fr} + \bigg(6 \bigg)  L_{qr} \bigg\} \,, \nonumber\\
c_{2,1}^q &= \bigg\{ \frac{511}{4} - 70 \zeta_2 + \frac{72}{5} \zeta_2^2  - 60 \zeta_3
+  \bigg(93 - 24 \zeta_2 - 48 \zeta_3 \bigg) \bigg( L_{fr} - L_{qr}\bigg) 
+ \bigg( -36 \bigg)L_{fr}L_{qr}
\nonumber\\& \quad
+  \bigg( 18 \bigg)\bigg( L_{fr}^2 + L_{qr}^2 \bigg)\bigg\} 
\,,  \nonumber\\
c_{2,2}^q &= \bigg\{ -\frac{1535}{12} + \frac{592}{9} \zeta_2  - \frac{12}{5} \zeta_2^2  + 28 \zeta_3 
+  \bigg( -\frac{17}{3} - \frac{88}{3} \zeta_2  + 24 \zeta_3 \bigg) L_{fr} 
+\bigg(\frac{193}{3} - 24 \zeta_3 \bigg) L_{qr} 
\nonumber\\& \quad
+  \bigg( 11 \bigg)\bigg( L_{fr}^2 - L_{qr}^2 \bigg) 
\bigg\}\,,  \nonumber\\
c_{2,3}^q &= \bigg\{ \frac{127}{3} - \frac{224}{9} \zeta_2 + 16 \zeta_3  
+  \bigg( \frac{4}{3} + \frac{32}{3} \zeta_2  \bigg) L_{fr} 
+\bigg(-\frac{68}{3} \bigg) L_{qr} 
+  \bigg( 4 \bigg)\bigg( L_{qr}^2 - L_{fr}^2 \bigg) 
\bigg\}\,,  \nonumber\\
c_{3,1}^q &= \bigg\{ -\frac{5599}{6} -\frac{130}{3}  \zeta_2 - \frac{612}{5} \zeta_2^2 + \frac{25696}{315} \zeta_2^3 - 460\zeta_3 + 80 \zeta_2 \zeta_3 + 32\zeta_3^2 + 1328\zeta_5 
	\nonumber\\&\quad
	+  \ \bigg( -\frac{1495}{2} - 24 \zeta_2 - \frac{48}{5} \zeta_2^2 
+ 992 \zeta_3 - 320 \zeta_2 \zeta_3 + 
 480 \zeta_5 \bigg) \bigg( L_{fr} - L_{qr}\bigg) 
	\nonumber\\&\quad
	+  \bigg( -270 + 288 \zeta_3 \bigg)\bigg( L_{fr}^2 + L_{qr}^2 \bigg) 
- 36 \bigg(L_{fr}^3  - L_{qr}^3\bigg)
+ \bigg( 540 - 576 \zeta_3 \bigg)L_{fr}L_{qr} 
	\nonumber\\&\quad	
	+ \bigg(-108 \bigg)\bigg(L_{fr}L_{qr}^2 - L_{fr}^2 L_{qr}\bigg)  \bigg\}\,,  \nonumber\\
c_{3,2}^q &= \bigg\{ \frac{74321}{36} - \frac{13186}{27} \zeta_2 + \frac{24064}{135} \zeta_2^2 - \frac{36944}{315} \zeta_2^3 - 
 \frac{34612}{27} \zeta_3 + \frac{3392}{9} \zeta_2 \zeta_3  + \frac{592}{3}\zeta_3^2 - \frac{5512}{9}\zeta_5
 \nonumber\\&\quad
+  \ \bigg( \frac{2348}{3} + \frac{908}{3} \zeta_2 
- \frac{1328}{15} \zeta_2^2 
- \frac{3344}{3} \zeta_3 + 
 160 \zeta_2 \zeta_3 - 240 \zeta_5 \bigg) L_{fr} 
 +\bigg( -\frac{3439}{2} + \frac{632}{3} \zeta_2
 \nonumber\\&\quad
  - \frac{256}{15} \zeta_2^2  
 + \frac{4664}{3} \zeta_3 
  - 160 \zeta_2 \zeta_3 
 + 240 \zeta_5 \bigg) L_{qr} 
 +  \bigg( -131 + 176 \zeta_2 + 32 \zeta_3 \bigg) L_{fr}^2 + \bigg( 551 
	\nonumber\\&\quad	
	- 320 \zeta_3 \bigg) L_{qr}^2 
+ \bigg( -420 - 176 \zeta_2 
+ 288 \zeta_3 \bigg)L_{fr}L_{qr} 
+\bigg( 66 \bigg) \bigg(L_{fr}^2L_{qr} + L_{qr}^2L_{fr} -L_{fr}^3 - L_{qr}^3   \bigg)\bigg\}
\,, \nonumber\\
c_{3,3}^q &= \bigg\{- \frac{1505881}{972} + 843\zeta_2 + \frac{14611}{135} \zeta_2^2 + \frac{13264}{315} \zeta_2^3 + \frac{82385}{81} \zeta_3 - 
 \frac{884}{3} \zeta_2 \zeta_3 - \frac{400}{3}\zeta_3^2 -204\zeta_5  
	\nonumber\\&\quad
	+  \ \bigg( \frac{1657}{18} - \frac{8992}{27} \zeta_2 + 4 \zeta_2^2 
+ \frac{3104}{9} \zeta_3 
- 80 \zeta_5 \bigg) L_{fr} 
+\bigg( \frac{3082}{3} - 240 \zeta_2 + \frac{68}{5} \zeta_2^2 - \frac{4952}{9} \zeta_3 
\nonumber\\&\quad
+ 80 \zeta_5 \bigg) L_{qr} 
+  \bigg( \frac{493}{9} + \frac{968}{9} \zeta_2 - 88 \zeta_3 \bigg) L_{fr}^2 
+ \bigg( -\frac{2429}{9} + 88 \zeta_3 \bigg) L_{qr}^2  
+\bigg( \frac{242}{9} \bigg) \bigg(L_{qr}^3 - L_{fr}^3   \bigg)\bigg\}
\,, \nonumber\\
c_{3,4}^q &= \bigg\{ -\frac{842}{3} + \frac{5264}{27} \zeta_2  - \frac{7136}{135} \zeta_2^2  - \frac{13904}{27} \zeta_3 - \frac{1792}{9} \zeta_2 \zeta_3 -\frac{1216}{9}\zeta_5  
+  \ \bigg( -\frac{550}{3} - \frac{112}{3} \zeta_2 
\nonumber\\& \quad
+ \frac{352}{15} \zeta_2^2  
+ \frac{256}{3} \zeta_3  \bigg) L_{fr} 
+\bigg( 460 - \frac{352}{3} \zeta_2 + \frac{224}{15} \zeta_2^2  - \frac{736}{3} \zeta_3  \bigg) L_{qr} 
+  \bigg( 40 - 64 \zeta_2 - 64 \zeta_3 \bigg) L_{fr}^2 
\nonumber\\& \quad
- \bigg( 184 - 64 \zeta_3 \bigg) L_{qr}^2 
+\bigg( 144 + 64 \zeta_2 \bigg) L_{fr}L_{qr} 
+ 24  \bigg(L_{qr}^3 + L_{fr}^3 - L_{qr}L_{fr}^2 - L_{qr}^2L_{fr} \bigg)\bigg\}
\,, \nonumber\\
c_{3,5}^q &= \bigg\{ \frac{221302}{243} - \frac{56264}{81} \zeta_2 - \frac{11512}{135} \zeta_2^2-\frac{12032}{81}\zeta_3  + \frac{416}{3}\zeta_2 \zeta_3 -16\zeta_5
+  \ \bigg( -80 + \frac{5344}{27} \zeta_2  
\nonumber\\&\quad
- \frac{16}{5} \zeta_2^2  - \frac{800}{9} \zeta_3 \bigg) L_{fr} 
+\bigg( -\frac{6104}{9} + \frac{640}{3} \zeta_2  - \frac{16}{5} \zeta_2^2  + \frac{416}{9} \zeta_3 \bigg) L_{qr} 
+  \bigg( -\frac{292}{9} - \frac{704}{9} \zeta_2 
\nonumber\\& \quad
+ 32 \zeta_3 \bigg) L_{fr}^2
+ \bigg( \frac{1700}{9} 
- 32 \zeta_3 \bigg) L_{qr}^2 
+\bigg( \frac{176}{9} \bigg) \bigg(L_{fr}^3 - L_{qr}^3 \bigg)\bigg\}
\,, \nonumber\\
c_{3,6}^q &= \bigg\{ -\frac{28324}{243} + \frac{9664}{81} \zeta_2  + \frac{512}{27} \zeta_2^2 - \frac{5056}{81}\zeta_3 
+  \ \bigg( \frac{136}{9} - \frac{640}{27} \zeta_2  + \frac{128}{9} \zeta_3  \bigg) L_{fr} 
+\bigg( \frac{880}{9} 
\nonumber\\& \quad
- \frac{128}{3} \zeta_2 + \frac{256}{9} \zeta_3 \bigg) L_{qr} 
+  \bigg( \frac{16}{9} + \frac{128}{9} \zeta_2  \bigg) L_{fr}^2 + \bigg( -\frac{272}{9}\bigg) L_{qr}^2 
+\bigg( \frac{32}{9} \bigg) \bigg(L_{qr}^3 - L_{fr}^3 \bigg)\bigg\}
\,, \nonumber\\
c_{3,4a}^q &= \bigg\{ \frac{3529}{9} - \frac{872}{9} \zeta_2 -\frac{96}{5}\zeta_2^2 - \frac{656}{3}\zeta_3 + \frac{128}{3} \zeta_2 \zeta_3 
+  \ \bigg( 74 + 32 \zeta_2 - 96 \zeta_3  \bigg) L_{fr} 
+\bigg( -138 
\nonumber\\&  \quad
+ 96 \zeta_3 \bigg) L_{qr}
+  \bigg(  12 \bigg) \bigg( L_{qr}^2 -  L_{fr}^2 \bigg)\bigg\} 
\,, \nonumber\\
c_{3,4b}^q &= \bigg\{ -\frac{6055}{9} + \frac{7880}{27} \zeta_2  - \frac{4544}{135} \zeta_2^2 + \frac{19808}{27}\zeta_3 - \frac{1216}{9} \zeta_5 - 
 \frac{2176}{9} \zeta_2 \zeta_3
+  \ \bigg( -\frac{772}{3} - \frac{208}{3} \zeta_2  
\nonumber\\& \quad
+ \frac{352}{15} \zeta_2^2  
+\frac{544}{3} \zeta_3   \bigg) L_{fr} 
+\bigg( 598 
- \frac{352}{3} \zeta_2  + \frac{224}{15} \zeta_2^2  - \frac{1024}{3} \zeta_3  \bigg) L_{qr} 
+\bigg( 144 + 64 \zeta_2 \bigg) L_{qr}L_{fr}
\nonumber\\& \quad
+  \bigg( 52 - 64 \zeta_2 - 64 \zeta_3  \bigg) L_{fr}^2 
+ \bigg( -196 + 64 \zeta_3 \bigg) L_{qr}^2  
+ 24  \bigg(L_{qr}^3 + L_{fr}^3  
\nonumber\\& \quad
- L_{fr}^2L_{qr} - L_{qr}^2L_{fr} \bigg)\bigg\} \,,
\nonumber\\
c_{3,7}^q & = 8 + 20\zeta_2 -\frac{4}{5}\zeta_2^2 + \frac{28}{3}\zeta_3 - \frac{160}{3}
\zeta_5
\end{align}
\end{small}
And for the bottom quark annihialtion $c_{i,j}^b$ at third order reads as,
\begin{small}
\begin{align}
\begin{autobreak}
   c_{1,1}^b  = 
   \big\{- 4
          + 8 \zeta_2 
          \big\}
       + L_{fr}  
        \big\{
          - 6
          \big\}
 \,,
\end{autobreak}
 \nonumber \\
\begin{autobreak}
   c_{2,1}^b =
       \bigg\{ 16
          - 60 \zeta_3
          + \frac{72}{5} \zeta_2^2 \bigg\}
       + L_{qr}   \big\{
           48 \zeta_3
          - 24 \zeta_2
          \big\}
       + L_{fr}   \big\{
           21
          - 48 \zeta_3
          - 24 \zeta_2
          \big\}
       + L_{fr}^2   \big\{
           18
          \big\} \,,
\end{autobreak}\nonumber \\
\begin{autobreak}
   c_{2,2}^b =
       \bigg\{ \frac{166}{9}
          - 8 \zeta_3
          + \frac{232}{9} \zeta_2
          - \frac{12}{5} \zeta_2^2 \bigg\}
       + L_{qr}   \big\{
          - 12
          - 24 \zeta_3
          \big\}
       + L_{fr}   \bigg\{
          - \frac{17}{3}
          + 24 \zeta_3
          - \frac{88}{3} \zeta_2
          \bigg\}
           + L_{fr}^2   \big\{
           11
          \big\} \,, 
\end{autobreak}\nonumber \\
\begin{autobreak}
   c_{2,3}^b =
       \bigg\{ \frac{16}{9}
          + 16 \zeta_3
          - \frac{80}{9} \zeta_2 \bigg\}
       + L_{fr}   \bigg\{
           \frac{4}{3}
          + \frac{32}{3} \zeta_2
          \bigg\}
       + L_{fr}^2   \big\{
          - 4
          \big\}  \,,
\end{autobreak}\nonumber \\
\begin{autobreak}
   c_{3,1}^b =
       \bigg\{\frac{1078}{3}
          -\frac{550}{3}\zeta_2
          - 1188 \zeta_3
          - 64 \zeta_2 \zeta_3
          + 32\zeta_3^2
          +848\zeta_5
          - \frac{104}{5} \zeta_2^2
          + \frac{25696}{315} \zeta_2^3
\bigg\}
       + L_{qr}   \bigg\{
          - 100
          - 480 \zeta_5
          - 56 \zeta_3
         + 132 \zeta_2
          + 320 \zeta_2 \zeta_3
          - \frac{384}{5} \zeta_2^2
          \bigg\}
       + L_{fr}   \bigg\{
          - 113
          + 480 \zeta_5
          + 416 \zeta_3
          - 156 \zeta_2
          - 320 \zeta_2 \zeta_3
         - \frac{48}{5} \zeta_2^2
          \bigg\}
       + L_{fr} L_{qr}   \big\{
          - 288 \zeta_3
          + 144 \zeta_2
          \big\}
       + L_{fr}^2   \big\{
          - 54
          + 288 \zeta_3
          \big\}
       + L_{fr}^3   \big\{
          - 36
          \big\} \,, 
\end{autobreak}\nonumber \\
\begin{autobreak}
   c_{3,2}^b =
       \bigg\{- \frac{982}{3}
          - \frac{6964}{27} \zeta_3
          + \frac{22106}{27} \zeta_2
          + \frac{2528}{9} \zeta_2 \zeta_3
          + \frac{592}{3}\zeta_3^2
          - \frac{3352}{9}\zeta_5
          - \frac{7348}{27} \zeta_2^2
          - \frac{36944}{315} \zeta_2^3
\bigg\}
       + L_{qr}   \bigg\{
           \frac{388}{3}
          + 240 \zeta_5
        + \frac{3296}{3} \zeta_3
          - 604 \zeta_2
          - 160 \zeta_2 \zeta_3
          - \frac{8}{3} \zeta_2^2
          \bigg\}
       + L_{qr}^2   \big\{
          - 176 \zeta_3
          + 88 \zeta_2
          \big\}
       + L_{fr}   \bigg\{
          - \frac{327}{2}
          - 240 \zeta_5
          - \frac{1832}{3} \zeta_3
          + \frac{572}{3} \zeta_2
          + 160 \zeta_2 \zeta_3
          - \frac{1328}{15} \zeta_2^2
          \bigg\}
       + L_{fr} L_{qr}   \big\{
           72
          + 144 \zeta_3
          \big\}
       + L_{fr}^2   \big\{
           1
          + 32 \zeta_3
          + 176 \zeta_2
          \big\}
       + L_{fr}^3   \big\{
          - 66
          \big\}  \,,
\end{autobreak}\nonumber \\
\begin{autobreak}
   c_{3,3}^b =
       \bigg\{ \frac{68990}{81}
          - 272 \zeta_2
          -\frac{14212}{81}\zeta_3
          - \frac{1064}{3} \zeta_2 \zeta_3
          + \frac{2528}{27} \zeta_2^2
          + \frac{13264}{315} \zeta_2^3
          -\frac{400}{3}\zeta_3^2 
          - 84\zeta_5 \bigg\}
       + L_{qr}   \bigg\{
          - \frac{1180}{3}
          + 80 \zeta_5
          - \frac{2576}{9} \zeta_3
           + \frac{160}{3} \zeta_2
          + \frac{68}{5} \zeta_2^2
          \bigg\}
       + L_{qr}^2   \big\{
           44
          + 88 \zeta_3
          \big\}
       + L_{fr}   \bigg\{
           \frac{1657}{18}
          - 80 \zeta_5
          + \frac{3104}{9} \zeta_3
          - \frac{8992}{27} \zeta_2
          + 4 \zeta_2^2
          \bigg\}
       + L_{fr}^2   \bigg\{
          \frac{493}{9}
          - 88 \zeta_3
          + \frac{968}{9} \zeta_2
          \bigg\}
       + L_{fr}^3   \bigg\{
          - \frac{242}{9}
          \bigg\} \,, 
\end{autobreak}\nonumber \\
\begin{autobreak}
   c_{3,4}^b =
       \bigg\{- \frac{140}{9}
          + \frac{17360}{27} \zeta_3
          - \frac{5200}{27} \zeta_2
          - \frac{1792}{9} \zeta_2 \zeta_3
          + \frac{16624}{135} \zeta_2^2
          -\frac{1216}{9}\zeta_5
\bigg\}
       + L_{qr}   \bigg\{
           \frac{16}{3}
          - \frac{1312}{3} \zeta_3
          + 144 \zeta_2
          + \frac{224}{15} \zeta_2^2  \bigg\}
       + L_{qr}^2   \big\{
           64 \zeta_3
          - 32 \zeta_2
          \big\}
       + L_{fr}   \bigg\{
          76
          + \frac{256}{3} \zeta_3
          - \frac{16}{3} \zeta_2
          + \frac{352}{15} \zeta_2^2
          \bigg\}
       + L_{fr}^2   \big\{
          - 8
          - 64 \zeta_3
          - 64 \zeta_2
          \big\}
       + L_{fr}^3   \big\{
           24
          \big\}
           \,, 
\end{autobreak}\nonumber \\
\begin{autobreak}
  c_{3,4a}^b =
       \bigg\{
        \frac{172}{9}
          - \frac{440}{9} \zeta_2
           + \frac{64}{3} \zeta_3
          + \frac{128}{3} \zeta_2 \zeta_3
\bigg\}
       + L_{fr}   \big\{
           74
          - 96 \zeta_3
          + 32 \zeta_2
          \big\}
       + L_{fr}^2   \big\{
          - 12
          \big\}
 \,, 
\end{autobreak}\nonumber \\
\begin{autobreak}
  c_{3,4b}^b =
       \bigg\{- \frac{104}{3}
          - \frac{3880}{27} \zeta_2
           + \frac{16624}{135} \zeta_2^2
           + \frac{16784}{27} \zeta_3
          - \frac{2176}{9} \zeta_2 \zeta_3
          -\frac{1216}{9} \zeta_5
         
\bigg\}
       + L_{qr}   \bigg\{
           \frac{16}{3}
          - \frac{1312}{3} \zeta_3
          + 144 \zeta_2
         + \frac{224}{15} \zeta_2^2
          \bigg\}
       + L_{qr}^2   \big\{
          64 \zeta_3
          - 32 \zeta_2
          \big\}
       + L_{fr}   \bigg\{
           2
          + \frac{544}{3} \zeta_3
          - \frac{112}{3} \zeta_2
          + \frac{352}{15} \zeta_2^2
          \bigg\}
       + L_{fr}^2   \big\{
          4
          - 64 \zeta_3
          - 64 \zeta_2
          \big\}
       + L_{fr}^3   \big\{
           24
          \big\}
 \,,
\end{autobreak} \nonumber \\
\begin{autobreak}
   c_{3,5}^b =
       \bigg\{- \frac{23080}{81}
          + \frac{6736}{81} \zeta_2
          + \frac{5104}{81}\zeta_3
          + \frac{416}{3} \zeta_2 \zeta_3
          - \frac{13456}{135} \zeta_2^2
          - 16\zeta_5
\bigg\}
       + L_{qr}   \bigg\{
          \frac{392}{3}
          + \frac{416}{9} \zeta_3
          - \frac{32}{3} \zeta_2
          - \frac{16}{5} \zeta_2^2
          \bigg\}
       + L_{qr}^2   \big\{
          - 16
          - 32 \zeta_3
          \big\}
       + L_{fr}   \bigg\{
          - 80
          - \frac{800}{9} \zeta_3
          + \frac{5344}{27} \zeta_2
          - \frac{16}{5} \zeta_2^2
          \bigg\}
       + L_{fr}^2   \bigg\{
          - \frac{292}{9}
          + 32 \zeta_3
          - \frac{704}{9} \zeta_2
          \bigg\}
       + L_{fr}^3   \bigg\{
          \frac{176}{9}
          \bigg\}
 \,,
\end{autobreak}\nonumber \\ 
\begin{autobreak}
  c_{3,6}^b =
       \bigg\{ 
       \frac{64}{27}
          - \frac{128}{81} \zeta_2
          + \frac{512}{27} \zeta_2^2
          - \frac{4480}{81}\zeta_3
\bigg\}
       + L_{qr}  
        \bigg\{
           \frac{256}{9} \zeta_3
          \bigg\}
       + L_{fr}   
       \bigg\{
           \frac{136}{9}
          + \frac{128}{9} \zeta_3
          - \frac{640}{27} \zeta_2
          \bigg\}
            + L_{fr}^2  
             \bigg\{
          \frac{16}{9}
          + \frac{128}{9} \zeta_2
          \bigg\}
       + L_{fr}^3   
       \bigg\{
          - \frac{32}{9}
          \bigg\} \,, 
\end{autobreak}\nonumber\\
\begin{autobreak}
    c_{3,7}^b =
    0\,.
\end{autobreak}
\end{align}
\end{small}

\bibliographystyle{JHEP}
\bibliography{QCDQED}
\end{document}